\documentclass[twocolumn]{aastex631}

\begin{document}

\title{On the evolutionary history of a simulated disc galaxy as seen by phylogenetic trees\footnote{Submitted on September, 11th, 2023}}

\author[0000-0002-2231-5113]{Danielle de Brito Silva}
\affiliation{Instituto de Estudios Astrof\'isicos, Facultad de Ingenier\'ia y Ciencias, Univesidad Diego Portales, Santiago de Chile}
\affiliation{Millenium Nucleus ERIS}
\correspondingauthor{Danielle de Brito Silva}
\email{danielle.debrito@mail.udp.cl}

\author{Paula Jofr\'e}
\affiliation{Instituto de Estudios Astrof\'isicos, Facultad de Ingenier\'ia y Ciencias, Univesidad Diego Portales, Santiago de Chile}
\affiliation{Millenium Nucleus ERIS}

\author{Patricia B. Tissera}
\affiliation{Instituto de Astrof\'sica, Pontificia Universidad Cat\'olica de Chile, Av. Vicuña Mackenna 4860, Santiago, Chile}
\affiliation{Centro de Astro-Ingenier\'ia, Pontificia Universidad Cat\'olica de Chile, Av. Vicuña Mackenna 4860, Santiago, Chile}
\affiliation{Millenium Nucleus ERIS}

\author{Keaghan J. Yaxley}
\affiliation{Ecology and Evolution, Research School of Biology, Australian National University}

\author{Jenny Gonzalez Jara}
\affiliation{Instituto de Astrof\'sica, Pontificia Universidad Cat\'olica de Chile, Av. Vicuña Mackenna 4860, Santiago, Chile}

\author{Camilla J. L. Eldridge}
\affiliation{Instituto de Estudios Astrof\'isicos, Facultad de Ingenier\'ia y Ciencias, Univesidad Diego Portales, Santiago de Chile}

\author{Emanuel Sillero}
\affiliation{Instituto de Astrof\'sica, Pontificia Universidad Cat\'olica de Chile, Av. Vicuña Mackenna 4860, Santiago, Chile}

\author{Robert M. Yates}
\affiliation{Centre for Astrophysics Research, University of Hertfordshire, Hatfield, AL10 9AB, UK}

\author{Xia Hua}
\affiliation{Mathematical Sciences Institute, Australian National University, Canberra ACT 0200 Australia}

\author{Payel Das}
\affiliation{Physics Department, University of Surrey, Guildford GU2 7XH, United Kingdom}

\author{Claudia Aguilera-Gómez}
\affiliation{Instituto de Astrof\'sica, Pontificia Universidad Cat\'olica de Chile, Av. Vicuña Mackenna 4860, Santiago, Chile}

\author{Evelyn J. Johnston}
\affiliation{Instituto de Estudios Astrof\'isicos, Facultad de Ingenier\'ia y Ciencias, Univesidad Diego Portales, Santiago de Chile}
\affiliation{Millenium Nucleus ERIS}

\author{Alvaro Rojas-Arriagada}
\affiliation{Departamento de F\'isica, Universidad de Santiago de Chile, Av. Victor Jara 3659, Santiago, Chile}
\affiliation{Millennium Institute of Astrophysics, Av. Vicu\~{n}a Mackenna 4860, 82-0436 Macul, Santiago, Chile}
\affiliation{Center for Interdisciplinary Research in Astrophysics and Space Exploration (CIRAS), Universidad de Santiago de Chile, Santiago, Chile}
\affiliation{Millenium Nucleus ERIS}

\author{Robert Foley}
\affiliation{Leverhulme Centre for Human Evolutionary Studies, Department for Anthropology and Archaeology, University of Cambridge, CB2 1QH, UK}

\author{Gerard Gilmore}
\affiliation{Institute of Astronomy, Cambridge, UK}
\affiliation{Institute for Astrophysics, FORTH, Crete}

\begin{abstract}

Phylogenetic methods have long been used in biology, and more recently have been extended to other fields – for example, linguistics and  technology – to study evolutionary histories. Galaxies also have an evolutionary history, and fall within this broad phylogenetic framework. Under the hypothesis that chemical abundances can be used as a proxy for interstellar medium's DNA, phylogenetic methods allow us to reconstruct hierarchical similarities and differences among stars – essentially a tree of evolutionary relationships and thus history. In this work, we apply phylogenetic methods to a simulated disc galaxy obtained with a chemo-dynamical code to test the approach. We found that at least 100 stellar particles are required to reliably portray the evolutionary history of a selected stellar population in this simulation, and that the overall evolutionary history is reliably preserved when the typical uncertainties in the chemical abundances are smaller than 0.08 dex. The results show that the shape of the trees are strongly affected by the age-metallicity relation, as well as the star formation history of the galaxy. We found that regions with low star formation rates produce shorter trees than regions with high star formation rates. Our analysis demonstrates that phylogenetic methods can shed light on the process of galaxy evolution.

\end{abstract}

\keywords{Galaxy abundances --- Galaxy stellar content ---  Interdisciplinary astronomy}

\section{Introduction} \label{sec:intro}

Several areas of evolutionary science investigate evolutionary histories with phylogenetic methods, including biology, language and astronomy \citep{Baum2005, gray2009language,ricker2014transiting, jofre2017cosmic, yaxley2019reconstructing, jackson2021using,bromham2022global}. Phylogenetic methods were originally developed in the context of biology studies, when Charles Darwin described patterns of descent among organisms as an evolutionary tree \citep{Darwin1859}. It was a century later that the DNA was identified as the information that is passed from one generation to the next, connecting the different life forms in the hierarchical way that Darwin had illustrated. This happens  because the DNA replication between progenitor and offspring is not perfect, e.g. the new DNA is modified. Modifications accumulate over time, causing the life forms to differ more with time. If one population is divided and each subgroup is isolated, their evolution and cumulative modification will occur independently. This process is named diversification and produces a hierarchy. Nowadays, DNA is widely used as input to build phylogenetic trees allowing the exploration of shared evolutionary histories of an immense variety of living organisms \citep{Bromham2008, yang2014molecular}.

This approach considers two main concepts. The first concept is {\it heritability} and the second is {\it descent with modification}. Heritability considers that there is information passed from one generation to the next one. Descent with modification stands for the knowledge that a characteristic transferred from one generation to the next one suffers small changes. These changes accumulate over time and if there is also diversification, an hierarchy in similarity is formed. Due to hierarchical similarity, related organisms have more similar characteristics.

Chemical evolution of galaxies respects both the concept of heritability and descent with modification. Chemical evolution in galaxies is linked to stellar nucleosynthesis \citep{Burbidge1957, Tinsley1979, matteucci2012chemical}. At the last stages of evolution, stars pollute the interstellar medium (ISM) with the chemical elements they synthesized during their lifetimes, causing the modification of the chemical composition of the ISM of their parent galaxy. The enriched ISM will later give origin to new generations of stars that are chemically altered with respect to the previous generation. A large fraction of the stars formed in each episode are low-mass objects, hence they live longer than this cycle of new stars forming and their atmospheres preserve the chemical composition of their birth environment. In this way, chemical abundances of low-mass stars can be considered as a proxy for the ISM's DNA \citep{Freeman2002} and are very important to unveil the history of the Galaxy.

Luckily, chemical abundances in industrial scale are now available, which is revolutionizing the field of Galactic archaeology, both due to direct discoveries from the data, but also because they are necessary to validate chemical evolution models. In particular, thanks to surveys such as GALAH \citep{buder2020galah+}, APOGEE \citep{majewski2017apache,Abolfathi2018,holtzman2018apogee} and Gaia (\citealt{GaiaCollaboration+2016a,GaiaCollaboration+2016b,gaia2018gaia,brown2021gaia,eyer2022gaia,recio2023gaia}), chemical abundances up to millions of stars are now available to better explore the processes that shaped the Galaxy.

As an example of the power of chemical abundances to unveil the past of the Milky Way it is possible to remark the on-going extensive search for the building blocks of the Milky Way. \cite{nissen2010two} found two different sequences in halo stars: one sequence containing stars enhanced in $\alpha$ elements (attributed to an ancient disc or bulge, which had its orbit heated due to a past merger event) and another sequence $\alpha$-poor (an accreted dwarf galaxy). \cite{hawkins2015using} found a population of $\alpha$-poor stars with abundances of Al, C+N and Ni which is different from $\alpha$-rich stars, indicating that the population had a different chemical enrichment history from the bulk of the Milky Way. Later works found evidences of a major merger event using, among other information, chemical abundances. This major merger event is believed to have occurred between the Milky Way and a galaxy whose remnant stellar population is now known as the Gaia Enceladus Sausage (GES) (\citealt{Helmi2018,Belokurov2018}). \cite{carrillo2022detailed} studied the chemical abundances of 62 stars accreted from GES, considering a wide wavelength range from the optical to the infrared. They report that accreted stars have enhanced neutron capture abundances when compared with Milky Way stars, in particular of Eu, indicating differences in the chemical evolution of GES when compared with the Milky Way (see also \citealt{matsuno2020star,aguado20,deBritoSilva2022_j01020100}, de Brito Silva in.prep.). \cite{buder2022galah} used GALAH chemical abundances to study accreted stars and concluded that they are chemically different from stars born in \textit{situ} in terms of Cu, Mg, Si, Na, Al, Mn, Fe and Ni. \cite{horta2022chemical} used data of Gaia and APOGEE to characterize 12 halo substructures, candidates to have accreted origins. We note that these are only a few examples, but other numerous works have done remarkable contribution to this topic.

It is undeniable how important chemical abundances are in order to understand the evolution of the Milky Way. However, several open questions still remain, such as the unknown number building blocks (i.e. accreted galaxies) that constitute the Milky Way. The building blocks are also not fully characterized. Their detailed chemical abundances distributions, masses, star formation histories and age-metallicity relations are still not defined. Some of the accreted stellar populations attributed to different progenitor galaxies could actually be from the same galaxy, considering the caveats associated with their selection (see \citealt{horta2022chemical,buder2022galah}). Currently, multiple works are starting to approach these questions using numerical simulations (e.g. \citealt{bignone2019gaia, monachesi2019auriga,amarante2022gastro,carrillo2023can}). In this paper we resort to a novel approach to contribute to answering open questions in Galactic Archaeology by applying phylogenetic concepts to galaxy formation.

Phylogeny applied to chemistry of low-mass stars can be referred to as stellar phylogeny. It was proposed in \cite{jofre2017cosmic}, where the authors used 17 chemical elements to perform a phylogenetic study of 22 solar neighbourhood stars. They found three groups that had different chemical enrichment rates measured from the relation between the age and other phylogenetic properties. A second stellar phylogenetic study of the Milky Way was performed in \cite{jackson2021using}, where they used 78 solar neighbourhood stars and 30 chemical elements to explore the  Milky Way disc. The goal of that study was to test if more stars and elements would help to understand how the three groups found in \cite{jofre2017cosmic} were related to each other. With the aid of new Gaia data \citep{GaiaCollaboration+2016a,GaiaCollaboration+2016b,gaia2018gaia,brown2021gaia,eyer2022gaia}, they proposed that one of the three groups was an ancestral population of the groups associated to the thin disc, having a significantly higher star formation rate due to perhaps a starburst during the first epochs of the thin disc formation.

While studies have explored stellar phylogenies in observed data, using simulated data has become key to help the interpretation of trees. The advantage of working with numerical simulations for these purposes is that they provide the full evolution of baryons as the gas is transformed into stars and chemical elements are produced and injected into the interstellar medium where the stars evolve. Since the chemical evolution is known and the simulated stellar populations can be traced back in time,  phylogenetic trees built from the simulated stellar populations can be directly compared to the true evolution, to learn which particular features of the trees can be related to events in the formation and evolution of galaxies. In this paper we propose to use simulated galaxies to advance in the development of stellar phylogeny.

In addition, simulations allow the assessment of the maximum chemical abundance uncertainties for which phylogenetic signal is sufficiently preserved to provide phylogenetic trees that portray reliable evolutionary histories. Furthermore, with simulated data it is possible to assess for selection effects, since we have information about the entire galaxy. 

Stellar phylogeny is still a very new approach and multiple questions about its applicability and interpretation remain open. Some of these questions can be best addressed by using simulations of galaxies.
In this work, we use for the first time phylogenetics applied to a simulated disc galaxy in order to answer three specific questions: First, how many stellar particles are required to build phylogenetic trees that robustly portray the evolutionary history of this simulated galaxy? Second, how do the uncertainties in the chemical abundance data impact the robustness of the evolutionary history represented by phylogenetic trees?  
And third, can phylogenetic trees from different regions of a simulated galaxy, which have different histories of formation, illustrate the different evolutionary histories? 

In Section \ref{sec:phylogenetic_methodology} we describe how the phylogenetic trees are built and how we compare them. In Section \ref{sec:simulated_data} we describe the simulation used in this work as well as the selection of stellar particles used to approach the different specific questions proposed. In Section \ref{sec:results}, we present the results and interpretation of our findings. Finally, in Section \ref{sec:discussion_conclusion} we present our summary and conclusions.

\section{Phylogenetic tree construction and analysis}{\label{sec:phylogenetic_methodology}}

In this section we describe how the phylogenetic trees are built and compared. An exhaustive analysis of the suitability of phylogenetic trees for the reconstruction of the ISM history is given by Eldridge et al (in prep).

\subsection{Tree concepts}{\label{sec:tree_interpretation}}

To interpret the phylogenetic trees presented in this paper, we focus on key concepts from the trees which involve the branching pattern, the root, and the branch lengths. Extensive explanations of these concepts and their applicability can be found in the seminal books on trees and phylogenetics such as  \cite{FelsensteinBook2004, Hall2004,Lemey2004,Baum2005} and \cite{yang2014molecular}.

The branching pattern is related to the structure or topology of the tree. In biology the tips represent present-day species, while the internal nodes represent the last common ancestor of all the tips which descend from it. In our case, tips represent the stellar particles, which are stellar populations with a given age and chemical abundances. Most of these stellar particles are fossil records of an ISM which is now extinct.  

The ancestral form of all objects considered in a tree is the root. We note that building a tree with the algorithm we used does not provide a rooted tree, even if many tree reconstruction methods might display trees in rooted form. To root a phylogenetic tree is a delicate procedure, because depending on the root chosen, the ancestor-descendant temporal relationship of the tree changes and so the reconstruction of the history. There are few ways to find the root, but most of them rely on an evolutionary model developed for biology. As a consequence we need to consider an alternative approach. Since we are working with a simulated galaxy, and therefore we know the origin of each stellar particle, we can consider the most ancient ones which exist as soon as the ISM started evolving due to chemical enrichment for rooting. Therefore, we set the outgroup as the most ancient stellar particle in the simulation which is related to the ingroup (all other sampled particles) and  place the root in the branch that connects that ancient stellar particle with the rest of the tree.  

The length of a branch represents the amount of chemical change or chemical divergence between nodes. A tree showing only the topology without the branch length information can be referred to as a cladogram, while a tree which specifies the branch lengths can be referred to as phylogram. This is important here, because that differs from the usage of dendograms or some other mathematical tree graphs widely used in astronomy to perform data analysis such as clustering or classifications (HDBSCAN by \cite{campello2013density}, t-SNE by \cite{van2008visualizing}, random forest by \cite{ho1995random}, for example). We can associate a relation of branch length and the age between two tips or between the root and the tips as a measure to the chemical enrichment rate \citep[see also][]{jofre2017cosmic}. 

\subsection{Building phylogenetic trees}{\label{sec:build_trees}}

We use the same methodology thoroughly described in \cite{jackson2021using}, which was adapted from \cite{jofre2017cosmic}. Briefly, it consists of three steps: (i) selection of evolutionary traits; (ii) building the phylogenetic tree; (iii) evaluating its robustness.

Encoding evolutionary traits is fundamental, since this has a direct impact on the tree topology and its interpretation. In modern biology, most trees are inferred from sequences of DNA, with each site in the sequence acting as an independent and discrete observation \citep{Drummond2007,Maddison2009,Hall2013}. In our case, chemical abundances of stars are continuous. Fortunately, there are methods that uses distances matrices and it is possible to calculated distances from continuous data.

Distance matrices are used to quantify the differences of traits between observations. In the case of our study, our traits are the chemical abundances of each single stellar population as mentioned above (see also Section \ref{sec:simulations: UNDER CONSTRUCTION --}) which in the simulations is represented by a stellar particle. The distance matrix is formed by the difference in chemical abundance (or chemical distance) of all the stellar particles we used to build a tree in relation to all the other particles. In order to calculate the pairwise distance of the stellar particles, we used the Euclidean distance. The total chemical distance between the stellar particles \textit{i} and \textit{j} was calculated as $D_{\mathrm{i,j}}= \sum_{k=1}^{N}{\sqrt{\mathrm{([X_{k}/H]_{i})^{2} - ([X_{k}/H]_{j})^{2}}}}$. For more details about chemical distances and distance matrices we refer to \cite{jofre2017cosmic}. 

From the distance matrices, the phylogenetic trees are built with the Neighbor-Joining (NJ, \citealt{saitou1987neighbor,Gascuel2006}) algorithm, which assesses the distances to find the most probable evolutionary sequence. This algorithm, unlike others available in the literature, does not compel equal distance between the root of the tree and any of the tips. This is an important consideration, because it is known that chemical evolution differs from place to place and from chemical element to chemical element  (e.g. \citealt{matteucci2012chemical,maiolino2019re,johnson2022dwarf}). Apart from this assumption that agrees with our knowledge of chemical evolution of galaxies, NJ methods can be used to infer phylogenies from distance matrices \citep{atteson1997performance,Kuhner1994,Lemey2004,Mihaescu2009, jofre2017cosmic,jackson2021using}. The NJ method has the advantage to be very fast and simple to implement, which satisfies our needs, since we aim to empirically test phylogenetic approaches in a dataset which is not one governed by the biological law of evolution. For more fundamental discussion about the usage of NJ trees in galaxy evolution, we refer to Eldridge et al (in prep).

\subsection{Comparing phylogenetic trees}{\label{sec:compare_trees}}

\subsubsection{Robinson-Foulds Distance}

One common method to compare trees is the widely-used measure of topological distance between two trees is defined by \cite{robinson1981comparison}, which is referred to as Robinson-Foulds distance (hereafter RFD). 

The RFD evaluates how similar two trees are by matching the similarity between a partition or split in one tree and its pair on the second tree.  
The partition distance is defined as the total number of splits that exist in one tree but not on the other. It can be equivalently defined as the number of contractions and expansions needed to transform one tree into the other. Removing an internal branch by reducing its length to zero is a contraction, while creating an internal branch is an expansion. For a rooted tree with $n$ tips and $(n-2)$ internal nodes, the partition distance ranges between 0 and $D_{\mathrm{max}}=2(n-2)$ \citep[see][for extensive discussion]{yang2014molecular}.  The RFD considers a performance parameter $P = 1 -D/D_{\mathrm{max}}$  to assess the similarity between trees. We note that the RFD varies between 0 to 1, where the smaller the value, more similar two phylogenetic trees are.

There are a few limitations on using the RFD. First, as  it only focuses at splits in the trees, it does not consider the branch length as a information for similarity. 
Second, some deep relationships in the tree might be neglected for trees in which splits of outer nodes are different despite sharing internal nodes. This implies that while the performance of RFD ranges between zero and one, two random trees normally differ by 80\%.  

We comment that the RFD parameter can only be calculated for a tree built from the same set of objects. It serves thus to compare different input data, but not to compare different set of objects, since the identification of splits in different trees can not be matched.   In order to calculate the RFD we used the R library {\tt treedist}\footnote{see \url{https://cran.r-project.org/web/packages/TreeDist/TreeDist.pdf} for details.} \citep{Smith2020information,Smith2022robust,TreeDist} and the module {\tt TreeDistance}, which follows \cite{Smith2020information} and uses the concepts of entropy and information described in \cite{mackay2003information}. 

\subsubsection{Consensus tree}{\label{sec:consensus_tree}}
While tree distances are a measure of how different trees are, consensus trees summarise common features about a collection of trees. In the same way as the RFD, the consensus tree can be obtained when the set of objects used to build trees is the same. 

In this work we consider the majority-rule consensus tree, which shows the branches and splits that are present in the majority of the trees. Majority is defined as more than 50\%. A consensus tree is a summary tree which essentially selects the nodes that appear in at least half of the trees,  and rejects all other nodes. Rejected nodes are transformed in {\it polytomies}, e.g. there are more than 2 branches connecting a given node with a tip \citep{Baum2005}.  There are two types of polytomies: hard and soft. Hard polytomies are associated with multifurcations in the tree, while soft polytomies are associated with unresolved relationships in the tree. Soft polytomies are an indication of lower phylogenetic resolution in the tree.  Hence, polytomies can imply a particular extreme event which might give rise to several evolutionary paths but in a consensus tree they might illustrate lack of accuracy in the data to solve the branching pattern of the historical events. Therefore, while consensus trees are not ideal to study the evolutionary history of a galaxy, they are extremely useful to study the global properties of a set of phylogenetic trees, since they display their common features.

It is worth noting that polytomies in a consensus tree are a way to illustrate uncertainties, and do not represent a particular evolutionary event that could cause a large divergence of lineages.  It is therefore not encouraged to interpret evolutionary histories with consensus trees because the polytomies easily lead to wrong interpretations.

\begin{figure*}
\centering
\includegraphics[width=16cm]{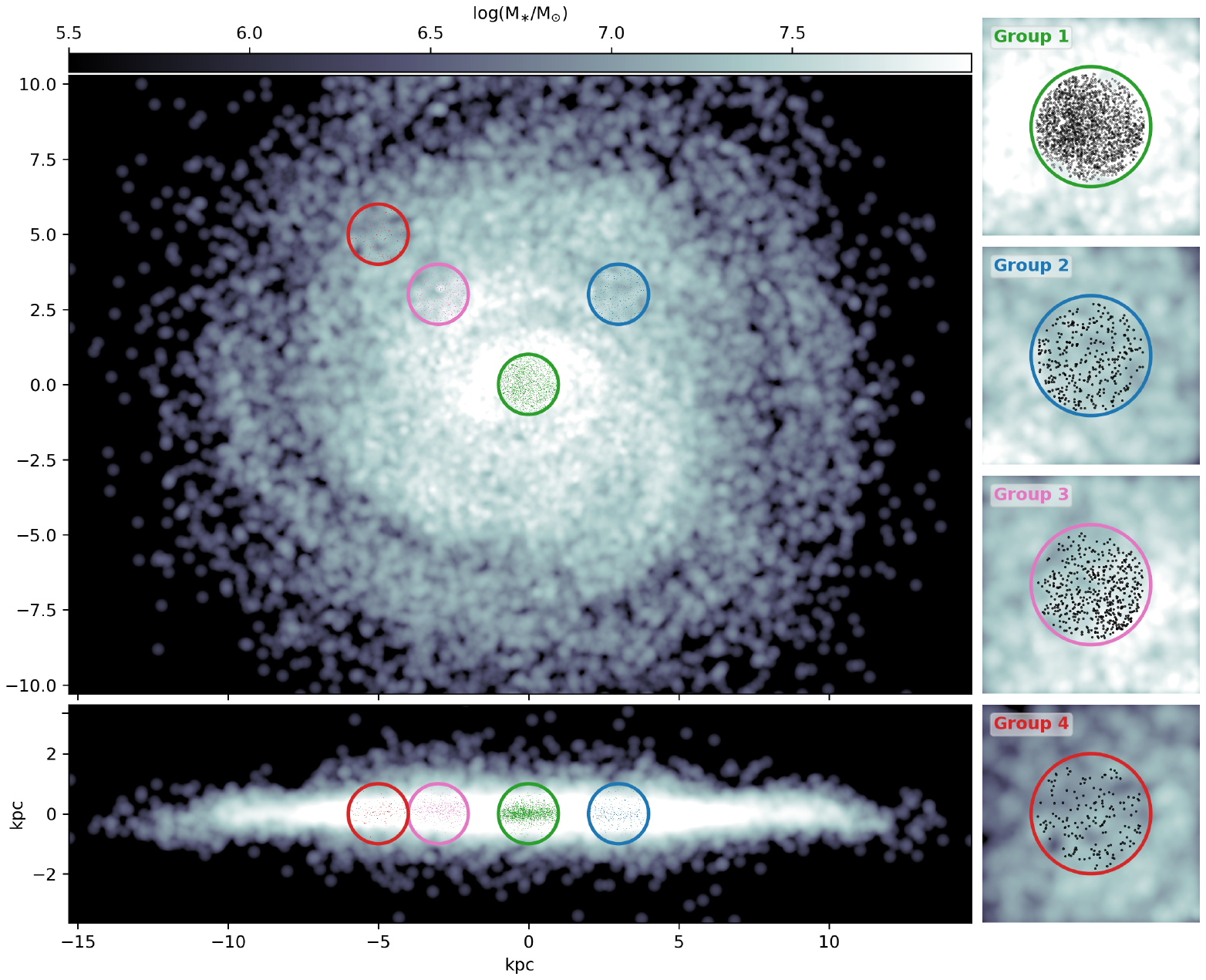}
\caption{Face-on (upper panel) and edge-on (lower panel) spatial distribution of the stellar populations in the four defined groups: Group 1 (green), Group2 (blue), Group 3 (pink) and Group 4 (red). The gray points represent the whole distribution of stellar populations in the simulated galaxy. We note that volumes mapped by the selected groups represent a sphere of 1kpc radius.}
\label{figure:position_regions}
\end{figure*}

\section{Simulated data}{\label{sec:simulated_data}}

In this work we use data of a simulated disc galaxy. The information available from the simulation will be used to characterize the level of agreement between the evolutionary history traced by the phylogenetic trees and the history of the simulated galaxy. This way we will take numerous advantages of the information provided by using hydrodynamical simulations. First, chemical abundances and ages for a large number of stellar particles are available. This allows the consideration of selection biases that are common when working with observed data. Second, it provides the opportunity to examine in detail the place and time different stellar particles were formed, which allow the assessment of the reliability of the phylogenetic trees to assign connections. Finally, the simulation provides information about the galaxy studied, from its star formation rate (SFR) through time, to its age-metallicity relation (AMR) and the nucleosynthetic channels that produce different chemical elements. Therefore, by using simulated data we can build phylogenetic trees for which reverse-engineering of the evolutionary history traced is possible.

\subsection{Simulations}{\label{sec:simulations: UNDER CONSTRUCTION --}}

For this paper, we use a pre-prepared simulation of an isolated disc galaxy. This simple initial condition allows us to perform the construction and analysis of the phylogenetic trees in a system which
does not receive material  (gas inflows or mergers) from the the surroundings. It is simple enough to be used as a first test-bed for phylogenetic trees. Therefore this simulated disc galaxy is not expected to represent a real galaxy. From this starting point, we will  build up more complex galaxy formation scenarios until reaching maturity in the technique to  adequately apply phylogenetic trees in a cosmological context in future works.

The analysed simulation was performed by using a version of P-GADGET-3 code \citep{springel2005}, which includes a multiphase model for the gas component, metal-dependent cooling, star formation and supernova feedback, as described in  \citet{scan05} and \citet{scan06}. A Chabrier Initial Mass Function is assumed with a lower and upper mass cut off of  $0.1$ and $40$ M$_\odot$ respectively, \citep{Chabrier2003}.

The chemical evolution model includes the enrichment by Type Ia (SNIa) and Type II (SNII) Supernovae \citep{mosconi2001,scan06}. The SNIa events are assumed to originate from CO white dwarf (hereafter CO WD) binary systems, in which the explosion is triggered when the primary star, due to mass transfer from its companion, exceeds the Chandrasekhar limit. For simplicity, the lifetime of the progenitor systems  (delay times) are assumed to be randomly distributed over the range [0.7, 1.1] Gyr. This simple model for the lifetime distribution produces consistent results with the single-degenerated model  \citep{Jimenez2015}. The nucleosynthesis yields of 
SNIa corresponds to \citet{iwamoto99}.
SNII originate from massive stars with lifetimes  estimated according to \citet{Raiteri1996}. Their nucleosynthesis products are derived from the metal-dependent yields of \citet{WW95}. 
The chemical model traces the following 12 different chemical elements: H (hydrogen), $^4$He (helium), $^{12}$C (carbon), $^{14}$N (nitrogen), $^{16}$O (oxygen), $^{20}$Ne (neon), $^{24}$Mg (magnesium), $^{28}$Si (silicon), $^{32}$S (sulfur), $^{40}$Ca (calcium),$^{56}$Fe (iron) and $^{62}$Zn (zinc). Initially, the gas component is assumed to have primordial abundances i.e. X$_{\rm H} = 0.76$, Y$_{\rm He} = 0.24$ and Z $=0$.

The initial conditions correspond to a disc galaxy composed of a dark matter (DM) halo, a stellar bulge component and an exponential disc, with a total baryonic mass of $\rm m_{\rm b} \sim 5.2 \times 10^{10} M_{\sun}$. The halo and bulge components were modelled by an NFW profile \citep{navarro1996structure} and a Hernquist profile \citep{hernquist1990analytical}, respectively. The gas component is distributed in the disc and accounts for 50\% of the total disc mass. The initial gas mass particle is  $\rm m_{\rm gas} = 1.96\times 10^{5} M_{\sun}$. The gravitational softening (i.e. a numerical length introduced to avoid unrealistic gravitational forces during particles close encounters) adopted is 200 pc for the gas and star particles and 320 pc for the dark matter component.

Each stellar particle represents a single stellar population with the same age and chemical abundances. Hereafter we will use the standard definition  $[X/H] = (log_{10} \mathrm{X_{*}/H_{*})} - (log_{10} \mathrm{X_{\sun}/H_{\sun})}$, where X and H are the abundance of the element X and H, respectively. Hence, for each stellar particles, abundances can be defined by combining the chemical elements described above.

\subsection{Data}{\label{sec:data}}

The simulated galaxy has a strong initial starburst, that while widely spread, is more intense in its central region. After the initial starburst, the star formation activity decreases. We chose to follow the evolution of the system until this time as this allows SNIa to take place in the simulation. Since the simulation starts with primordial gas, the first stellar particles that formed will have $Z=0$, where Z is the so-called metallicity which quantifies the abundances of elements heavier than He. However, this simulation does not include a model for the formation of such stellar particles, which are known to be different from second generation ones. Considering this and the fact that chemical abundances are the input parameters to build phylogenetic trees, we excluded the stellar particles that have been formed from primordial gas. The stellar particles selected for the analysis have ages  $\leq$ 1.5 Gyr and $-3.0 \leq \mathrm{[Fe/H]} \leq 0.5$ approximately. We used these particles to create different sub-samples that were used to explore the different specific questions concerning this analysis.

\begin{table*}
\centering
\caption{Description of the different samples of stellar particles used in this work. We note that the total number of stellar particles (sample size) refers to the global number of the entire sample and not the number of stellar particles used to build the phylogenetic trees. All of the spheres used to select Groups 01, 02, 03 and 04 and the deterministic sample have a radius of 1 kpc.}
\label{tab:basic_info_samples}
\begin{tabular}{lllll}
\hline
  \multicolumn{1}{|c|}{Sample name} &
  \multicolumn{1}{c|}{Description} &
  \multicolumn{1}{c|}{Sample size} &
  \multicolumn{1}{c|}{Used in} \\
\hline
\hline
Deterministic & \begin{tabular}[c]{@{}l@{}}Sphere of 1 kpc of radius centered \\ at the position (0,0,0). We only \\ consider stellar particles which \\ progenitor gas particles were \\ inside the sphere in the beginning \\ of the simulation and have \\ remained within  the same  \\ region since they were born.\end{tabular} & 761 & \begin{tabular}[c]{@{}l@{}}Section 4.2 Phylogenetic signal \\ in numerical simulations \\ Section 4.3 Phylogenetic signal \\ considering  uncertainties \end{tabular} \\
\hline
Noise & \begin{tabular}[c]{@{}l@{}} Built using chemical abundances \\ randomly created, without any \\ astrophysical meaning. The \\ synthetic chemical abundances \\ created respect the range of the \\ distribution as observed in the \\ simulation. \end{tabular}  & 10, 50, 100 and 200 & \begin{tabular}[c]{@{}l@{}} 4.2 Phylogenetic signal in \\ numerical simulations \end{tabular} \\
\hline
Group 01 & \begin{tabular}[c]{@{}l@{}}Sphere with center at (0,0,0), without \\ the other constraints considered in the \\ deterministic sample (birth place, \\ location of progenitor gas particle) \end{tabular} & 2365 & \begin{tabular}[c]{@{}l@{}}Section 4.4 Evolutionary history \\ considering different regions of \\ the galaxy \end{tabular} \\
\hline
Group 02 & \begin{tabular}[c]{@{}l@{}}Sphere with center at  (3,3,0) \end{tabular} & 324 & \begin{tabular}[c]{@{}l@{}} Section 4.4 Evolutionary history \\ considering different regions of \\ the galaxy \end{tabular}\\
\hline
Group 03 & \begin{tabular}[c]{@{}l@{}}Sphere with center at  (-3,3,0) \end{tabular} & 478 & \begin{tabular}[c]{@{}l@{}}Section 4.4 Evolutionary history \\ considering  different regions of \\ the galaxy \end{tabular}\\
\hline
Group 04 & \begin{tabular}[c]{@{}l@{}}Sphere with center at (-5,5,0) \end{tabular} & 159 & \begin{tabular}[c]{@{}l@{}} Section 4.4 Evolutionary history \\ considering different regions of \\ the galaxy \end{tabular}\\
\hline
\hline
\end{tabular}
\end{table*}

\subsubsection{Stellar samples}{\label{sec:stellar_samples}}

For our study, we perform different selections of the stellar particles from different regions of the simulated disc galaxy as described above. We refer to them as deterministic, noise, Group 01, Group 02, Group 03 and Group 04. They are summarised in Table~\ref{tab:basic_info_samples} and explained below.

The deterministic sample is our primary sample and was created to explore the phylogenetic signal based on the number of stellar particles  used to build the phylogenetic trees (see Section \ref{sec:results_number_particules}) and also the impact that uncertainties on the chemical abundances have in this kind of study (see Section \ref{sec:results_uncertainties}). We wanted this sample to have a history in which  older populations directly contributed to the chemistry of the younger populations. In order to select these particles, we defined a sphere of 1 kpc of radius around the galaxy's centre of mass at the snapshot that corresponds to 1.5 Gyr. The radius of the sphere is larger than three gravitational softening lengths but small enough to maximize the possibility that the stellar particles represent populations that have a common chemical history of evolution. Then we chose only the stellar particles whose progenitor gas particle was also in the same region since the beginning of the simulation. We adopt a time of 0.016 Gyr  which corresponds to the first snapshot available of the simulation. Finally, we chose only the stellar particles whose birth radii were also inside the sphere. The central location of the deterministic sample also considers that the particles have low probabilities to experience significant migration, since they are located at the centre of the gravitational potential well.

The noise sample was built by replacing the chemical abundances of the deterministic sample by random chemical abundances. The random chemical abundances were generated within the range of the deterministic sample. Therefore the noise sample has stellar particles whose chemical abundances have no astrophysical meaning. This sample was included in this study in order to compare how phylogenetic trees from data compare to trees from random chemical abundances and evaluate the presence of phylogenetic signal.

Finally, groups 01, 02, 03 and 04 are used to explore the evolutionary histories of different regions of the galaxy (see Section \ref{sec:results_regions}).  We selected stellar particles in four different spheres at different galactic radii. All the spheres have 1 kpc of radius like the deterministic sample. Unlike the deterministic sample, however, here we perform no further selections on the birth radii or the location of their progenitor gas particles, hence allowing the particles to come from outside the corresponding sphere. Group 01 was built from a sphere centred at $(x,y,z) = (0,0,0)$. Group 02 was built around the position $(x,y,z) = (3,3,0)$ kpc. Group 03 is from a sphere centered at $(x,y,z) = (-3,3,0)$ kpc. Group 04 was selected around the position $(x,y,z) = (-5,5,0)$ kpc. Groups 02 and 03 were selected to assess possible azimuthal variations, in which only Group 03 select stellar particles from a spiral arm. 

Figure \ref{figure:position_regions} shows the spatial distribution of the four regions studied. Group 01 (green) contains 2516 stellar particles. Group 02 (blue) contains 563 stellar particles. Group 03 (pink) has 506 stellar particles. Finally, Group 04 has 100 stellar particles. The colors associated with each group is respected in the rest of this work. In grey we show the spatial distribution of all stellar particles at 1.5 Gyr. The difference in number of particles in these regions is due to the different gas densities in the simulation, which follows an exponential profile. This has an impact on the star formation history, and therefore the chemical enrichment.

The deterministic and noise samples are used to assess the dependence of the phylogenetic signal on the number of stellar particles selected to built the trees. Hence  we created subsamples containing 10, 50, 100 and 200 stellar particles. Groups 01 to 04 are used to study the physical information that can be retrieved by the phylogenetic trees. For these groups, we selected 100 stellar particles to represent the stellar population of the corresponding region, based on the results of the analysis of the deterministic and noise samples. We applied a Kolmogorov–Smirnov test (hereafter KS test) to guarantee that every sub-sample of 100 stellar particles provided a fair representation of the properties of their parent sample. We rejected the null hypothesis if the $p$-value was lower than 0.05. The KS test considered the distributions of [Fe/H], [O/Fe] and star formation time.

\subsubsection{Input information for trees}{\label{sec:input_for_trees}}

We used chemical abundances of ten chemical elements in order to build the phylogenetic trees. The chemical elements are: O, Mg, Ca, Si, Ne, S, Fe, Zn, C, and N. They trace different nucleosynthetic channels and provide important information about chemical evolution processes in the simulation. O, Mg, Ca, Si and Ne, for example, are $\alpha$ elements, produced mainly by SNII, while Fe and Zn are iron-peak elements produced mainly by SNIa. In the case of C and N in this simulation, the production is done only by SNIa and SNII as AGB winds are not included in our simulation. 

The chemical abundances were defined in relation to hydrogen and the Sun, in the format [X/H] as defined in Section \ref{sec:simulations: UNDER CONSTRUCTION --}. We chose this format in order to have a more direct parallel between the abundances in this work and the observational works on chemical evolution of the Milky Way. Another reason for this choice is to have Fe as an independent element to build phylogenetic trees. We note that a galaxy that might experience inflow  of pristine gas can have a trend of [X/H] which is not monotonic. In the case of this simulated galaxy, there is no inflow of pristine gas, therefore the ratio [X/H] can be used without that concern.

The chemical abundances provided by the simulation do not have intrinsic uncertainties, therefore each tree we build is the result of one distance matrix which is the result of the simulated abundances. When studying the impact of uncertainties in the  evolutionary history provided by the phylogenetic trees, we varied the original abundance value considering a normal distribution. In order to do so, we created a normal distributions where their mean was the original abundance value and the  uncertainties ($\sigma$)  were 0.01, 0.05, 0.08, 0.1, 0.2 and 0.3 dex. The widths of the normal distribution were chosen in order to investigate uncertainties found in standard observational studies (e.g. 0.1, 0.2 and 0.3 dex) and also in high-precision studies (e.g. 0.01 and 0.05 dex), while considering intermediate cases to better delimitate the maximum uncertainties possible for which phylogenetic signal is mostly preserved (e.g. 0.08 dex).

\section{Results and interpretation}{\label{sec:results}}

In this Section we present the results we obtained in three tests, performed using the different samples discussed in Section~\ref{sec:stellar_samples}.  The astrophysical properties of the samples used here are discussed in Section~\ref{sec:astrophysical_properties}. In our first test we explore the phylogenetic signal provided by trees when we vary the number of stellar particles (Section \ref{sec:results_number_particules}). Then, we investigate the impact of chemical abundances uncertainties on the evolutionary history traced by the trees and in the phylogenetic signal (Section \ref{sec:results_uncertainties}). Finally we explore the evolutionary history found in different regions of the simulated galaxy and its connection with the AMR and SFH of the location (Section \ref{sec:results_regions}).

\begin{figure}
\centering
\includegraphics[width=6.5cm]{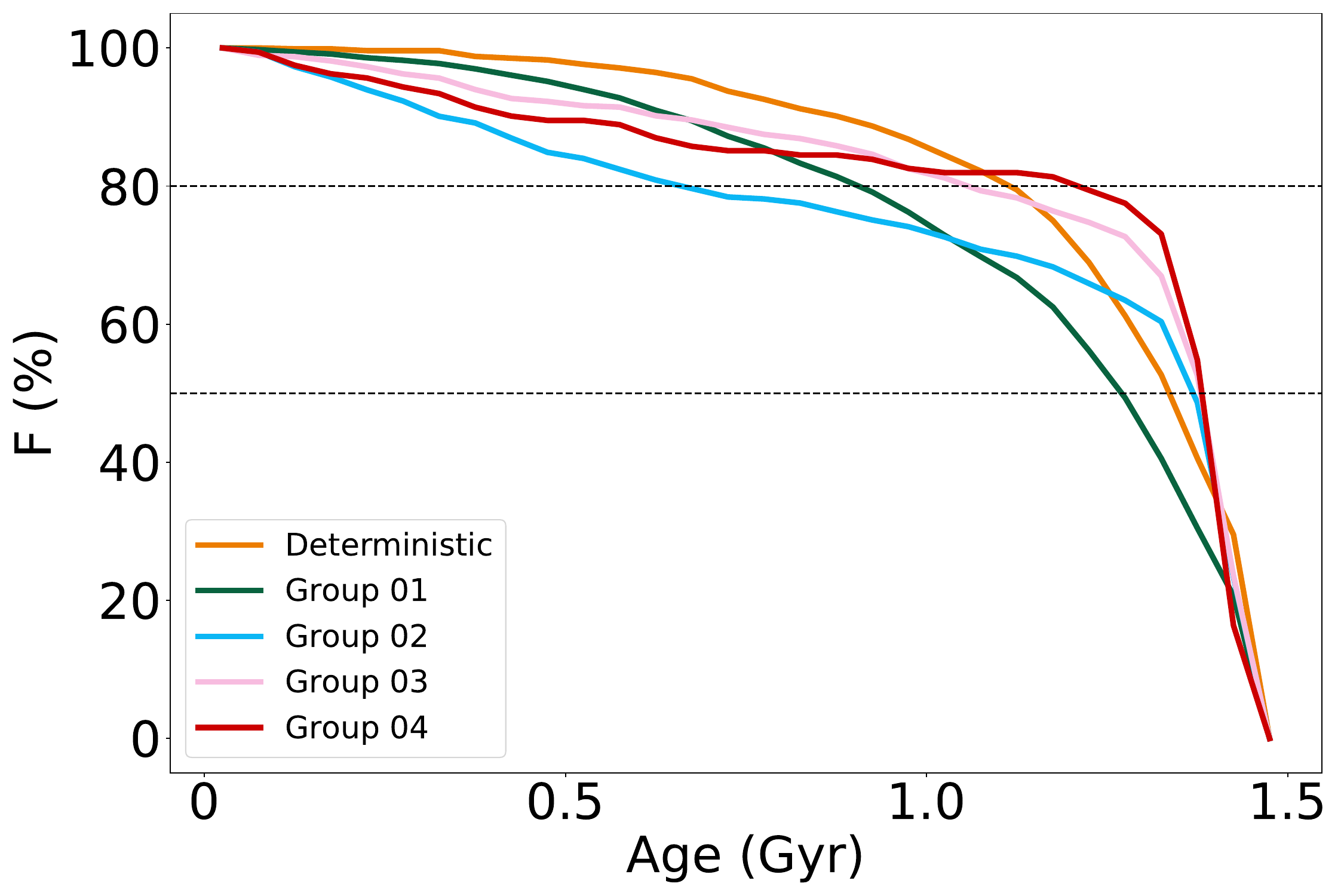}
\caption{Cumulative stellar mass fraction as a function of age of all samples considered in this work (see Table 1). The orange, green, blue, pink and red lines refers to deterministic, Groups 1, 2, 3 and 4 samples. The horizontal lines indicates where 50 and 80 percentiles of the stellar mass contribution.}
\label{figure:mass_fraction}
\end{figure}

\subsection{Astrophysical properties of the different samples used}{\label{sec:astrophysical_properties}}

In order to explore the astrophysical properties of the different samples used in this work, their star formation histories (SFH), age-metallicity relation (AMR) and [O/Fe] \textit{vs.} [Fe/H] distribution are considered. Oxygen is a chemical element mostly deposited in the ISM due to SNII which are the explosions of massive stars, while the production of Fe by SNIa and SNII varies according to the yields adopted. Hence, the deviation of [O/Fe] from what is that typically found in SNII ejecta represents the contribution from low-mass stars. As a consequence, the ratio [O/Fe] is a powerful diagnosis of the low and high mass stars contribution to the chemical evolution of the ISM, which happen over different timescales because of stellar evolution. The AMR relation is also key, since it shows how the metallicity of the environment changes with time. Finally from the SFH we can identify when star formation, hence chemical enrichment, has been most prominent in the simulation, or how star formation might vary in the different samples studied. We therefore use [O/Fe] \textit{vs.} [Fe/H] distribution, the AMR and the SFH  to guide the interpretation of the evolutionary history traced by the phylogenetic trees.

\begin{figure*}
\centering
\includegraphics[width=6.5cm]{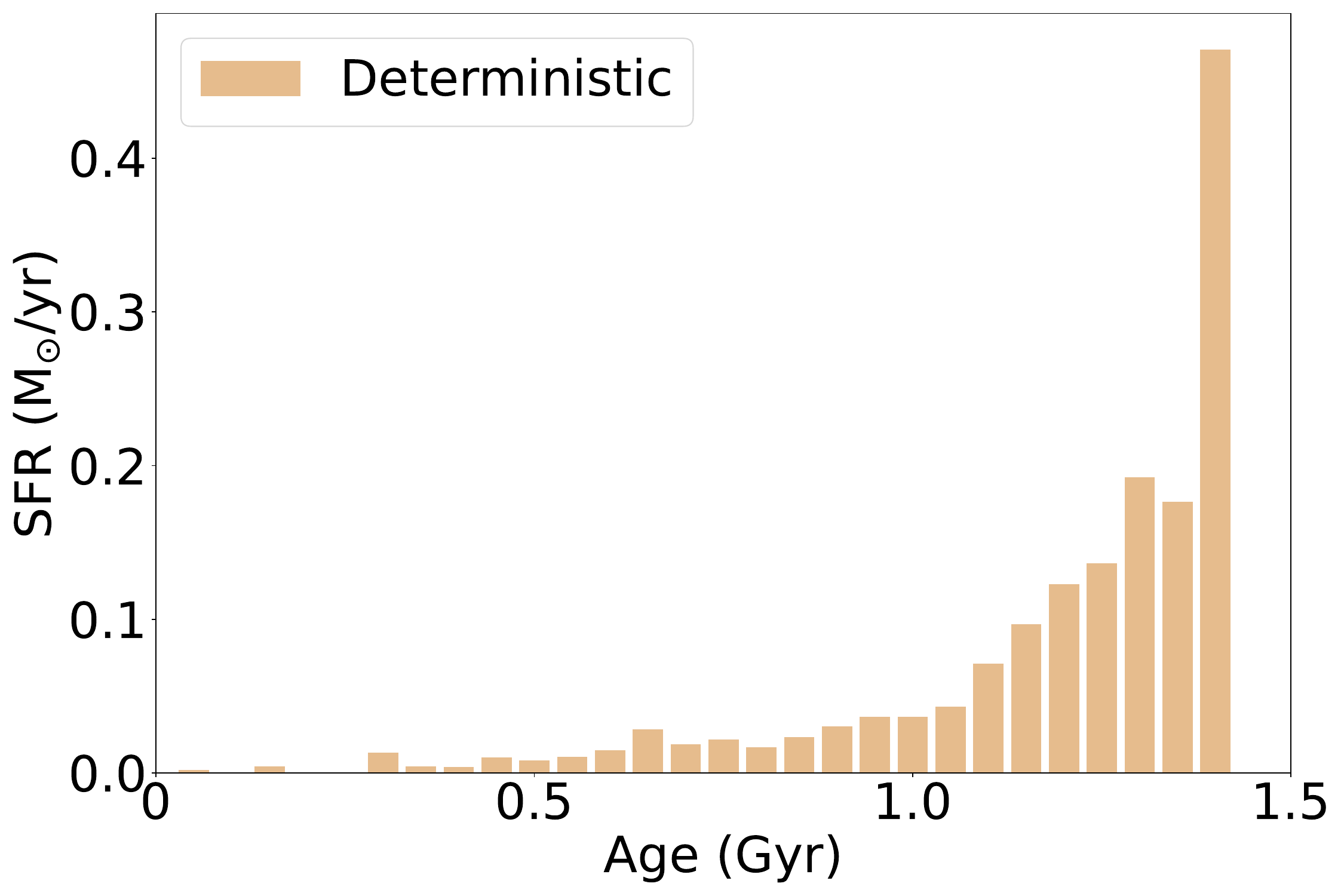}
\includegraphics[width=4.5cm]{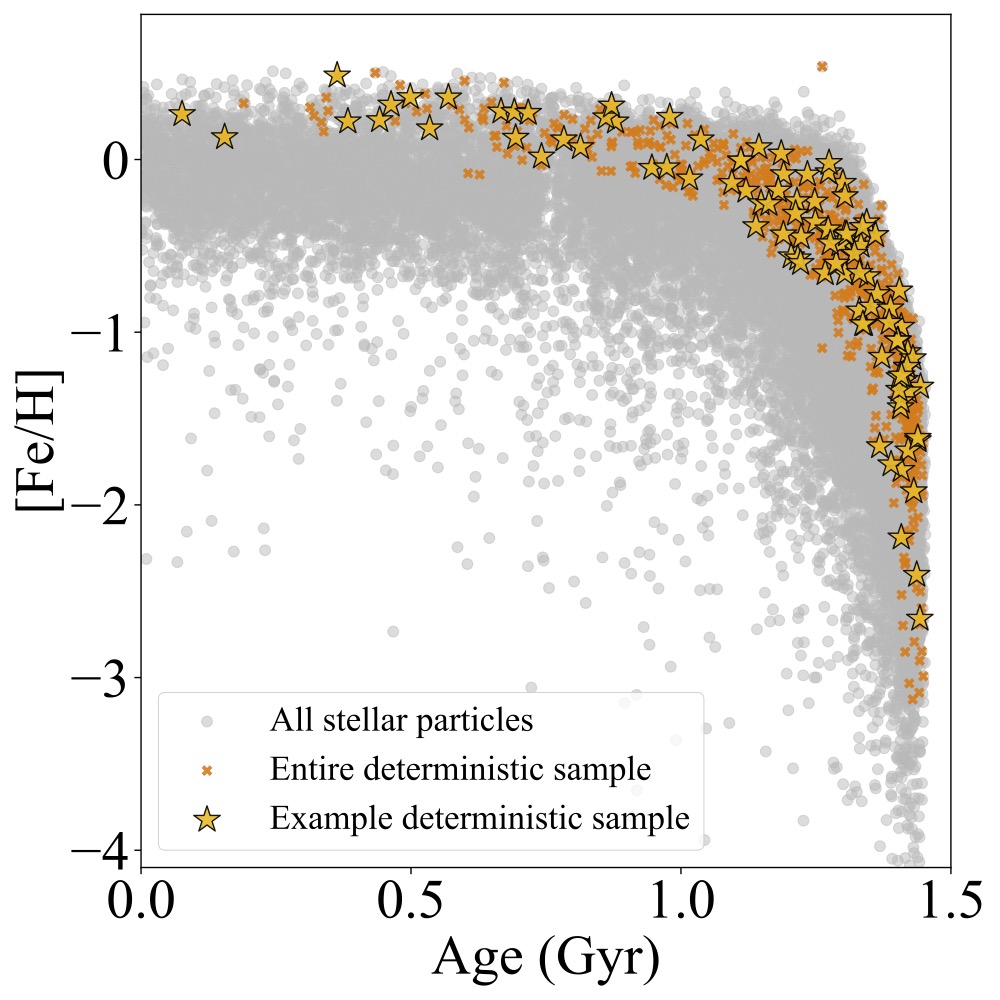}
\includegraphics[width=4.5cm]{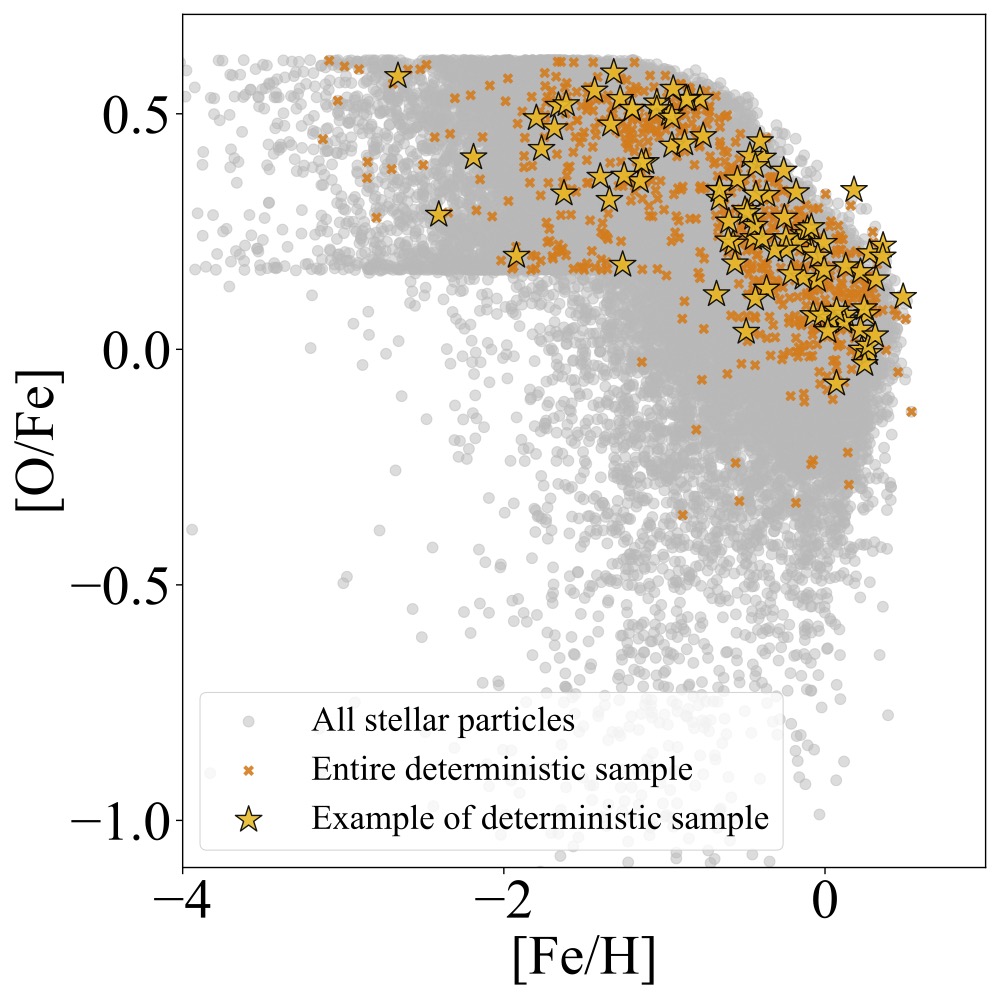}
\newline
\includegraphics[width=6.5cm]{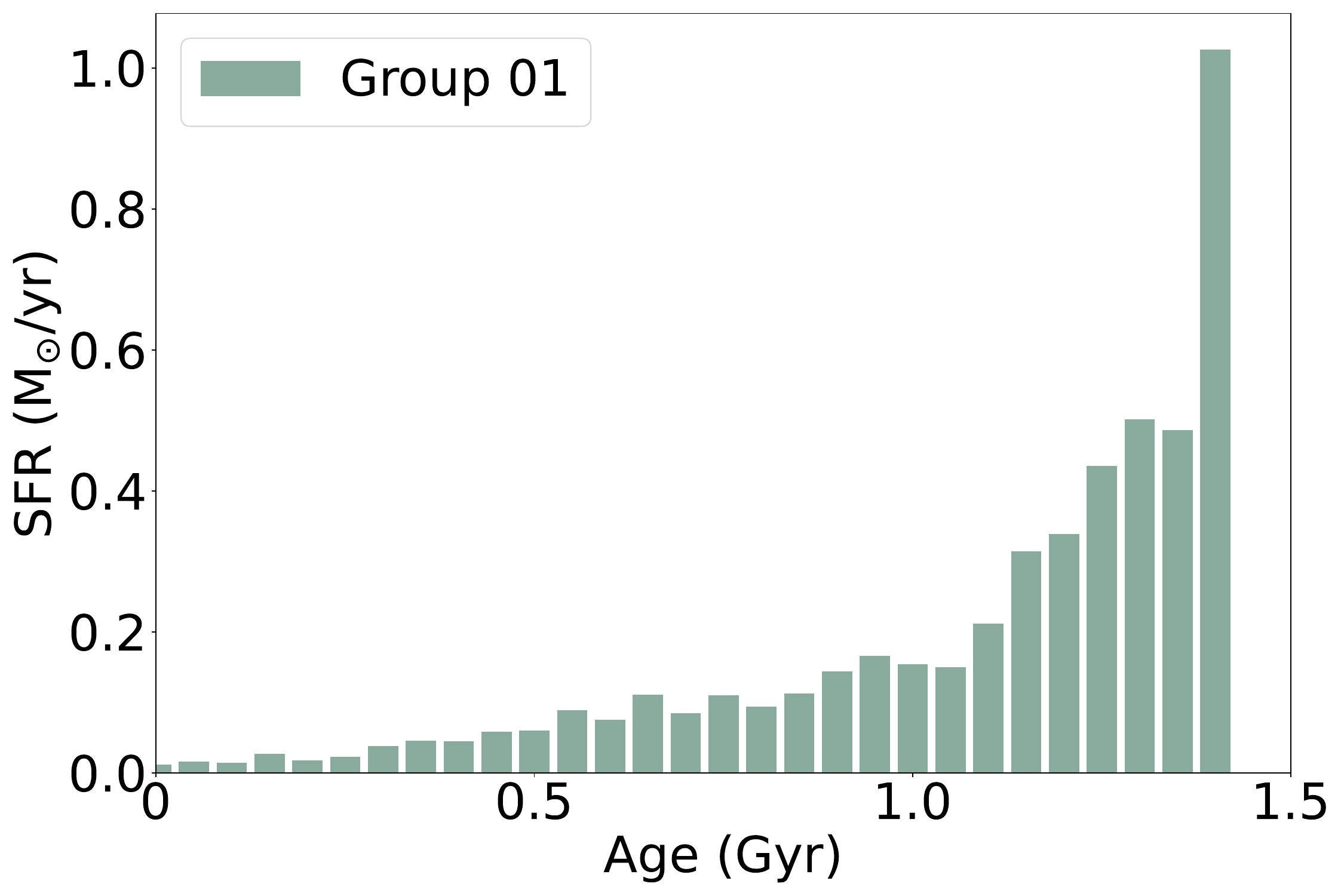}
\includegraphics[width=4.5cm]{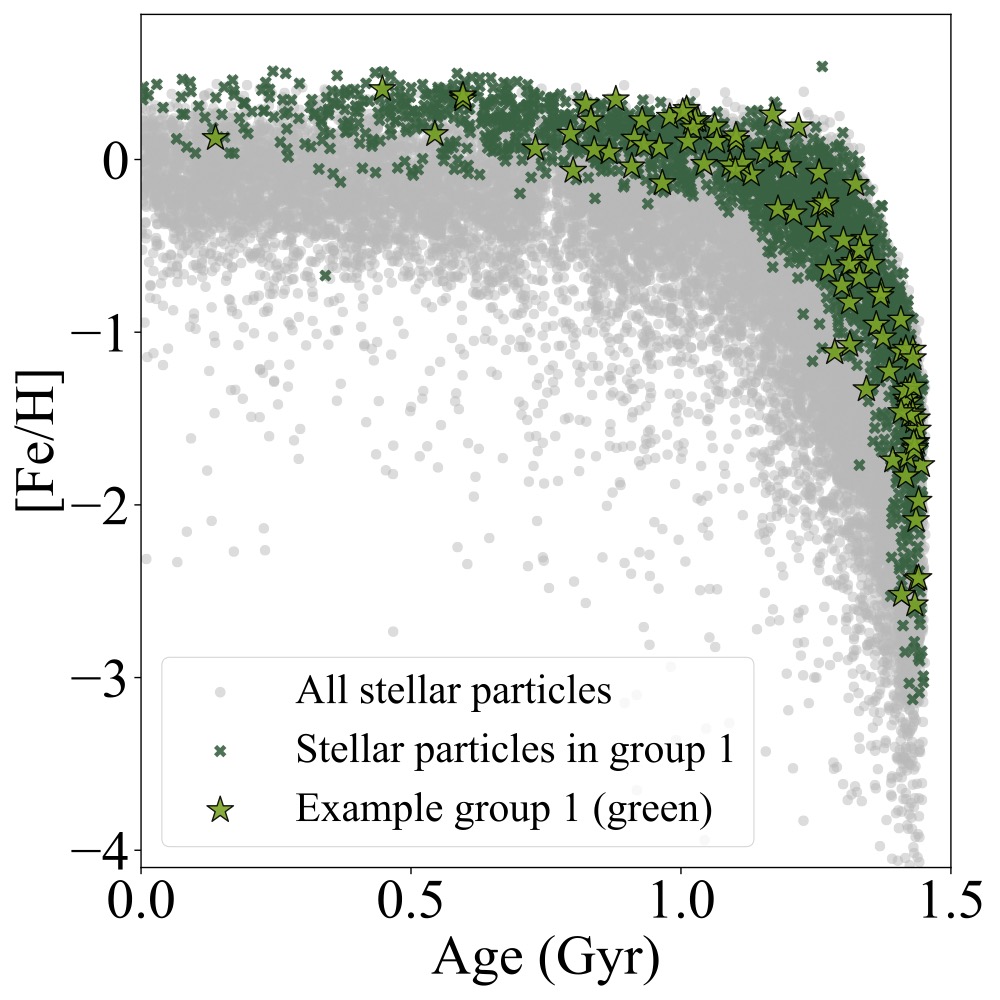}
\includegraphics[width=4.5cm]{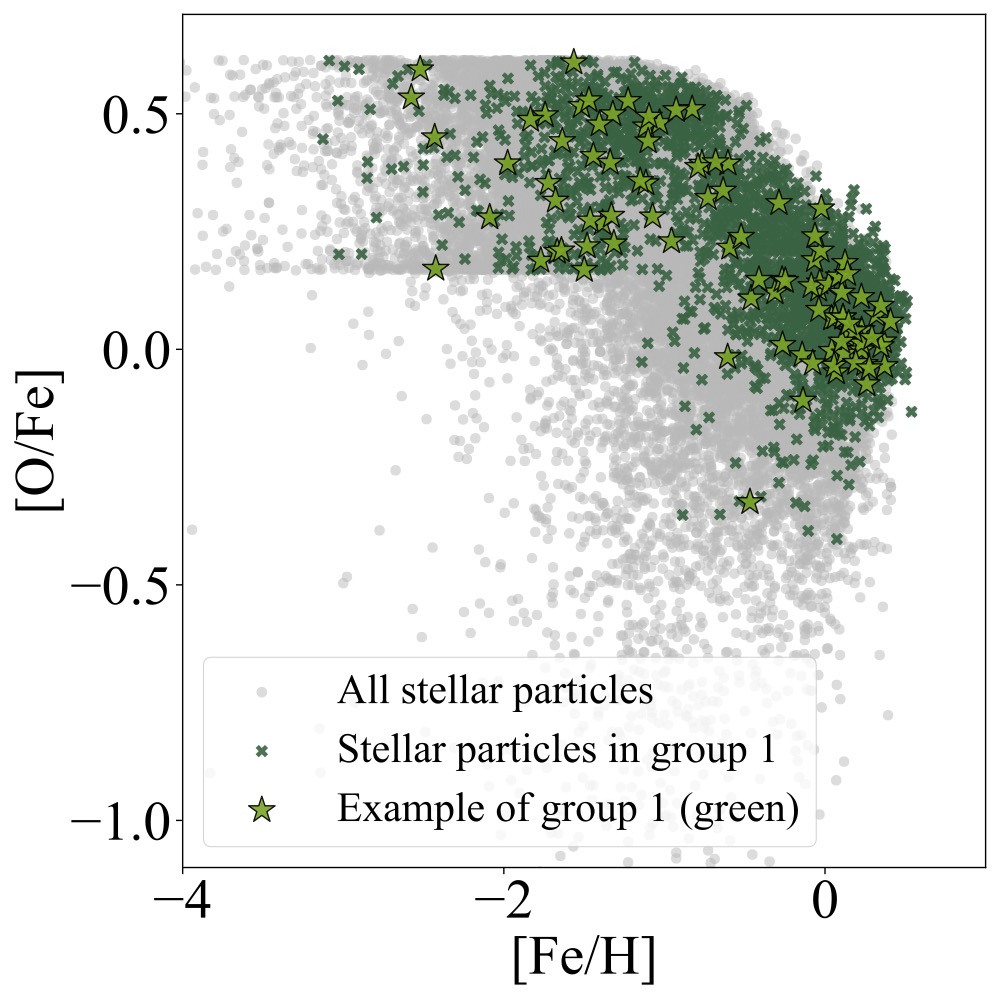}
\newline
\includegraphics[width=6.5cm]{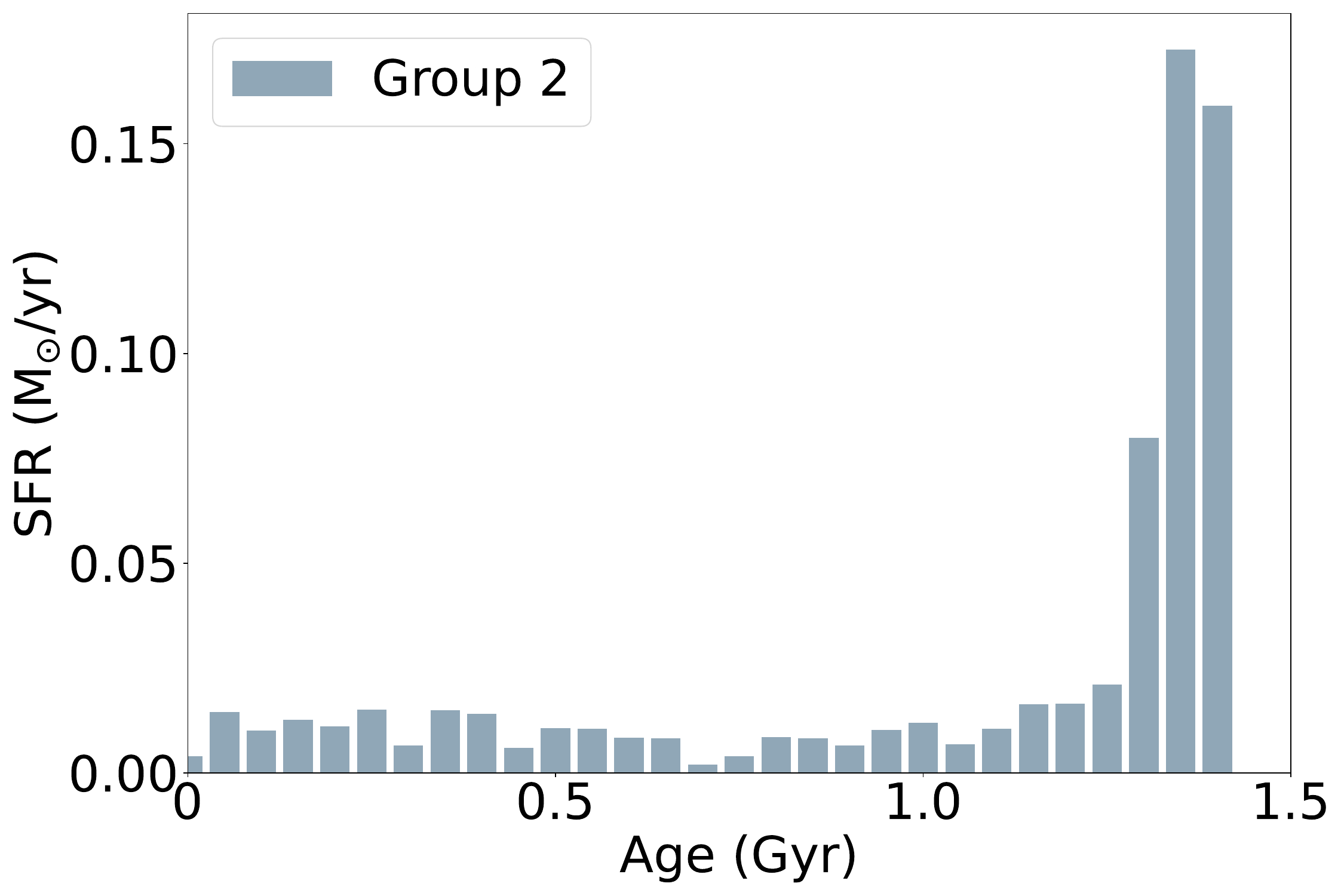}
\includegraphics[width=4.5cm]{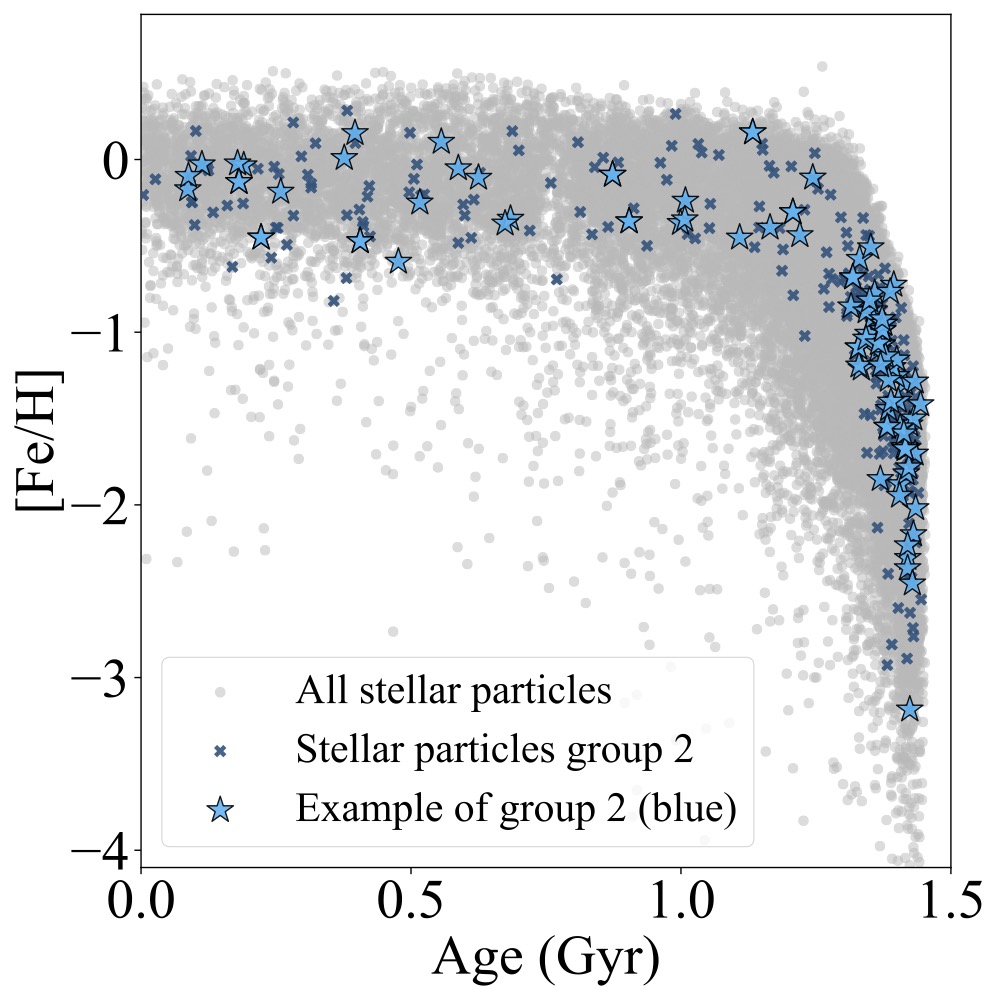}
\includegraphics[width=4.5cm]{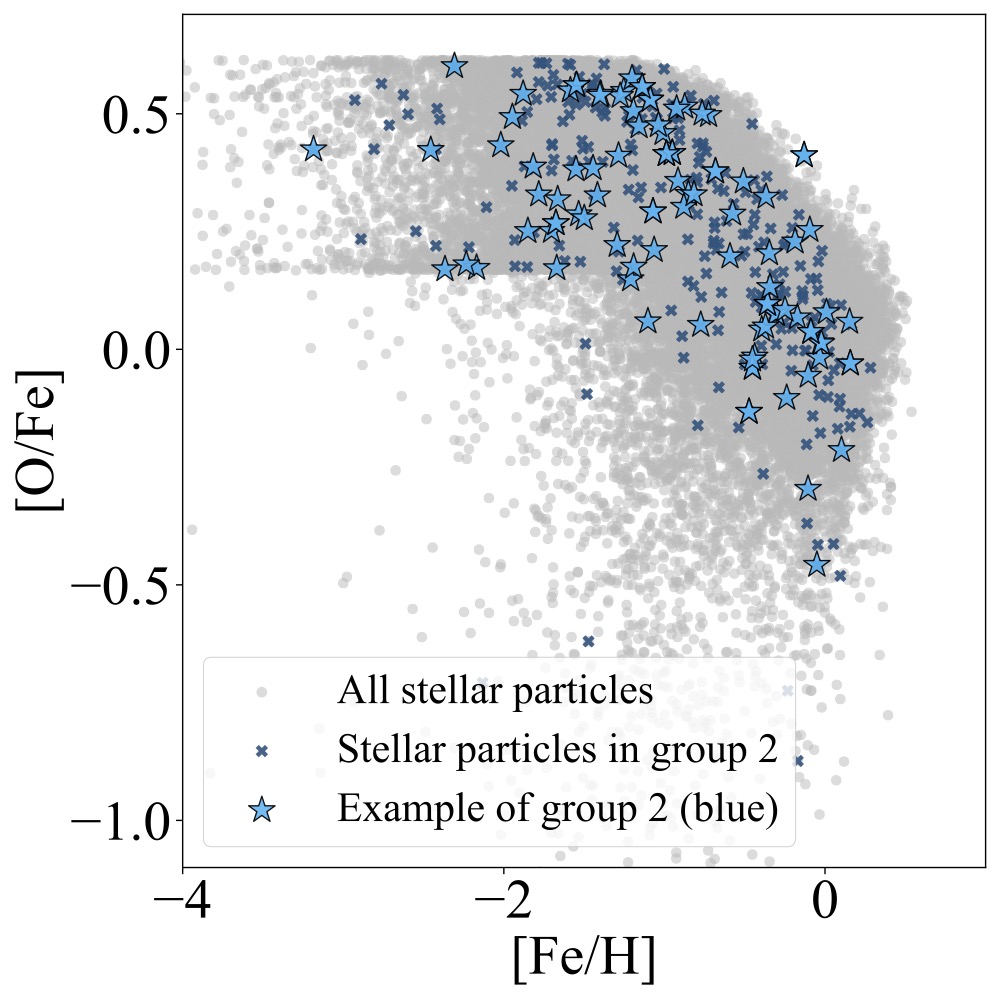}
\newline
\includegraphics[width=6.5cm]{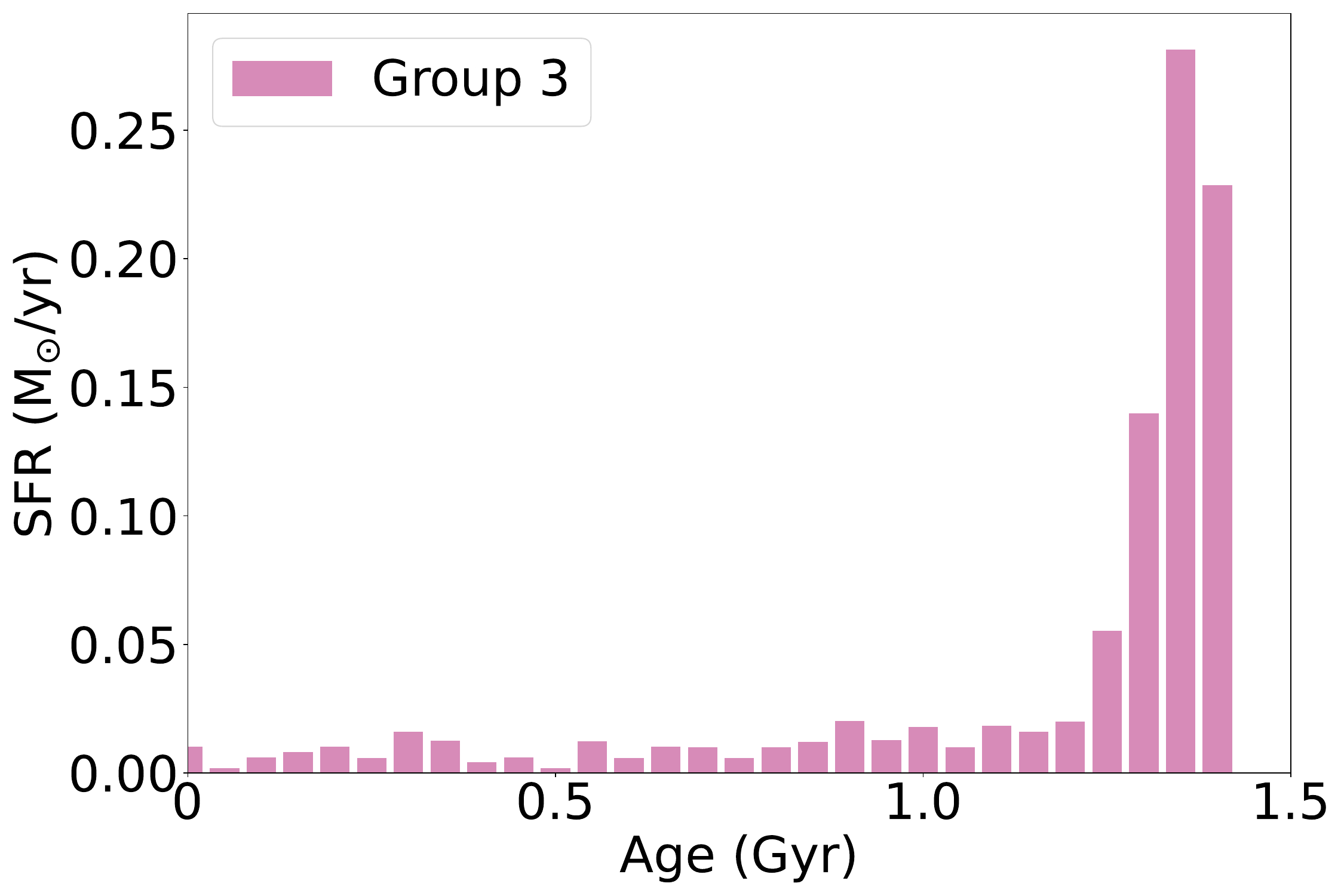}
\includegraphics[width=4.5cm]{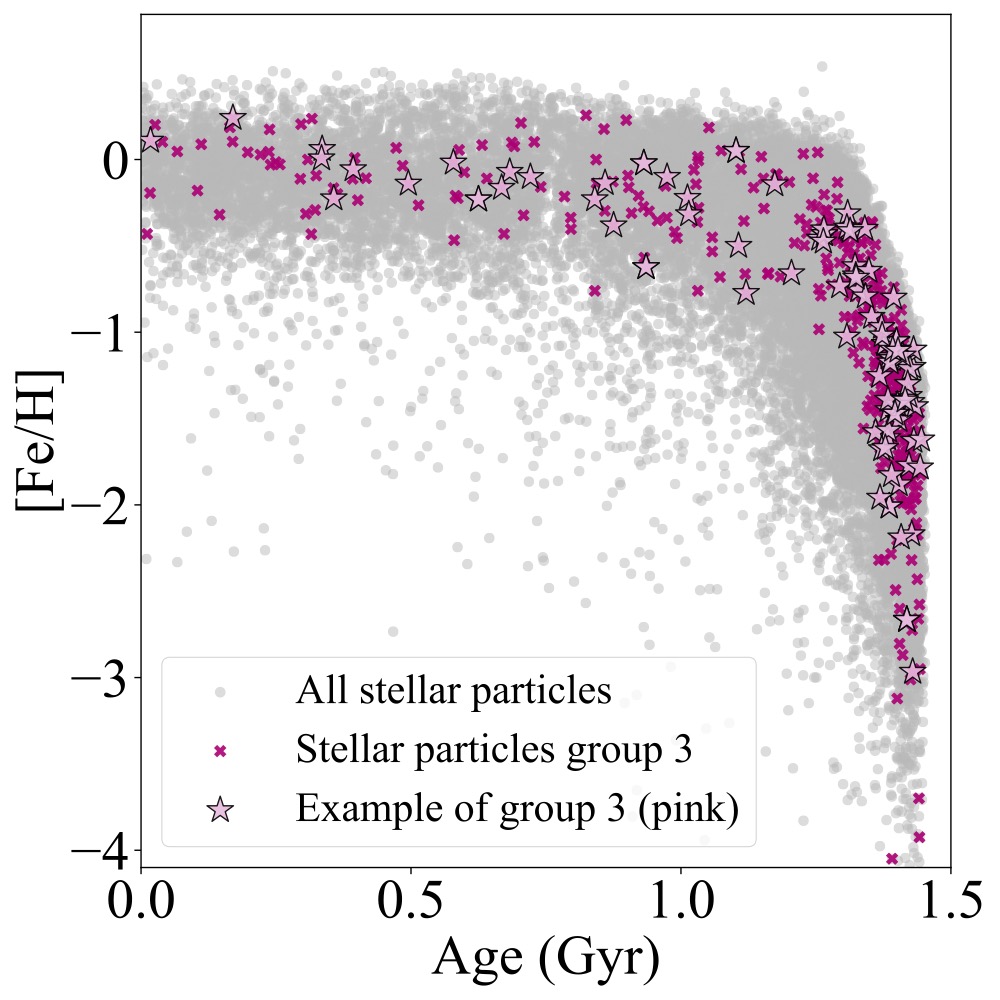}
\includegraphics[width=4.5cm]{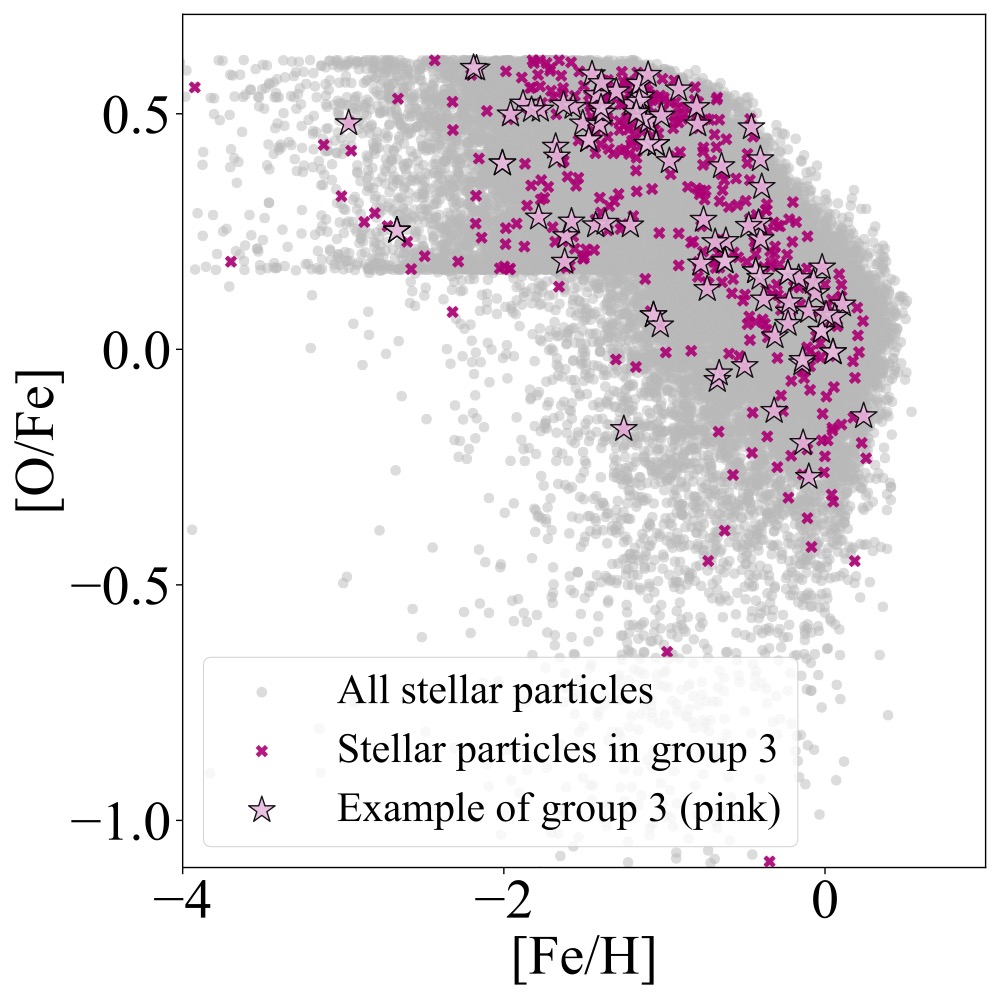}
\newline
\includegraphics[width=6.5cm]{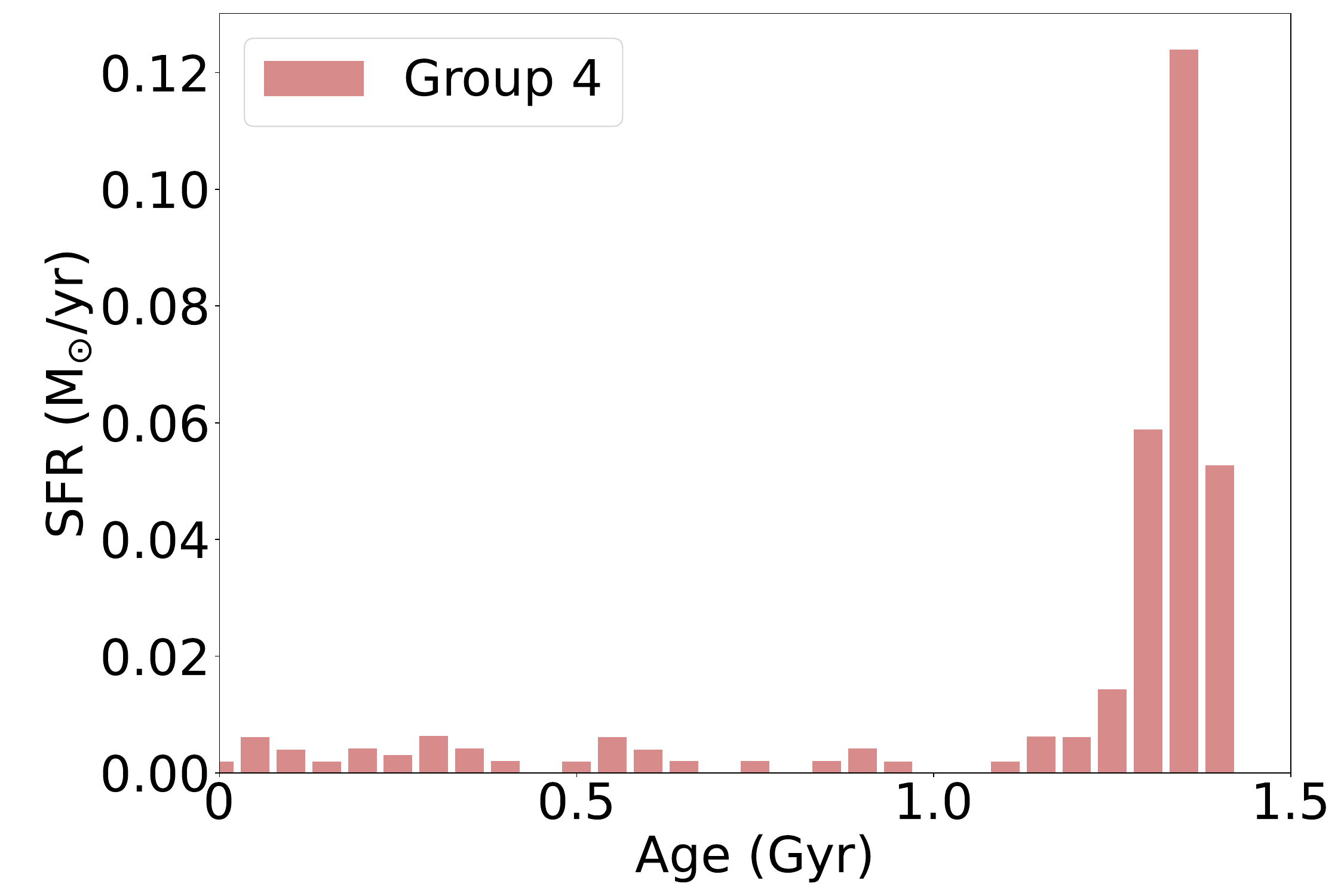}
\includegraphics[width=4.5cm]{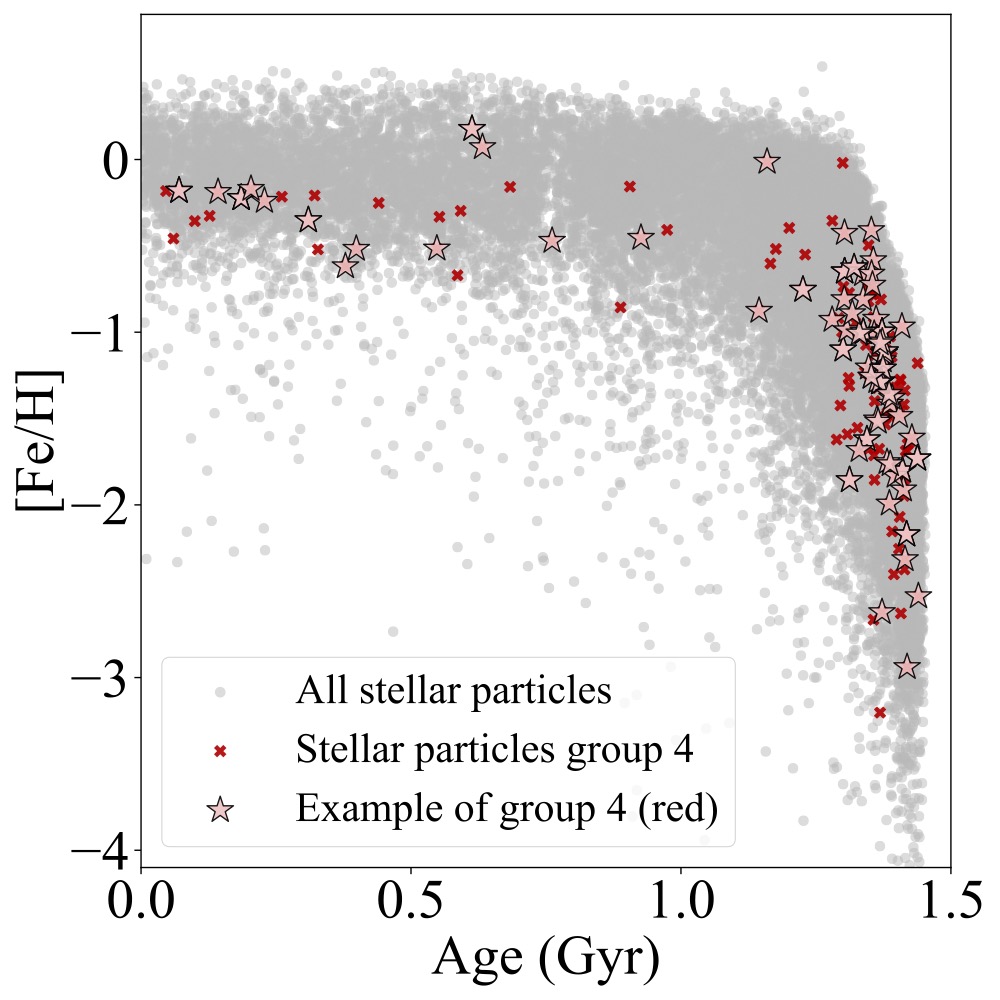}
\includegraphics[width=4.5cm]{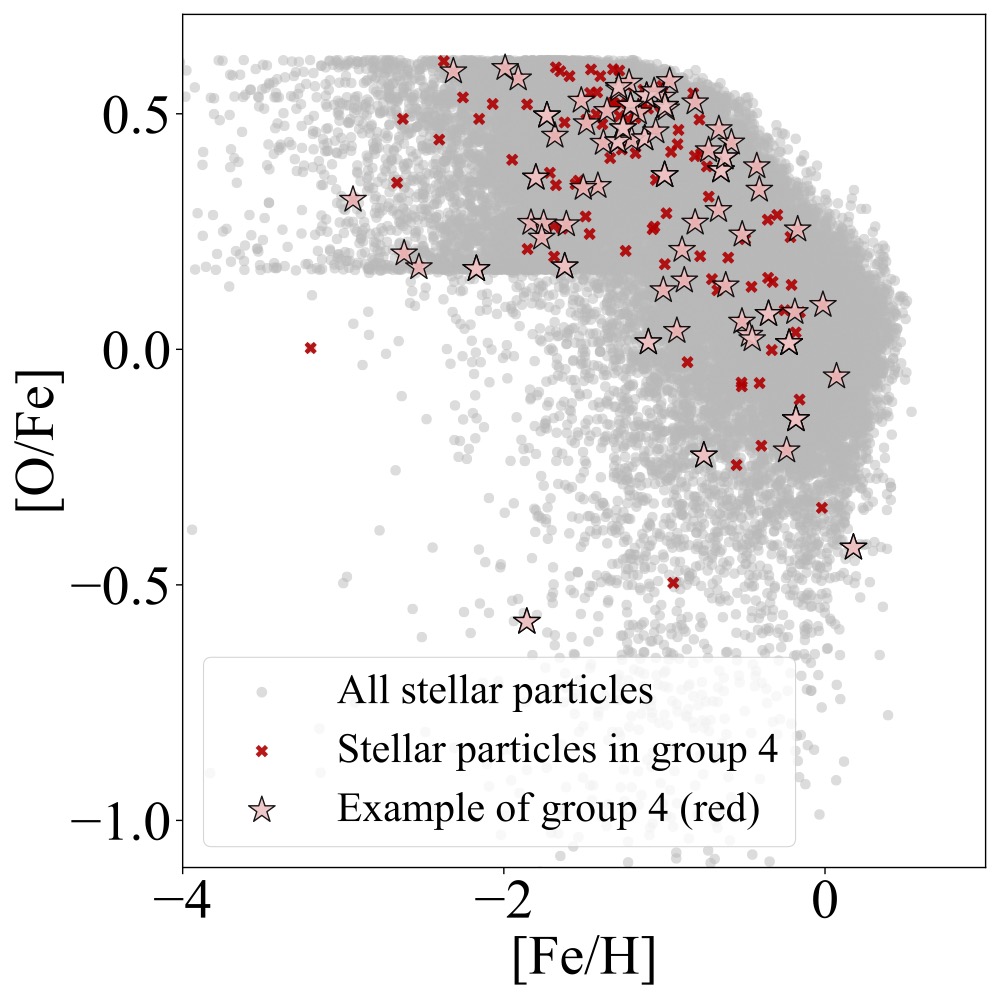}
\newline
\vspace{-0.5cm}
\caption{Example of astrophysical properties of the samples studied. Each line represents respectively the samples: deterministic, Group 01, Group 02, Group 03 and Group 04. Left: Star Formation History (SFH). Center: Age metallicity relation (AMR). Right: [O/Fe] \textit{vs} [Fe/H] relation. Gray: all stellar particles with chemical abundances available in the simulation at 1.5 Gyr. Dark colors represent all possible stellar particles from each sample. Star symbols represent the chosen 100 particles used to build phylogenetic trees in this work.}
\label{figure:astrophysical_properties}
\end{figure*}

Figure~\ref{figure:mass_fraction} shows the cumulative stellar mass fraction as a function of the stellar ages of the populations within each analysed sample. 
In yellow the deterministic sample, and  Groups 01, 02, 03, and 04 in green, blue, pink and red, respectively. Dashed horizontal lines represent the 50  and 80 percentiles of the stellar mass contribution. This figure allows us to compare the star formation histories of the different samples. We note that Group 04 forms 80\% of its stellar mass in a considerably shorter time scale than Group 01, reflecting that the outskirts of the galaxy formed the majority of its stellar mass faster than the center of the galaxy at the given time. We also observe that Group 03 also creates 80\% of its stellar mass faster than Group 02. 

In Figure \ref{figure:astrophysical_properties} we show the SFH, the AMR and the [O/Fe] \textit{vs.} [Fe/H] diagrams for the different samples studied here.  Each row of the figure is a different sample. The gray background points correspond to all the 31 807 stellar particles at 1.5 Gyr that passed our first selection criterion (i.e. have Z higher than 0) and is therefore the same in all rows. In color we show all stellar particles selected for each sample. The stellar symbols enclosed correspond to a random selection of 100 particles which are referred to as {\it example samples}. We make this selection because we can only build trees with limited number of stellar particles to avoid visual cluttering, therefore we need to assess if these selections are a good representation of the entire sample. These are the 100 particles selected considering a K-S test and displayed in the trees of the following sections, and we can see that in every sample, they are well distributed with respect to the main sample. 

Looking at the left columns of Fig.~\ref{figure:astrophysical_properties}, we see that the peak of the star formation happened at the start of the galaxy's evolution. This peak is seen across the different samples, although it lasts for longer at the central part of the galaxy. One can see that both the SFH from the deterministic and the Group 01 samples have a SFH that peaks at 1.4 Gyr and decreases gradually over approximately 0.3 Gyr, while the Groups 02, 03 and 04 have peaks that lasts only for about 0.1 Gyr. There is still star formation happening during the rest of the history of this galaxy across all regions, but at a much lower rate.  

It is expected that a galaxy that evolved in isolation would not present further enhancement of the star formation after the first peak, which is driven by formation of the arms in this simulation. The new born stellar populations tend to be concentrated in the central regions following the initial gas density distribution, but they will also populate the denser regions of arms. After this, the star formation self-regulates consuming the remaining gas (recall that there are no external gas inflows or mergers in this isolated case) into stars which, subsequently, injects supernova feedback into the ISM. The energy increases the temperature and pressure and contributes to regulate the star formation activity, producing a more continuous star formation activity with same weak star formation bursts.

The AMR relations in the middle panels show the relation between chemical enrichment and the SFH. Since a lot of stellar particles are formed at the beginning of the galaxy's history, it is expected that chemical enrichment will happen quickly, particularly at the central regions where the gas density is highest. The AMR is therefore expected to be steep for stellar particles formed at the epoch of the star formation peak. We observe that happening in all regions. Once the star formation has slowed down, the metallicity slightly increases. We can note some  differences among regions. The AMR relation at the central regions increases more monotonically, which is an effect of a more significant star formation happening over a longer period of time with respect to the outer regions.  This can also be seen from the cumulative mass ratio of Fig.~\ref{figure:mass_fraction}  where the central region forms 50\% or 80\% of its stellar particles later than the outer regions. The level of metal enrichment reached by each stellar populations is also different, being the central regions systematically more enriched, as expected.

The AMR of the Groups 02, 03 and 04 show a breaking point around 1.3 Gyr, which is related to the abrupt change of star formation activity at that time. The AMR of the Group 04 has very few stellar particles with ages younger than about 1.2 Gyr. In fact, from Fig.~\ref{figure:mass_fraction}, we see that 80\% of the stellar mass in that region are formed 1.2 Gyr ago. It is thus more difficult to attribute these stellar particles as a population that is following one chemical evolution path through a ancestral-descendant relationship.

In all the analysed samples, we observe a decrease of [O/Fe] with the increase of [Fe/H], as expected according to chemical evolution of galaxies. In the first stages of evolution of the simulation, multiple SNII occur producing O in great quantity. SNIa progenitors have longer lifetimes, therefore only at later stages,  Fe is deposited in the ISM in a more substantial way, decreasing [O/Fe].  This is seen in every panel.

It is worth commenting the differences between the deterministic and the Group 01 samples, since both concern the same region in the galaxy, namely the central one. We note that the deterministic sample is a subset of Group 01, since we impose that both the stellar and the gas particles residing at the end of the simulation must have stayed in the inner region. This results in removing most of the younger particles of Group 01, which shows how much gas flow is on-going in the central region of the simulation. In Fig.~\ref{figure:mass_fraction} it is possible to see how the deterministic sample assembles its 80\% of stellar particles around 1.2 Gyr while Group 01 does it about 0.3 Gyr later. 

Groups 02 and 03 are also worth commenting on, since they are selected to study possible asymmetric effects in the disc. It is customary to assume that because of the galactic rotation, discs are asymmetric, and therefore only the galactic radius is considered as a variable to study variations in galactic structure and evolution, but the presence of the spiral arms might cause some asymmetries. Here we see that SFH, AMR and [O/Fe]  vs [Fe/H] have very similar distributions in Fig.~\ref{figure:astrophysical_properties}. But we also note that the total number of stellar particles in both regions is different, which is related to the different densities across the arms. Group 03 is located on a spiral arm. This has an impact on the star formation rate, as seen from the cumulative mass fraction of Fig.~\ref{figure:mass_fraction}  where Group 02 assembles 80\% of its stellar particles about 0.4 Gyr later than Group 03. 

As a consequence of the star formation histories, [Fe/H] has a quick increase during the first half Gyr, but after 1.2 Gyr it is approximately constant with a weak increase in some regions depending on the star formation history and local characteristics of the ISM. That delayed enrichment of SNIa relative to SNII causes [O/Fe] to decrease as metallicity increases across the entire galaxy as a result of the interplay between the chemical production of O and Fe caused by stars of different lifetime. We also show in Figure \ref{figure:astrophysical_properties} that our selection of 100 particles from our samples is a fair representation of the particles in that sample.  We build the trees to explore the impact of these different SFH and AMR in these regions in the following sections. 

\begin{figure*}
\centering
\includegraphics[width=7cm]{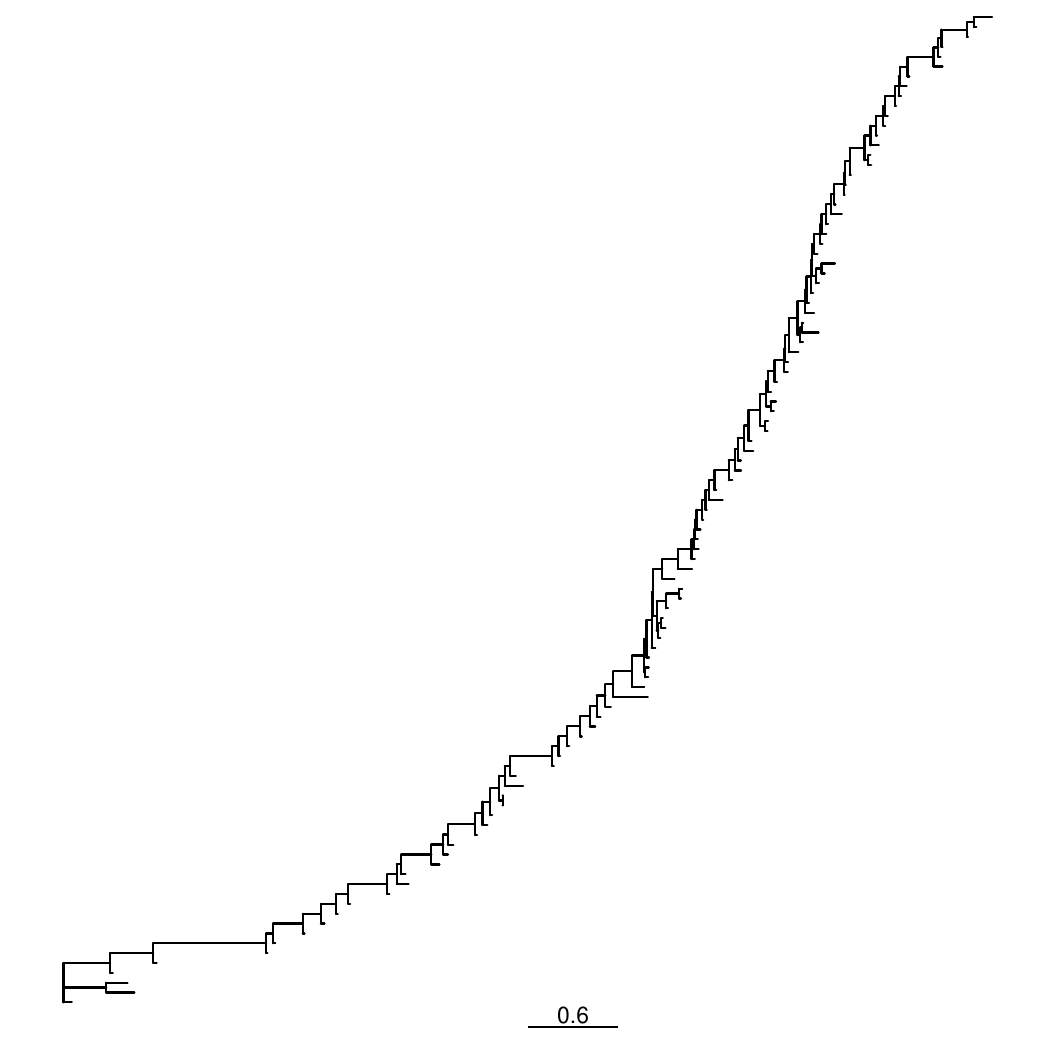}
\includegraphics[width=7cm]{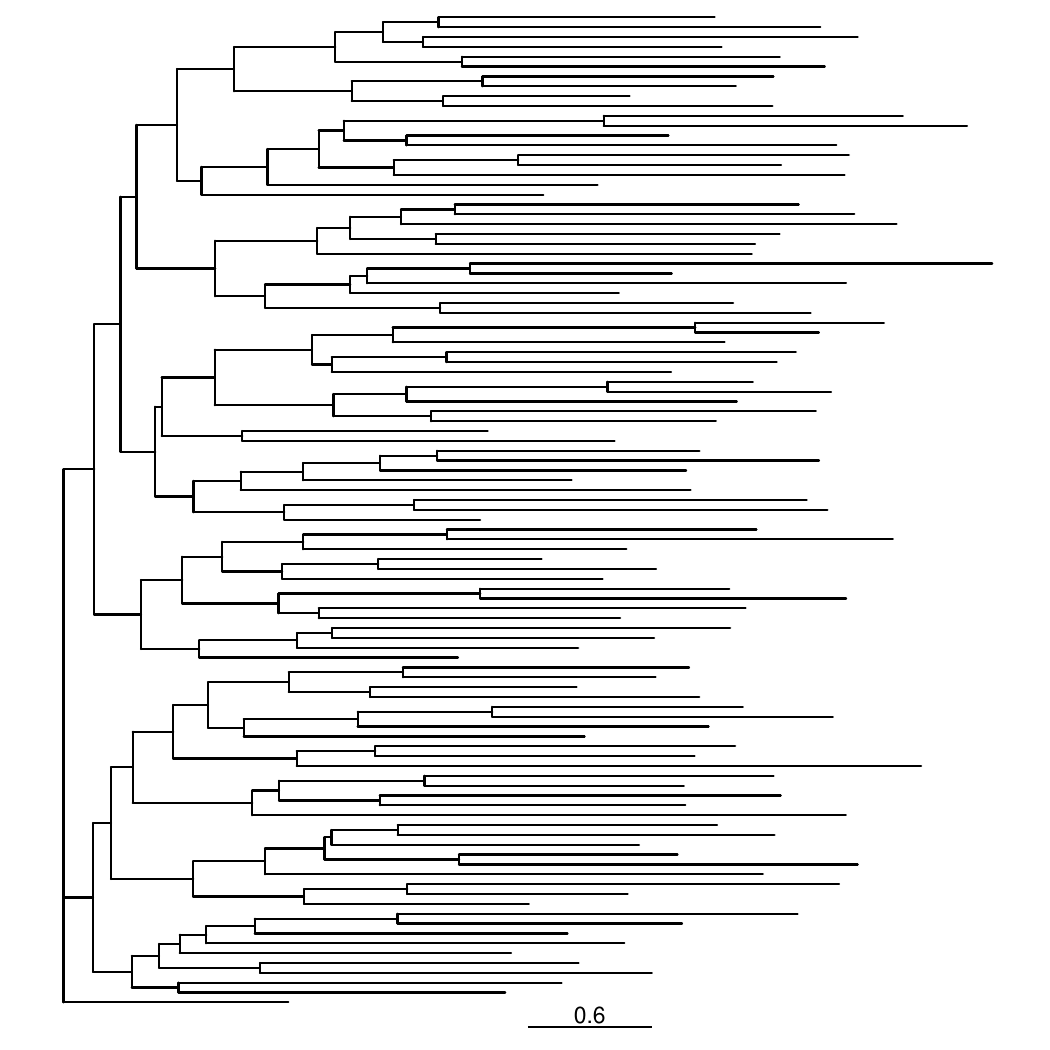}
\caption{Left: example tree built from deterministic selected stellar particles. Stellar particles represents a single stellar population. Right: example tree built from noise sample (see Tab.~\ref{tab:basic_info_samples}). In order to build both trees, we included an outsider stellar particle that corresponds to the oldest stellar particle in the simulation for which chemical abundances are available. Both the trees presented in the left and right panels were rooted in this stellar particle, for better comparison. The scale at the bottom of each panel refers to the branch length (total chemical difference).}
\label{figure:example_tree_deterministic_noise}
\end{figure*}

\subsection{Phylogenetic signal in numerical simulations}{\label{sec:results_number_particules}}

In this section, we focus on the deterministic sample to study if there is phylogenetic signal in our simulation. To do so, we first compare our trees with the noise sample to ensure we are obtaining results that are different than a random distribution, and then interpret our tree in the context of historical reconstruction. We used as the root of the trees the oldest stellar particle for which chemical abundances were available, as discussed in Section \ref{sec:tree_interpretation}.

\subsubsection{Trees from chemical abundances obtained from simulated data or from a random distribution}

In this section, we investigate the dependence of the phylogenetic signal on the population density, specifically the number of stellar particles used to construct it within a given volume. This analysis is highly relevant as it allows us to determine the minimum number of stellar particles necessary to extract a signal that surpasses numerical noise in the simulation as well as natural stochasticity. %Furthermore, these findings can be extended to observational analyses as well.

Figure~\ref{figure:example_tree_deterministic_noise} shows an example of a tree built using the deterministic sample and one tree built using the noise sample. Both trees were built by using sub-samples of 100 stellar particles selected at random from the corresponding volume (see Table 1 for a detailed definition of these sub-samples). In this figure, we can see that the two trees are very different from each other in their general aspect.

\begin{figure*}
\centering
\includegraphics[width=6cm]{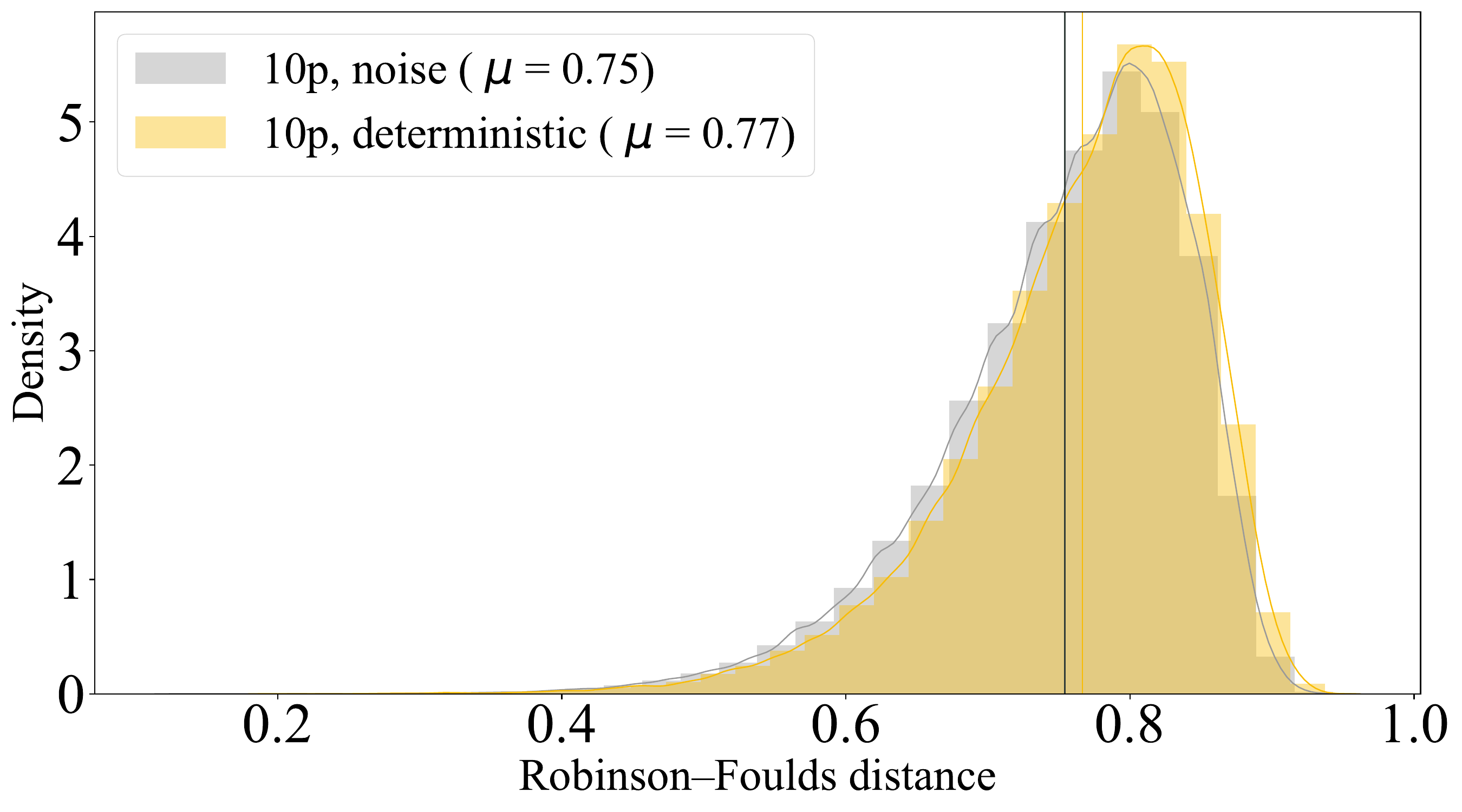}
\includegraphics[width=6cm]{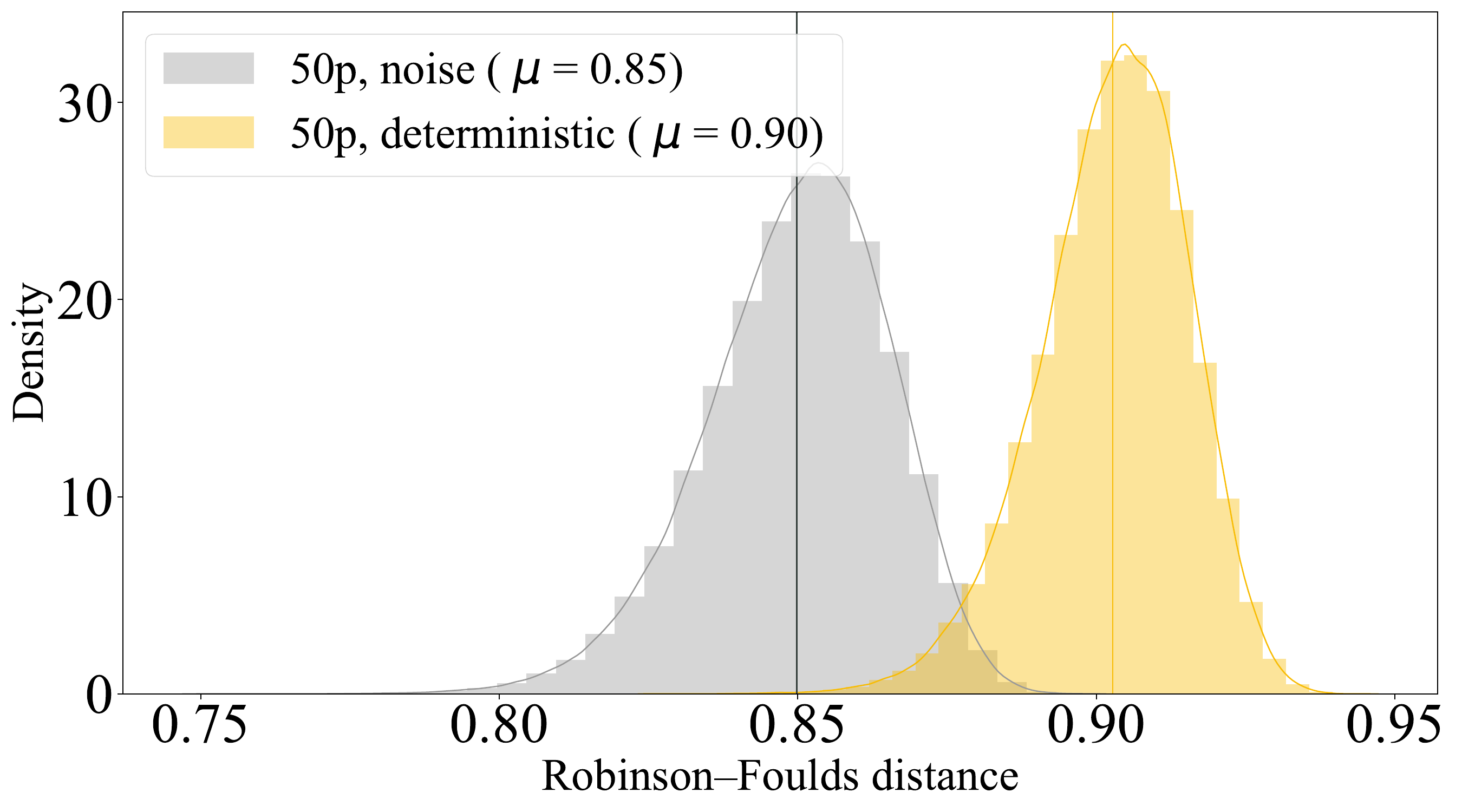}
\includegraphics[width=6cm]{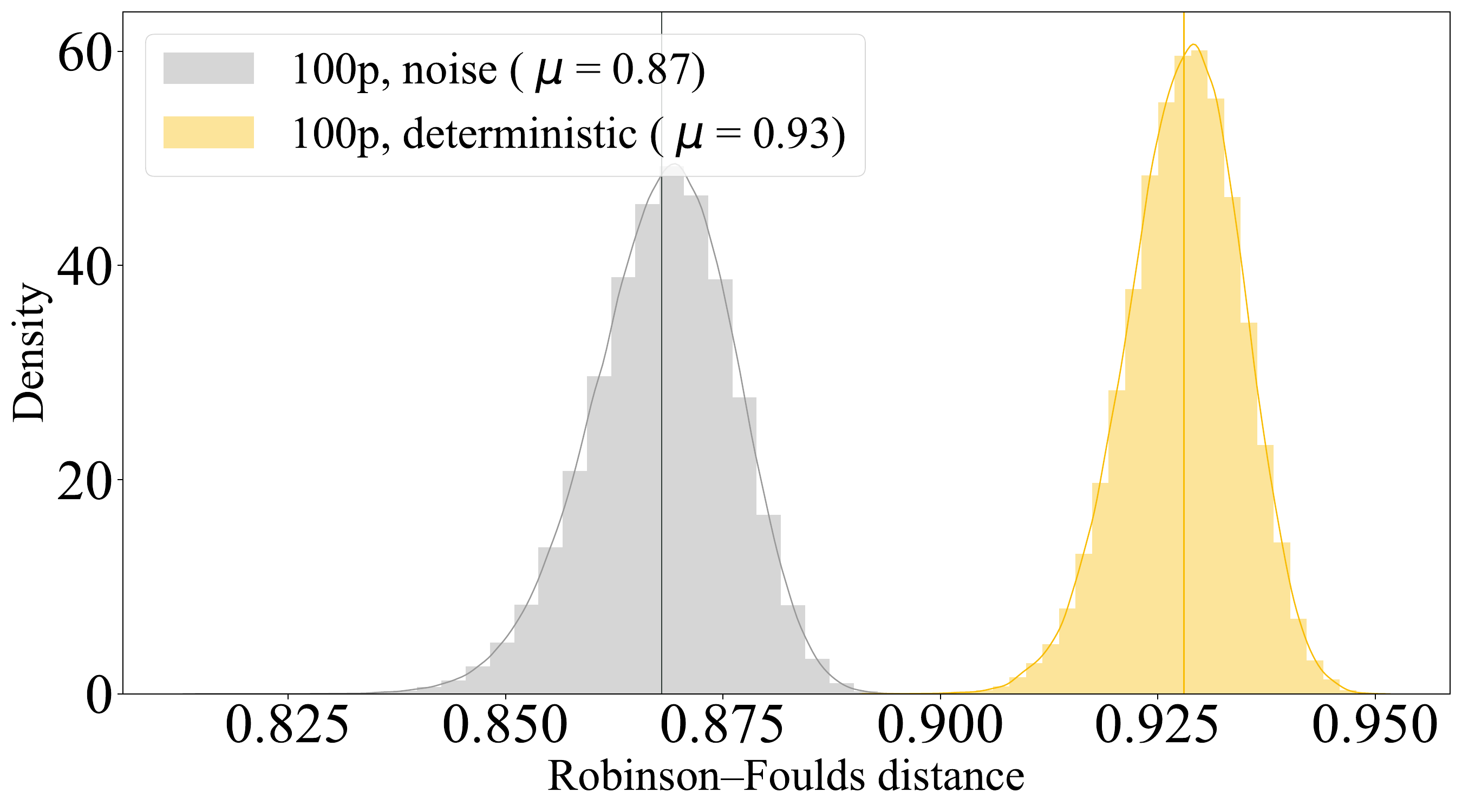}
\includegraphics[width=6cm]{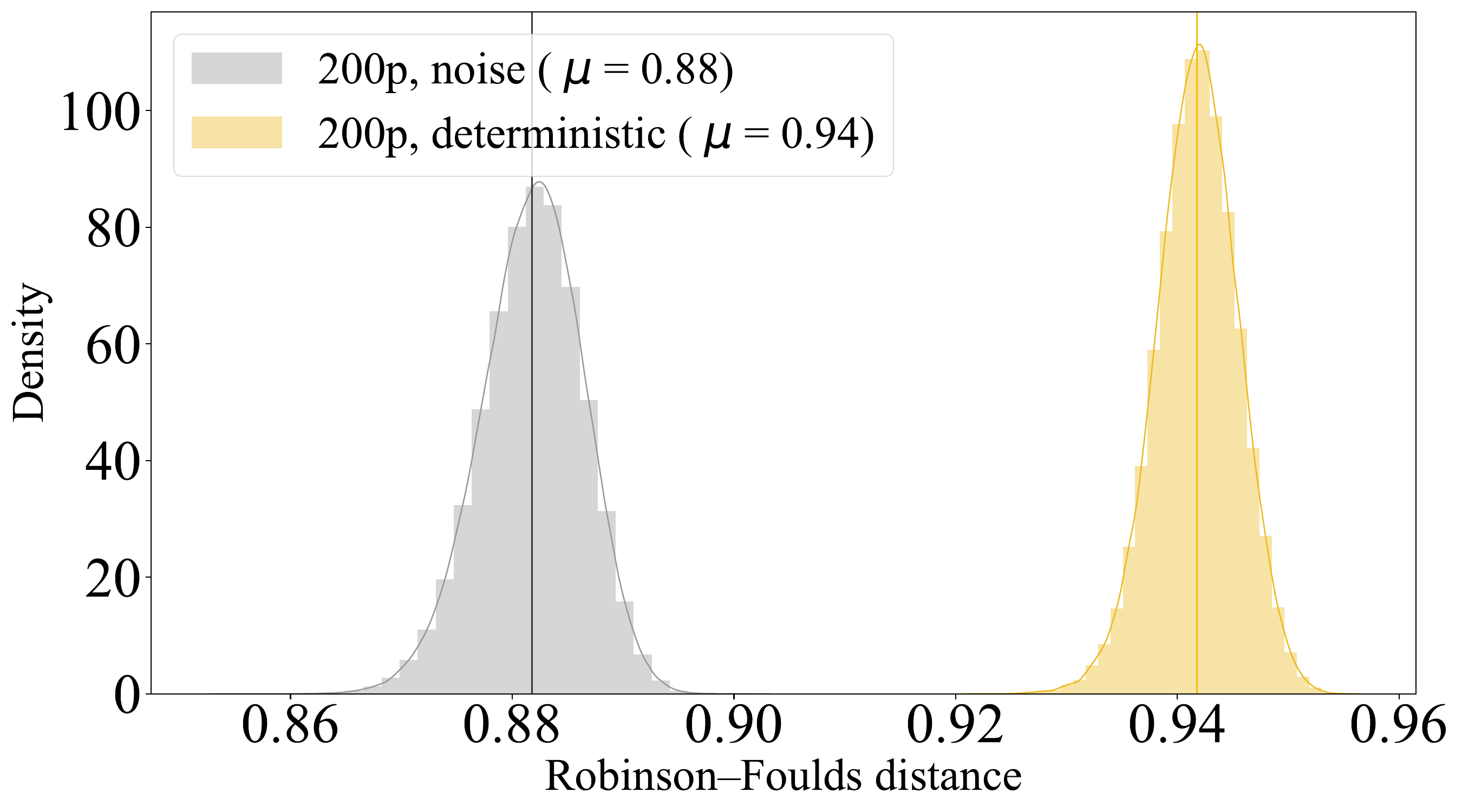}
\caption{Robinson-Foulds distance (RFD) distributions for the trees built with  stellar particles selected  from the deterministic sample compared to noise (yellow histograms) and noise only (grey histograms) samples. Top left, top right, bottom left and bottom right represent respectively the cases considering 10, 50, 100 and 200 stellar particles. Each panel contains the mean ($\mu$) RFD of the distributions. The largest the RFD, more different the trees are from each other.}
\label{figure:distance_number_particles}
\end{figure*}

The most notable difference between the trees is the branching pattern, in particular, the number of main branches. The tree from the deterministic sample shows one main branch, e.g. the tree is very asymmetric or {\it imbalanced}. Moreover, the branch lengths that connect tips to nodes are very short. We recall that nodes in biology reflect the last common ancestor of the descendant lineages. Here, since almost all the nodes have at least one descendant linage that connects directly to a tip, one might attribute that we are sampling the ancestral states and directly tracing the ancestral-descendant relationship of the stellar particles. 

The noise tree, on the contrary, has long branches, especially at the tips. All nodes are therefore a representation of a state that is very different to the tips and not directly sampled in the data.  Moreover, the internal branches are shorter than the external ones, which is a reflection that the differences in this sample are driven by randomness and not by an internal hierarchical structures since this branching pattern shows that much of the chemical distance between the stellar particles is not explained by the inferred phylogenetic relationship and it is then deposited in the tips. The tree shows an even distribution of branches which bifurcate from nodes from the root to the tips (e.g. it is a symmetric or {\it balanced} tree). 

As discussed in \cite{jackson2021using}, imbalanced trees happen when there is gradual evolution of a single lineage through time. Differences between traits can therefore be traced as information passed through generations but they still might represent the evolution of the same population. Balanced trees might reflect rather the differentiation of populations and processes which cause populations to evolve independently from each other. In astronomy so far stars or stellar particles whose chemical abundances are the result of a shared chemical evolution history produce very imbalanced trees. That was found in \cite {jackson2021using}, in Walsen et al. (submitted) and Yaxley et al. (in prep)  (both with solar twin observed data), in Eldridge et al. (in prep), and throughout this article.

Based on these findings, we can report that the trees constructed from the simulated chemical abundances successfully capture a discernible phylogenetic signal that deviates from noise. We will now investigate the minimum number of members required in the sample to attain this objective and hence, justify the use of 100 members as adopted above.

Figure \ref{figure:distance_number_particles} shows the RFD (see Sect.~\ref{sec:compare_trees}) between the deterministic and noise sample. Here we attempt to quantify the difference between a tree built from the deterministic sample and the noise sample (e.g. comparing the trees displayed in Fig.~\ref{figure:example_tree_deterministic_noise}). We consider trees built using 10, 50, 100 and 200 stellar particles. We compare 1000 times this difference by randomly selecting particles from the deterministic sample and the noise sample. The yellow distribution represents these 1000 RFD estimates. This figure also shows the RFD obtained between two noise samples. In the same fashion as with the deterministic sample, we select randomly particles 1000 times from the noise sample and compare them. The RFD distribution in this case is  represented with the grey color. We recall that the higher the RFD, the more different the trees are from each other. Therefore, when the mean RFD of the yellow distribution is larger than the mean RFD of the gray distribution, we consider we have phylogenetic signal. Also, to have trees that are generally different from noise, it is preferable that both distributions do not overlap.

In the case of building trees with 10 stellar particles, the distribution of RFD of deterministic and noise samples  overlap. The mean RFD for the deterministic sample is 0.77, while  for the noise sample alone, the mean is 0.75. The standard deviations (SDs) are respectively 0.08 and 0.09. Hence, we interpret that trees built from noise containing 10 stellar particles are not more similar to each other than they are to trees built from simulated data.  When using 50 stellar particles to build trees, the distributions of RFD become more different, but the tails in the distributions still overlap. The mean RFD for the comparison between deterministic sample is 0.90, while the mean for the noise sample alone is 0.85. The SDs are 0.01 and 0.02, respectively. With 50 stellar particles, we interpret that phylogenetic trees built from noise are more similar among each other than they are compared to a tree built using abundances from simulated data.

In the cases of 100 and 200 stellar particles, distributions of RFD do not overlap but the mean of the deterministic distributions increases. In both cases the RFD are on average larger in the comparison of trees made from the deterministic and noise samples, than among trees from only the noise sample. This indicates that phylogenetic trees built from noise are more similar to each other than they are to phylogenetic trees from simulated data containing 100 and 200 stellar particles. In the case of 100 stellar particles, the mean of the RFD between random and deterministic is 0.93 and in the case of 200 particles that mean is 0.94. The SDs are respectively 0.01 and 0.01. The mean RFD of noise against noise particles are 0.87 (with a SD of 0.01) and 0.88 (with a SD of 0.01) for 100 and 200 stellar particles  respectively.

We conclude that the more particles we consider, the more our trees are different from a random distribution, but using 50 particles or less might still produce some phylogenetic trees whose topologies are comparable with a random tree. When using 100 particles however, we obtain trees that are always different from noise, therefore we use 100 particles from  now on to interpret the phylogenetic signal of our data and reconstruct the history of our simulated galaxy. We note that this result might differ when considering more complex cases or a different resolution for the simulation and it is possible that more stellar particles might be required in those scenarios to reliably represent the evolutionary history of the system.

\begin{figure}
\centering
\includegraphics[width=9cm]{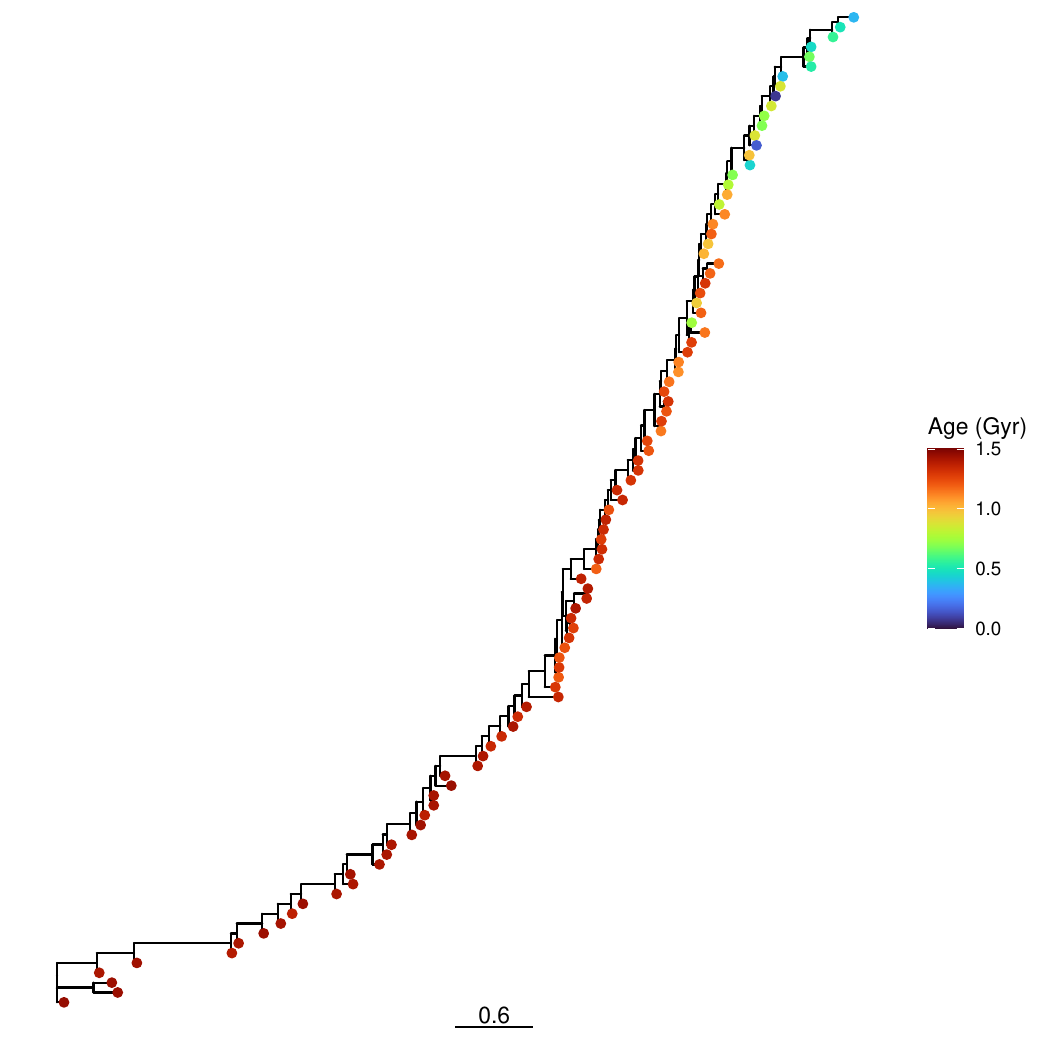}
\caption{Example of phylogenetic tree of deterministic sample (same as presented in Figure \ref{figure:example_tree_deterministic_noise}) color-coded according to age.}
\label{figure:derministic_trees_colorcoded}
\end{figure}

\subsubsection{Phylogenetic signal from the deterministic sample}

We consider the difference between the tree built from abundances resulting of a simulation and the tree built from noise as a proxy for phylogenetic signal. We now investigate if our tree can help us to reconstruct the history of the deterministic sample. Figure~\ref{figure:derministic_trees_colorcoded} shows the example of deterministic tree color-coded according to age. This is the same tree as the one showed in Fig.~\ref{figure:example_tree_deterministic_noise}.  We can see  that this tree has its internal nodes rank-ordered according to the ages.  Moreover,  the oldest particles are closer to the root, but that is expected if we have used the most ancient particle to root the tree.  Given the AMR of this sample (see top middle panel of Fig.~\ref{figure:astrophysical_properties}), it does not come as a surprise that the tree will have clear directional evolution, since our tree uses [Fe/H] as one of the traits in the distance matrix. The AMR is flat below ages of approximately 1.2 Gyr and this lack of sensitivity leads to a worst age-ranking at the top of the phylogenetic trees relative to the base of the trees. Figure \ref{figure:trees_deterministic_other} shows this tree but with particles are coloured by their [O/Fe].

There is a section in the tree where the neighbouring particles do not necessarily have very similar ages. This coincides with the sector on which [O/Fe] mixes. It is possible that this is related to the moment in which SNIa events start to occur, which changes the overall chemical enrichment rate. Considering that the distance matrix uses a mix of elements coming from SNII and SNIa, if the rates of their production varies during the history of the galaxy, it might cause that particles of different ages to be chemically more similar than coeval particles. This section in the tree corresponds to ages below 1.0 Gyr, which is when the star formation slows down, the AMR becomes flatter and the [O/Fe] reaches solar values.

It is further interesting to note the branch lengths between nodes in this tree, becomes shorter along the path of the tree. This might be related to the star formation history. At earlier stages of the history, when star formation is at its peak, there is a notable change in chemical abundances, which is represented by the steep AMR (see Fig.~\ref{figure:astrophysical_properties}). In the beginning the gas is very metal-poor, therefore any enrichment is significant compared to its surroundings. This causes long branches. As the star formation slows down, the difference in chemistry becomes smaller, the AMR flatter, and the branch lengths shorter. Therefore, the branching pattern illustrates that the rate of chemical enrichment declines.

Because our tree is asymmetric (e.g. it presents only one main branch), and that has rank-ordered ages, it reflects the result of one single history.  This is consistent with the fact that our simulated galaxy did not experience  interaction with another chemical-enriched galaxy, causing mixing of pre-processed gases or inflow of pristine gas from filaments. 

\subsection{Phylogenetic signal considering uncertainties}{\label{sec:results_uncertainties}}

In the previous section, we defined the minimum number of stellar particles necessary in order to have enough phylogenetic signal to have trees that represent the evolutionary history of our studied galaxy. In this section, we investigate the maximum uncertainties on the chemical abundances for which the phylogenetic trees are evolutionary informative. Using the example tree of the deterministic sample (Section \ref{sec:results_number_particules}), we explore the effect chemical abundances uncertainties have on the phylogenetic signal and how they affect the evolutionary history we can interpret from the tree. 

For the purpose of this analysis, we perturbed the chemical abundances of the 100 stellar particles from the deterministic sample considering six uncertainty values: 0.01, 0.05, 0.08, 0.1, 0.2 and 0.3 dex. Abundances with precision below 0.05 dex fall in the high-precision domain and are rather obtained when analysing very high resolution and high signal-to-noise spectra \citep[e.g.][]{ Nissen&Gustafsson2019} or using machine learning tools when large samples of reference stars are available for training a good model (\citealt{ness-15, LeungBovy,Wheeler-20,Ambrosch-23}, Walsen et al. sumitted). Standard spectral analyses have abundance precision that are rather of the order of $0.1-0.2$ dex. A precision of 0.3 dex is understood as a large uncertainty, but is unfortunately still very common for studies, particularly for faint stars for which the signal-to-noise is not very high, as for example for halo stars. 

\begin{figure}
\centering
\includegraphics[width=10cm]{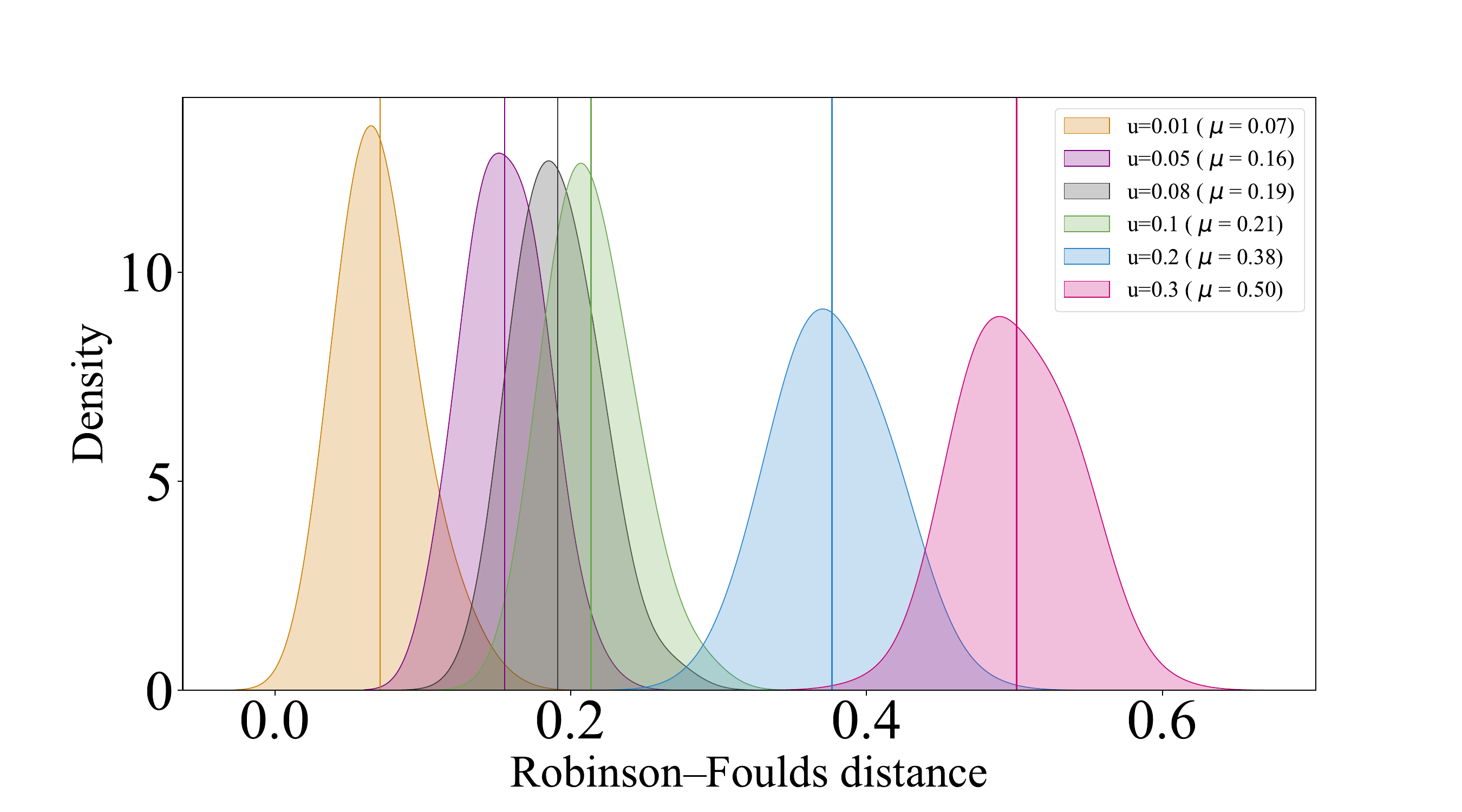}
\caption{Robinson-Foulds distance (RFD) when comparing the deterministic tree and those with chemical abundances uncertainties of 0.01, 0.05, 0.08, 0.1, 0.2 and 0.3 dex. The means of these distributions are shown in the legend. The larger the uncertainty, the more different trees become from the original.}
\label{figure:treedist_uncertainty}
\end{figure}

In order to account for these uncertainties, we created new values of chemical abundances for each stellar particle. The new values consider a normal distribution with the mean as the original value and the standard deviation as the corresponding uncertainty considered. Using the perturbed chemical abundances we built new trees for the deterministic sample, which we compare with the original tree. In Figure \ref{figure:treedist_uncertainty}, we show the RFD between the original tree and the the trees built considering uncertainties of 0.01, 0.05, 0.08, 0.1, 0.2 and 0.3 dex with different colours.

\begin{figure*}
\centering
\includegraphics[width=5cm]{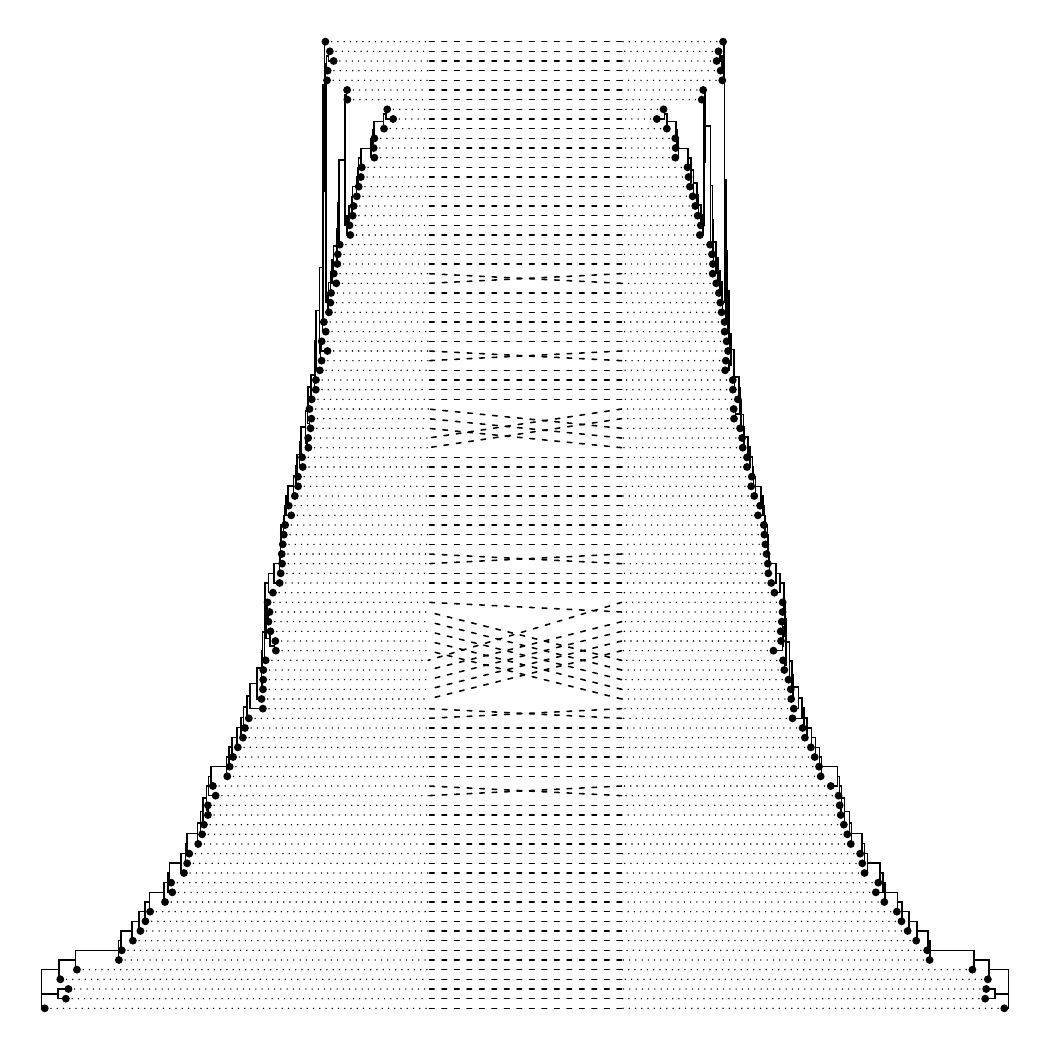}
\includegraphics[width=5cm]{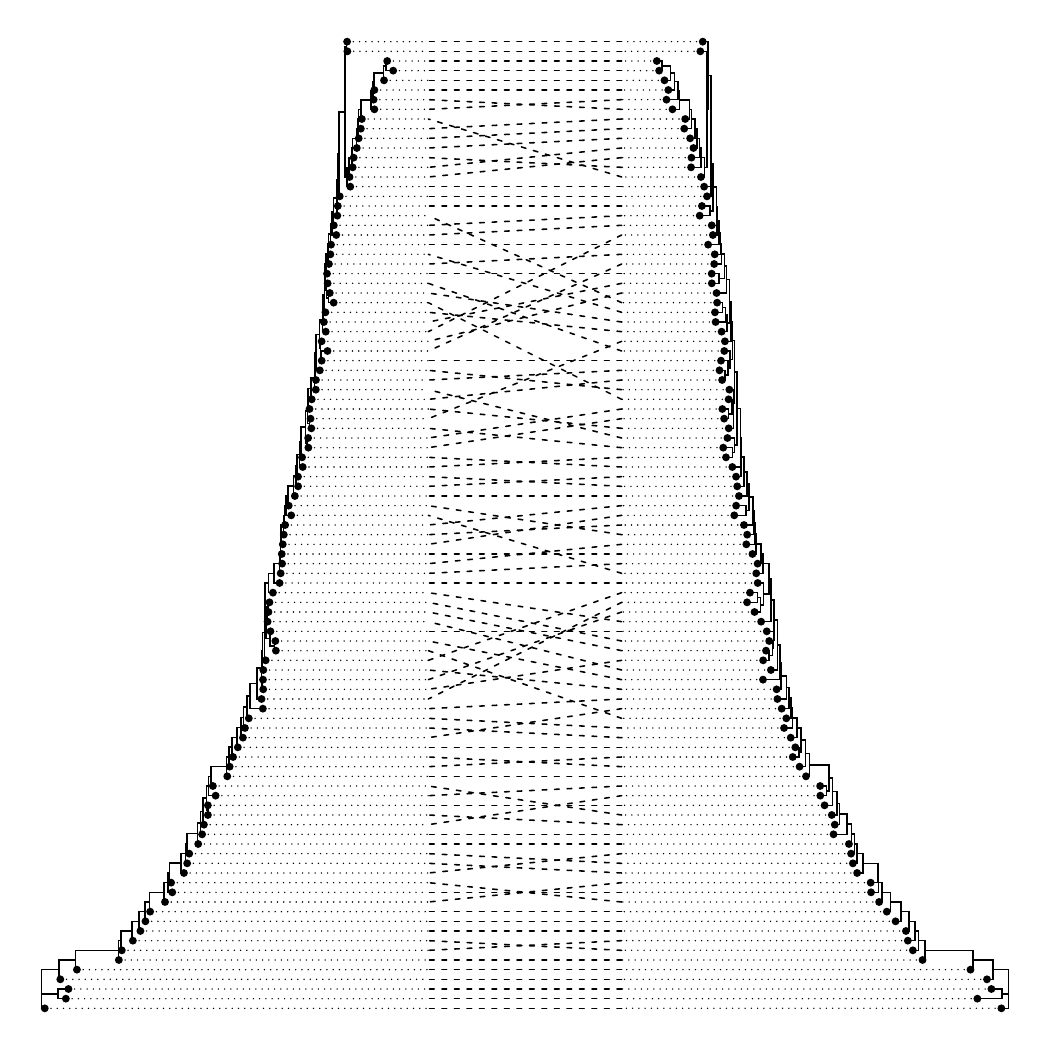}
\includegraphics[width=5cm]{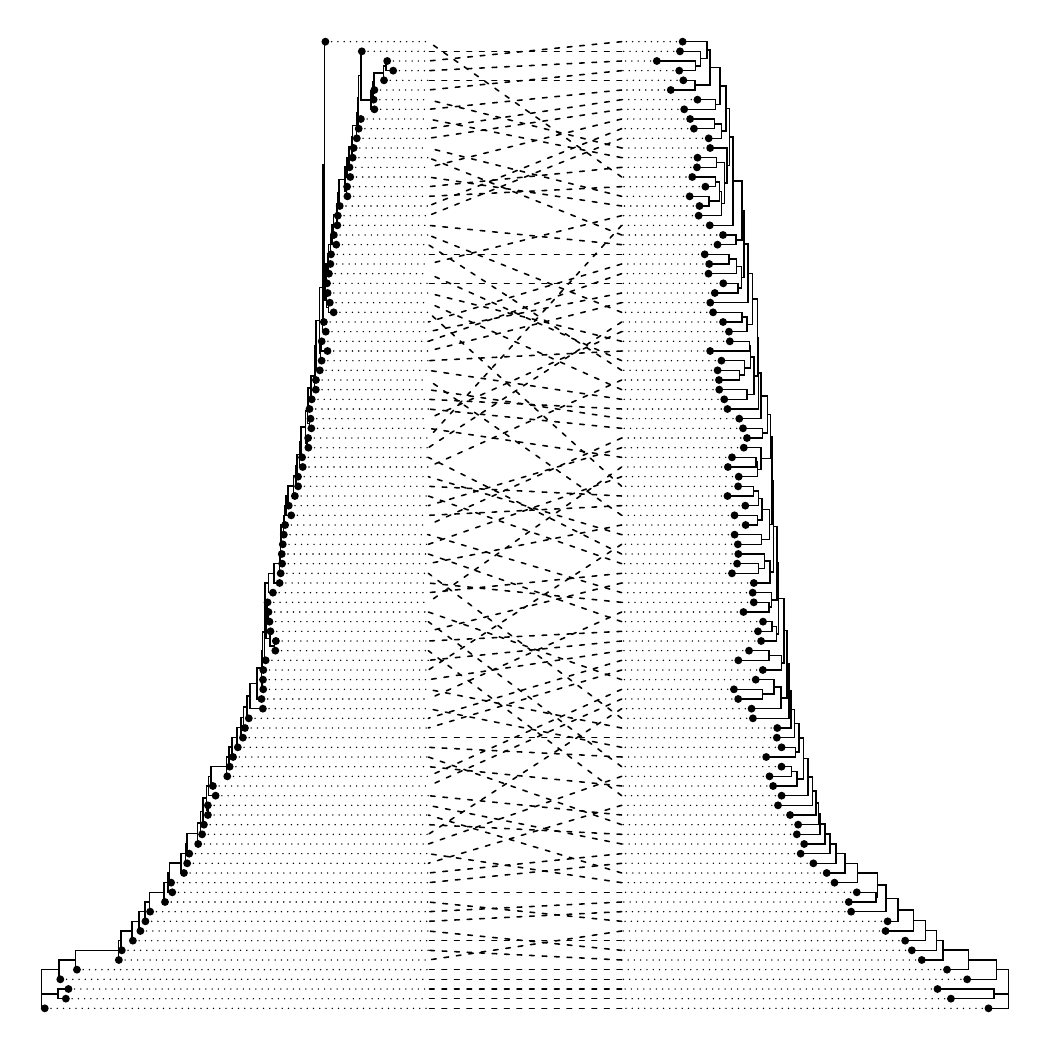}
\caption{Comparison between the deterministic tree (left tree in all three examples) and a new tree built by perturbing the abundances (right trees in all three examples) within a range of 0.01 (left), 0.1 (middle) and 0.3 (right) dex. Dashed lines connect same particle in each tree. }
\label{figure:cophylo_uncertainties}
\end{figure*}

The RFD shows that the yellow distribution has a mean of 0.07 (and SD of 0.03), indicating that considering uncertainties within 0.01 dex do not significantly change the trees. As uncertainties increase, the RFD increase as well, which is expected. For an uncertainty of 0.3 dex, trees deviate from the original one reaching a mean RFD of 0.50 (and SD of 0.04). We note that this value is still lower than the mean RFD of  0.93 for the comparison of the deterministic and the noise sample when 100 particles are considered. This suggests that while the trees with uncertainties of 0.3 dex differ among each other, there is still some phylogenetic signal as they are still distinct from pure noise. Additionally, to have a better idea of how the trees change when abundances are modified,  we show in Fig.~\ref{figure:cophylo_uncertainties} the link between two example trees for 3 cases of abundances. The left-hand trees are always the deterministic tree with no uncertainties in the chemical abundances, and the right-hand trees correspond to one example tree obtained by perturbing the abundances, considering 0.01, 0.1 and 0.3 dex, respectively.  The dashed lines in each case connect the same particle in each tree. 

From the lines showed in Fig.~\ref{figure:cophylo_uncertainties}, we can see that when abundances have an uncertainty of 0.01 dex, 24 of the 100 particles change their labelling order (locations in the tree). Their new location is relatively close from the original tree, as expected if the change in abundance is small. In the case of an uncertainty distribution of 0.1 dex, 60 out of 100 stellar particles change their places. Finally in the case of 0.3 dex uncertainties, 90 stellar particles change place in the tree. The new positions are quite far from the original tree.  It is interesting to note the gradual increase of the branch lengths when uncertainties increase. This is also seen in the noise tree (see Fig.~\ref{figure:example_tree_deterministic_noise}) and in Walsen et al. (submitted), who compared trees built from observed stars whose abundance measurements have different uncertainties. This tree, however, is different to the noise tree,  as expected from the different RFD value obtained here and in Sect.~\ref{sec:results_number_particules}. The tree with 0.3 dex uncertainty is in fact still very imbalanced, unlike the noise tree. 

\begin{figure*}
\centering
\includegraphics[width=18cm]{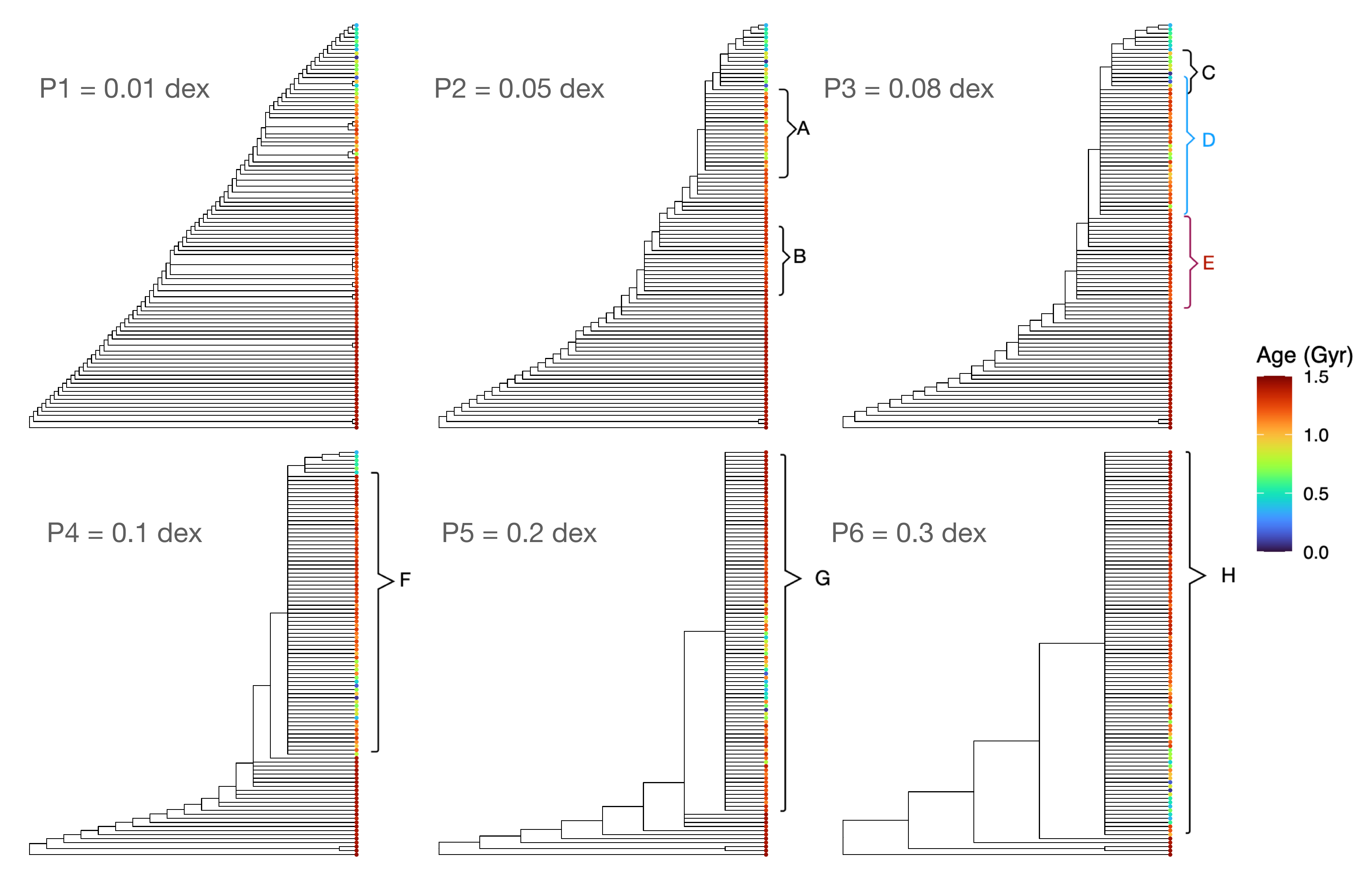}
\caption{Consensus trees topologies color-coded according to ages. Top: trees built considering chemical abundances uncertainties of the order of 0.01, 0.05 and 0.08 dex respectively. Bottom: trees built considering chemical abundance uncertainties of the order of 0.1, 0.2 and 0.3 dex respectively. The polytomies in each trees are indicated as A, B, C, D, E, F, G and H.}
\label{figure:consensus_trees}
\end{figure*}

While Figure \ref{figure:cophylo_uncertainties} shows the displacement of stellar particles when considering uncertainties, it is fundamental to evaluate if different chemical abundances would still carry evolutionary information to reconstruct the shared history of the selected stellar particles. It is thus necessary to study the support of a tree with uncertain abundances has in this context.  To do so, we computed 1000 trees by perturbing the abundances and collected these trees in a  majority rule consensus tree (see Sect.~\ref{sec:compare_trees}).  Figure \ref{figure:consensus_trees}  shows the consensus trees when considering the uncertainties of 0.01, 0.05, 0.08, 0.1, 0.2. 0.3 dex. We note that only the tree topology is shown, since the branch length of consensus trees can not be directly related to the branch length of an actual sampled tree, which is the result of a distance matrix. 

 In this work, we aim to focus on the branching pattern of the nodes and the age ranking of the selected nodes, we do not focus on the branch lengths. By collapsing nodes into multifurcations when nodes are conflicting in a sample of phylogenetic trees, we are reducing the number of total nodes in a tree, which essentially means reducing the resolution in which the shared history can be extracted. It is not trivial to define a limit of the maximum number of nodes that can be reduced from a sampled tree to a consensus tree that means a significant loss of the phylogenetic signal, but it is clear that if we allow multifurcations in our trees, they should be somehow distributed along the tree such that groups of stellar particles can be distinguished in e.g. their mean ages.  That means, a polytomy that contains more than 50\% of the particles which span the entire age range is not evolutionary informative. 

 Figure \ref{figure:consensus_trees}   shows consensus trees made with sampled trees that consider different abundance uncertainties. The top left panel (P1)  considers an abundance uncertainty of 0.01 dex, and shows that overall most nodes are present in more than 50\% of the sampled trees. That tree has very few multifurcations, with four branches rising from a node at most. Moreover, these polytomies are at a significant distance from the root. Overall the age-ranking of the branches remains, thus we conclude that uncertainties of 0.01 dex do not affect the phylogenetic signal of an evolutionary tree of these properties. 

 When focusing on the middle top panel of Fig.~\ref{figure:consensus_trees} (P2), we see the consensus tree topology obtained from trees sampled considering an uncertainty of 0.05 dex. As expected, the number and size of the polytomies increase. In this case, we find two significant multifurcations, labelled as A and B. The particles in the polytomy B are mainly old stellar particles, while the stellar particles in the polytomy A are intermediate-age particles. The age-ranking in the tree is kept, even if the relation of age and distance from the root is not as tight as in the deterministic tree (see Figure \ref{figure:derministic_trees_colorcoded}). The polytomy B is closer to the root than the polytomy A. 
 
 The top right panel (P3) shows the consensus tree with uncertainties of 0.08 dex. Close to the root, the tree is still resolved, but it becomes less resolved with further out from the root. We label three significant polytomies, C, D and E. As in the previous case, these polytomies contain stellar particles that overall have different ages, with the polytomy C containing young stellar particles, D containing intermediate-age stellar particles, and E containing old stellar particles. Polytomy E is at a comparable distance from the root than Polytomy B in the tree of Panel 2. We thus conclude that while the age-ranking of the nodes has a large scatter, the ranking is still present and therefore with uncertainties of 0.08 dex, we are still able to reconstruct a history from a phylogenetic tree. 
 
 The situation with uncertainties above 0.1 dex is more critical. Consensus trees with uncertainties of 0.1, 0.2 and 0.3 are shown in the lower panel of Fig.~\ref{figure:consensus_trees}, in Panels 4, 5 and 6 (P4, P5 and P6). Here we are able to label only one significant polytomy per tree, that is, F, G and H, respectively. They contain stellar particles of all ages, and contain a significant fraction of the particles of the sample. In these consensus trees it is not possible to arrange the star particles according to their ages in the tree, and therefore it is not possible to reconstruct the evolutionary history of this galaxy. We further find that as the uncertainty increases the polytomy becomes deeper in the tree. For an uncertainty of 0.3 dex, the polytomy is a few nodes away from the root. 
 
The fact that only close to the root we are able to resolve the tree in these cases is due to the significant change in metallicity at old ages (see AMR in Figure \ref{figure:astrophysical_properties}), which is related to the peak in SFH. When the star formation is less extreme, and the AMR does not present a significant change arriving to a plateau, uncertainties above 0.1 dex in abundance measurements do not allow us to study the evolution of that system using phylogenetic trees.  

\subsection{Evolutionary history considering different regions of the galaxy}{\label{sec:results_regions}}

While in Section \ref{sec:results_number_particules} we investigated the dependence of the phylogenetic signal on the population density and in Section \ref{sec:results_uncertainties} we explored the dependence of phylogenetic signal on the uncertainties in the chemical abundances, in this section we explore how the AMR and SFH of different regions of the galaxy impact the properties of phylogenetic trees.

In Section \ref{sec:results_number_particules}, we discussed the evolutionary history traced by phylogenetic trees from the deterministic sample. In this section we repeat that analysis using phylogenetic trees from different regions of the galaxy. We thus analyze the trees built from the example samples of Groups 01, 02, 03 and 04, whose spatial distributions are shown in Fig.\ref{figure:position_regions} and astrophysical properties in Fig.~\ref{figure:astrophysical_properties} with the colours green, blue, pink and red, respectively. The chosen 100 stellar particles are used to build and analyse the trees of this section. 

\begin{figure*}
\includegraphics[scale=0.28]{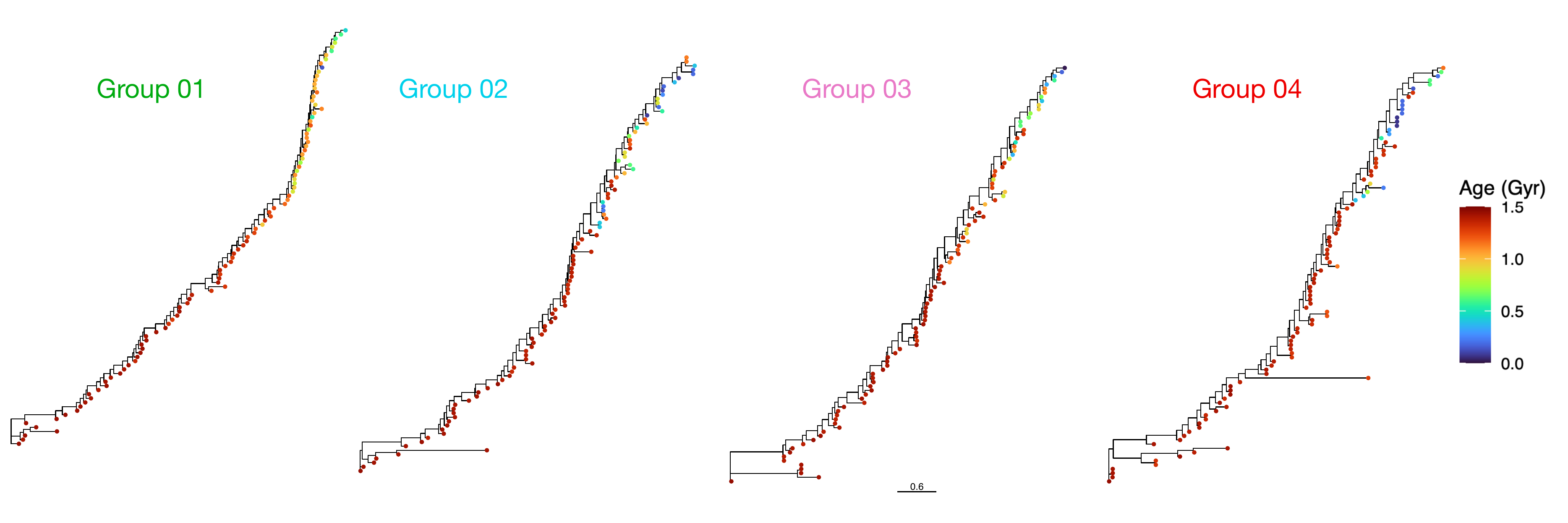}
\includegraphics[scale=0.28]{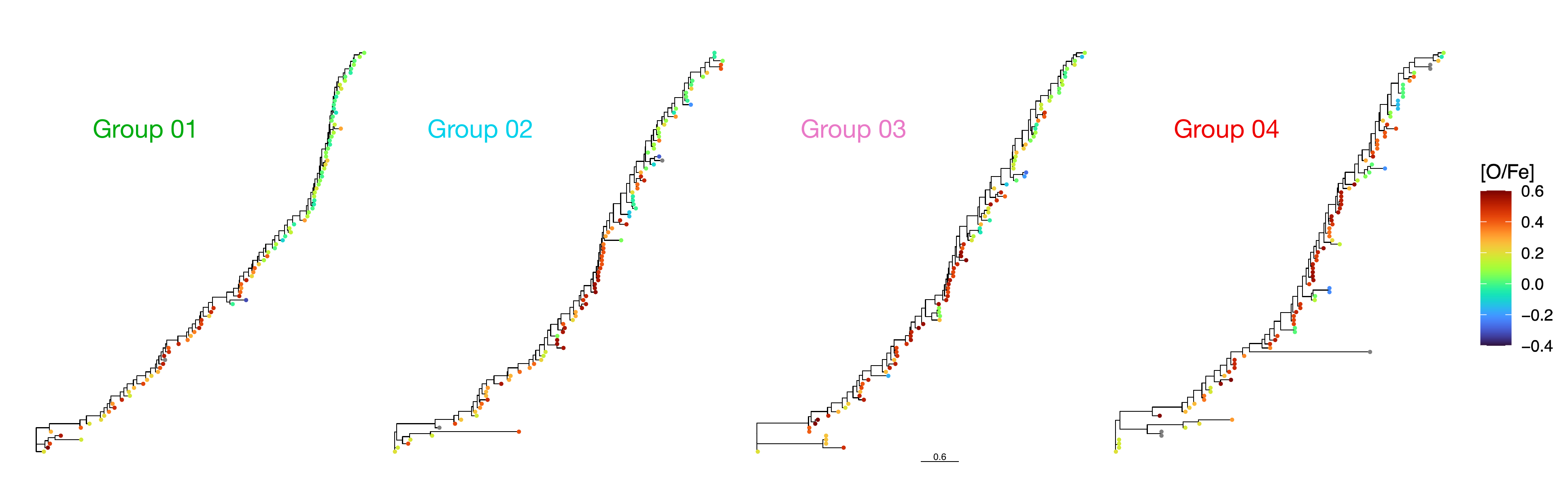}
\caption{Phylogenetic trees of a selection of 100 stellar particles from the groups selected in the different regions shown in Figure \ref{figure:position_regions}. Tips are color-coded according to ages (upper panels) and to [O/Fe] (lower panels).}
\label{figure:area_trees_age}
\end{figure*}

Figure~\ref{figure:area_trees_age} shows the trees of each group, with the stellar particles color-coded according to age in the top row, and according to [O/Fe] in the bottom row. Similarly to the tree built using the deterministic sample, these trees are imbalanced, and show rank-ordered ages, implying that everywhere in the galaxy we can reconstruct history. The branching order of the ages, however, becomes weaker from Group 01 to Group 04. This might be an effect of the SFH, whose peak becomes narrower towards the edge of the galaxy (see Fig.~\ref{figure:astrophysical_properties}). This translates into a flatter AMR for stellar particles younger than about 1.2 Gyr. 

\begin{figure*}
\centering
\includegraphics[width=12cm]{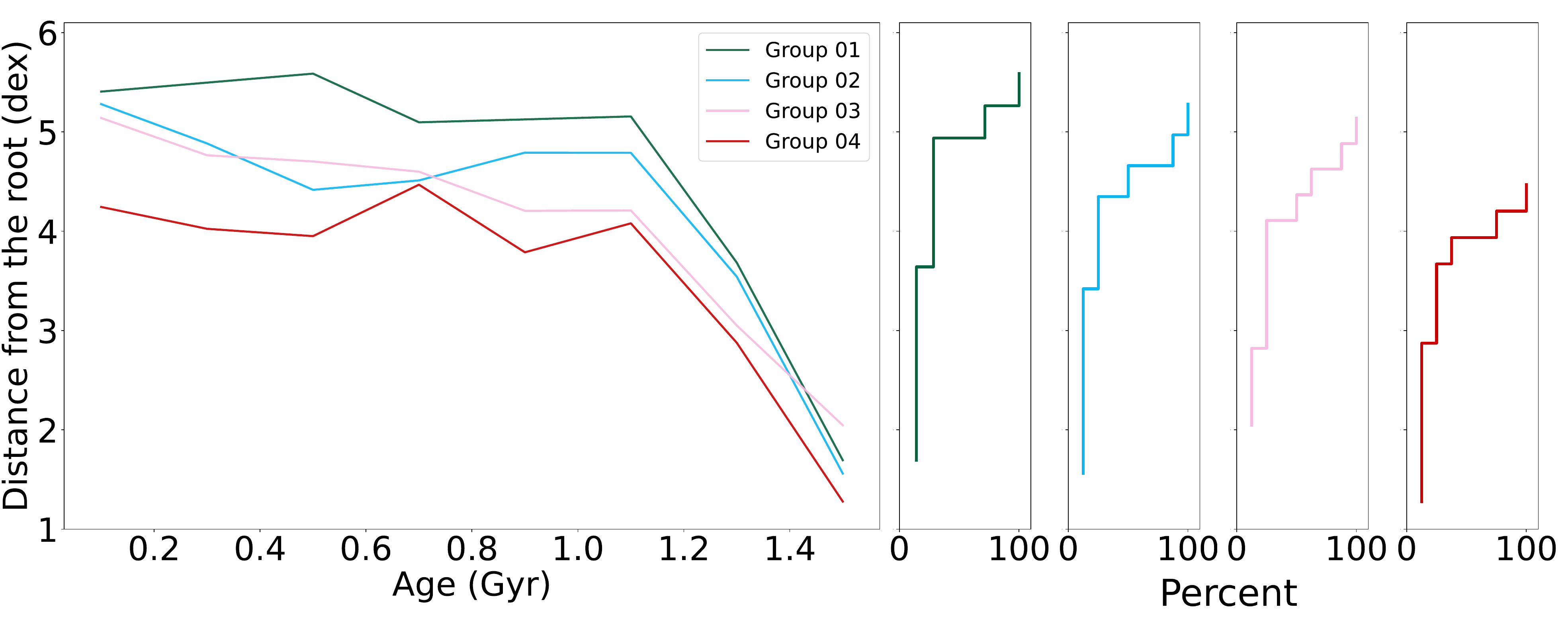}
\caption{Left: Cumulative distances from the root to the tip as a function of the ages of stellar particles. Groups 01, 02, 03 and 04 are represented as green, blue, pink and red lines, respectively. Right: Cumulative percentage of stellar particles contained in bins of distance from the root. In the left panel it is shown that the distance from the root reaches a plateau or a region with slow increase in the same region that contains the majority of stellar particles according to the right panel.}
\label{figure:distance_from_root}
\end{figure*}

All trees show the  presence of an apparent second branch of very old stellar particles, which are close to the root. The trees here have been rooted using the oldest star particle, but that does not imply that this particular stellar particle is a common ancestor to the rest of the stellar population. At the beginning of the simulation, there is significant homogeneity  in the distribution of metals in the gas that reflects the local distribution of the cold gas from which stars are formed. This has an impact in how chemical evolution due to the first supernovae enriches the ISM. At the very first stages of evolution, the metallicity of the ISM is strongly heterogeneous. As star formation progresses the regions became more chemically enriched and mixed and the exchange of enriched material between regions could take place (e.g SN outflows, radial migration). However, as we moved from the central to the outer regions, the level of enrichment systematically decreases even though the AMR shapes are similar. This decrease in the global metallicity with radius is expected for galaxies with an exponential gas density distribution as the simulated galaxy used in this study.

In order to better quantify the different trees and so discuss the rate of change in the chemical distance from the root to each tip, we calculate the distances of each tip to the root.  Figure \ref{figure:distance_from_root}  shows the cumulative chemical distance from the root of stellar particles as a function of their ages in the left panel and the distribution of distances for each sample in the remaining panels. We can first observe that in all the four groups there is a sharp increase of the distance for the oldest stellar particles, with few tips having short distances from the root. At around 1.2 Gyr, the distance reaches a more or less constant value, which ranges between 3.5 dex and 5.5 dex approximately depending on the group. The point when the sharp increase in the distance from the root stops is related to when the peak of star formation ends in each region according to their SFH (see Figure \ref{figure:astrophysical_properties}).

Considering that Group 01 corresponds to the galactic center, that Group 04 corresponds to an outskirt of the galaxy, and that Group 02 and 03 are in the middle, it is encouraging to notice that the largest maximum distance is reached by the tree built from stellar particles in Group 01 and that the shortest maximum distance is built from the Group 04. From Fig.~\ref{figure:astrophysical_properties} we know that the SFH between Group 01 and Group 04 is different, in the sense that the central region experienced a long peak of star formation and continued forming stellar particles until present date, while the outer region experienced a short star formation peak, with an abrupt stop and almost no recent star formation. This translates into an AMR of a population that increases in metallicity until present day for the Group 01, while for Group 04 the AMR is rather flat.

It is thus expected that a tree path which is drawn from a population with more star formation will be longer. From our results we find that indeed trees can be used to learn about the SFH of galaxies, since the difference in the total length-path of the tree (i.e. distance from the root) is large (2 dex), even if the AMR or the [O/Fe]-[Fe/H] planes are comparable.This shows that the tree enhances the differences. We note that another advantage of using tree path length to study the efficiency of star formation is that it is not necessary to know accurately the ages of the particles. This is an advantage because determining stellar ages is a challenging task. 

The right-hand panel of Fig.~\ref{figure:distance_from_root} shows how the distribution of distances from the root is different for the different groups. The group with higher and more extended SFH reaches higher lengths than the group with lower SFH. The latter has a wider distribution of lengths between 2 and 4 dex, reflecting also the scatter in the AMR. In the case of Groups 02 and 03, the SFH is very similar in both cases. There are more stellar particles formed in Group 03 than 02 due to the higher gas density in Group 03, which is where the spiral arm lies. From the AMR or the [O/Fe] vs [Fe/H] diagrams the impact on the gas density is difficult to identify, and the same can be said considering the length of the tree.

Figure~\ref{figure:distance_from_root} can be related to the AMR, since the metallicity is one of the traits in the tree distance matrix. It is therefore not surprising that the age-branch length relation will be very similar to the AMR. The tree branch lengths incorporate the other chemical abundances, in addition to the Fe, which is why it covers a larger range in chemistry. Since we are using all abundances relative to hydrogen, all elements are expected to increase with time, making the chemical distance increase in a way that directly relates to the increase in metallicity. This is valid for the studied system, which does not experience infall of pristine gas. Moveover, the distance matrix uses [Zn/H], which are also produced by SNIa. They follow a comparable evolution to [Fe/H], and cause the relation between branch length and age observed in Fig.~\ref{figure:distance_from_root}. 

\section{Prospects and limitations of stellar phylogeny}

As previously mentioned, stellar phylogeny had already being applied to observational data (\citealt{jofre2017cosmic,jackson2021using}, Walsen et al. submitted). However, this is the first time it is have been applied to simulations. As this is the first study of its kind, using an isolated disc galaxy simulation serves as an ideal test case and a fundamental step to mature the method before applying it to more complex systems, that can better represent real galaxies.

Interactions play an important role in the evolution of galaxies \citep{toomre1977mergers,efstathiou1990galaxy, barnes1992dynamics}. When a galaxy undergoes mergers, for instance, both its stellar population and gas content experience alterations \citep{torrey2012metallicity,
monachesi2019auriga}. Additionally, such events can trigger episodes of star formation, further impacting the galaxy's chemical composition and stellar populations (as illustrated in \citealt{di2007star}). Consequently, the environment becomes more complex, making the application of stellar phylogeny more delicate. We expect that in more complex systems that experience interactions, the results regarding stellar phylogeny can possibly be impacted by the mass ratio of the galaxies and also their amount of available gas. While a comprehensive investigation into how mergers affect phylogenetic trees is currently a work in progress, we anticipate that a meticulous selection of stellar particles or stars (in the case of observational studies) will be fundamental for conducting stellar phylogeny in more complex systems. The selection will be crucial both to build phylogenetic trees that are evolutionary informative, but also in order to have a robust interpretation of the results.

Another factor that requires further characterization is how stars born from the same molecular cloud, but having different masses can be addressed in stellar phylogenetic studies. The inclusion of stars with a wide range of masses can introduce an additional layer of complexity, since different stellar evolution processes rule stars with difference masses, potentially altering chemical abundances in the atmospheres of stars. Using massive stars might complicate the application and interpretation of stellar phylogeny, due to potential alterations in their chemical abundances resulting from internal processes, such as mass loss and mixing \citep{meynet2000stellar, langer2012presupernova,martins2015mimes}. However, low-mass stars can also have their chemical composition altered by processes such as atomic diffusion and rotation \citep{deal2020chemical}.

A better characterization of the limits of chemical tagging would benefit the development of stellar phylogeny. We acknowledge the significance of exploring the effects of studying stars from the same molecular cloud but with different masses to better characterize this method. However, such an investigation falls outside the scope of this work, since here the stellar particles represent stellar populations, where such effects are not included.

\section{Summary and conclusions}{\label{sec:discussion_conclusion}}

In this study, we investigated phylogenetic signal within a simulated disc galaxy, addressing three specific questions. First, we explored the dependence of phylogenetic signal on population density. Second, we investigated the dependence of phylogenetic signal on the uncertainties associated with the chemical abundances. Third, we studied the dependence of the properties of the phylogenetic trees with different regions of the simulated disc galaxy.

Approaching the first question, we explored the minimum number of stellar particles required to obtain phylogenetic signal and reconstruct the galaxy's evolutionary history. This was done because it is fundamental to be able to differentiate phylogenetic signal from noise and stochasticity. In this analysis we varied the number of stellar particles from 10 to 200 and found that using 100 stellar particles allowed for the reconstruction of this galaxy's history. For 100 stellar particles, the distributions of Robinson-Foulds distances (RFD) did not overlap when considering trees built from simulated and random data. The mean RFD considering trees from simulated samples was 0.93 with a standard deviation of 0.01, while the RFD considering  trees from random chemical abundances was 0.87 with a standard deviation of 0.01. We also observed that the topologies of the trees built using the simulated and random data were different, supporting the conclusion that phylogenetic trees from simulated data were significantly different from random noise.

In the second question, we studied the impact of uncertainties in the chemical abundances on the evolutionary history portrayed by the phylogenetic trees. In order to do so, we perturbed the chemical abundances of reference phylogenetic trees considering uncertainties in the range of 0.01 and 0.3 dex. As the uncertainty in abundances increases, the RFD between the original trees and perturbed trees also increases. Trees with uncertainties of 0.01 dex remain similar to the original tree having a mean RFD of 0.07 with standard deviation of 0.03, while those with 0.3 dex uncertainties deviate significantly, having a mean RFD of 0.50 and standard deviation of 0.04. However, even with uncertainties as high as 0.3 dex, there was still retrievable phylogenetic signal when considering trees built from random chemical abundances. We report that the resolution of phylogenetic trees decreased with higher uncertainties and that the displacement of stellar particles within the trees becomes more pronounced as uncertainties increase. Finally we observed that for uncertainties below 0.08 dex, we could successfully reconstruct the galaxy's history, since the uncertainties do not significantly affected the age-ranking of nodes in the tree and the polytomies are not the domineering structure of the trees.

In the final question approached in this work, we analyzed whether the evolutionary histories inferred from phylogenetic trees constructed using stellar particles from different regions of the galaxy were consistent with their age-metallicity relations (AMR) and star formation histories (SFH). We observed that the trees displayed one primary branch, indicating a gradual evolution of a single lineage over time. Also, the trees from the different regions displayed rank-ordered ages, with older particles closer to the root. However, there are differences between regions. Cumulative distances from the root to stellar particles revealed that the path lengths in the phylogenetic trees were related to the SFH. Regions with higher and more extended star formation activity had longer tree path lengths, while regions with lower and shorter star formation activity exhibited shorter tree path lengths. The observed differences of the cumulative distances achieved a value of 2 dex. The aspect of the path length as a function of ages were also related to the AMR of the system, with a sharp increase of the distance from the root associated with periods of rapid chemical enrichment. These findings highlight the potential of phylogenetic trees to capture variations in the SFH and AMR across different regions of the simulated disc galaxy, providing insights into its chemical and star formation history.

In summary, this work demonstrated that it is possible to use phylogenetic trees to reconstruct the evolutionary history of a simulated disc galaxy. It highlighted the relationship between phylogenetic tree properties and the AMR and SFH. This parallel between the phylogenetic trees and the global properties of a galaxy will be particularly useful when applying phylogeny to observed data of stars when the method is more mature, since usually the SFH as well as AMR of real galaxies are not fully known. We also note that a natural next step to continue this work is to explore phylogenetic trees in more realistic simulated galaxies. These results open doors for exploring several other exciting questions about archaeology of galaxies and their evolution, both with simulated and observed data applied to stellar phylogeny. 

\begin{acknowledgments}
We thank the anonymous referee for their comments. This work has been funded by Millennium Nucleus ERIS NCN2021\_017. DDBS thanks Jorge Gonz\'alez L\'opez and Kurt Walsen for the constructive discussions that made this project even more intriguing. DDBS also thanks ANID (Beca Doctorado Nacional, Folio 21220843) and Universidad Diego Portales for the financial support provided. PJ thanks the Stromlo Distinguished Visitor Program at the ANU. PBT acknowledges partial funding by Fondecyt-ANID 1200703/2020, ANID Basal Project FB210003 and ERIS Millenium Nucleus. We acknowledge the use of  the Ladgerda Cluster (Fondecyt 1200703/2020), and the National Laboratory for HPC  (NLHPC). JGJ acknowledges support from CONICYT/ANID-PFCHA/Doctorado Nacional/2021-21210846 and the CONICYT Basal project AFB-170002. E.J.J. acknowledges support from FONDECYT Iniciaci\'on en investigaci\'on 2020 Project 11200263. 
\end{acknowledgments}

\appendix

\section{Tree coloured by [O/Fe]}

\begin{figure}
\centering
\includegraphics[width=8cm]{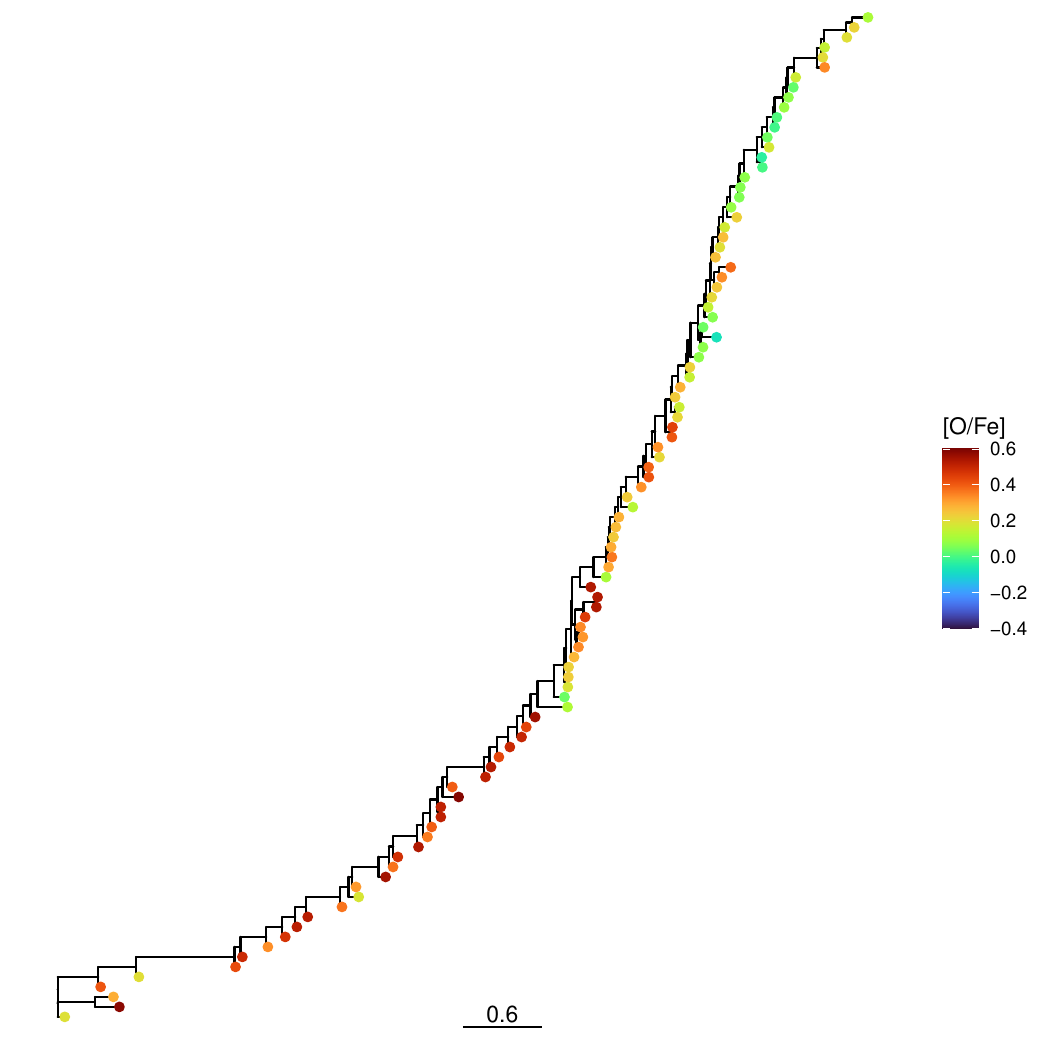}
\caption{Deterministic tree of Fig.~\ref{figure:derministic_trees_colorcoded}. Tree color coded with [O/Fe].}
\label{figure:trees_deterministic_other}
\end{figure}

Previously in this work we showed the phylogenetic tree of the deterministic sample color-coded according to the ages of stellar particles (see Figure \ref{figure:derministic_trees_colorcoded}). In the Figure \ref{figure:trees_deterministic_other} we present the same tree, but color-coded according to  [O/Fe]. In this tree it is possible to see that there is a section where [O/Fe] is mixed, which might be related to the moment in which SNIa events start to occur. This is the same region were the age-ranking in Figure \ref{figure:derministic_trees_colorcoded} is weaker.

\bibliography{bib}{}

\begin{thebibliography}{}
\expandafter\ifx\csname natexlab\endcsname\relax\def\natexlab#1{#1}\fi
\providecommand{\url}[1]{\href{#1}{#1}}
\providecommand{\dodoi}[1]{doi:~\href{http://doi.org/#1}{\nolinkurl{#1}}}
\providecommand{\doeprint}[1]{\href{http://ascl.net/#1}{\nolinkurl{http://ascl.net/#1}}}
\providecommand{\doarXiv}[1]{\href{https://arxiv.org/abs/#1}{\nolinkurl{https://arxiv.org/abs/#1}}}

\bibitem[{Abolfathi {et~al.}(2018)Abolfathi, Aguado, Aguilar, Prieto, Almeida, Ananna, Anders, Anderson, Andrews, Anguiano, {et~al.}}]{Abolfathi2018}
Abolfathi, B., Aguado, D., Aguilar, G., {et~al.} 2018, The Astrophysical Journal Supplement Series, 235, 42

\bibitem[{{Aguado} {et~al.}(2021){Aguado}, {Belokurov}, {Myeong}, {Evans}, {Kobayashi}, {Sbordone}, {Chanam{\'e}}, {Navarrete}, \& {Koposov}}]{aguado20}
{Aguado}, D.~S., {Belokurov}, V., {Myeong}, G.~C., {et~al.} 2021, \apjl, 908, L8, \dodoi{10.3847/2041-8213/abdbb8}

\bibitem[{Amarante {et~al.}(2022)Amarante, Debattista, Silva, Laporte, \& Deg}]{amarante2022gastro}
Amarante, J.~A., Debattista, V.~P., Silva, L. B.~E., Laporte, C.~F., \& Deg, N. 2022, The Astrophysical Journal, 937, 12

\bibitem[{{Ambrosch} {et~al.}(2023){Ambrosch}, {Guiglion}, {Mikolaitis}, {Chiappini}, {Tautvai{\v{s}}ien{\.{e}}}, {Nepal}, {Gilmore}, {Randich}, {Bensby}, {Bayo}, {Bergemann}, {Morbidelli}, {Pancino}, {Sacco}, {Smiljanic}, {Zaggia}, {Jofr{\'e}}, \& {Jim{\'e}nez-Esteban}}]{Ambrosch-23}
{Ambrosch}, M., {Guiglion}, G., {Mikolaitis}, {\v{S}}., {et~al.} 2023, \aap, 672, A46, \dodoi{10.1051/0004-6361/202244766}

\bibitem[{Atteson(1997)}]{atteson1997performance}
Atteson, K. 1997, in International Computing and Combinatorics Conference, Springer, 101--110

\bibitem[{Barnes \& Hernquist(1992)}]{barnes1992dynamics}
Barnes, J.~E., \& Hernquist, L. 1992, Annual review of astronomy and astrophysics, 30, 705

\bibitem[{Baum {et~al.}(2005)Baum, Smith, \& Donovan}]{Baum2005}
Baum, D.~A., Smith, S.~D., \& Donovan, S. S.~S. 2005, Science Perspectives, 310

\bibitem[{Belokurov {et~al.}(2018)Belokurov, Erkal, Evans, Koposov, \& Deason}]{Belokurov2018}
Belokurov, V., Erkal, D., Evans, N., Koposov, S., \& Deason, A. 2018, Monthly Notices of the Royal Astronomical Society, 478, 611

\bibitem[{Bignone {et~al.}(2019)Bignone, Helmi, \& Tissera}]{bignone2019gaia}
Bignone, L.~A., Helmi, A., \& Tissera, P.~B. 2019, The Astrophysical Journal Letters, 883, L5

\bibitem[{Bromham(2008)}]{Bromham2008}
Bromham, L. 2008, Reading the story in DNA: a beginner's guide to molecular evolution (Oxford University Press on Demand)

\bibitem[{Bromham {et~al.}(2022)Bromham, Dinnage, Skirg{\aa}rd, Ritchie, Cardillo, Meakins, Greenhill, \& Hua}]{bromham2022global}
Bromham, L., Dinnage, R., Skirg{\aa}rd, H., {et~al.} 2022, Nature ecology \& evolution, 6, 163

\bibitem[{Brown {et~al.}(2021)Brown, Vallenari, Prusti, De~Bruijne, Babusiaux, Biermann, Creevey, Evans, Eyer, Hutton, {et~al.}}]{brown2021gaia}
Brown, A.~G., Vallenari, A., Prusti, T., {et~al.} 2021, Astronomy \& Astrophysics, 649, A1

\bibitem[{Buder {et~al.}(2020)Buder, Sharma, Kos, Amarsi, Nordlander, Lind, Martell, Asplund, Bland-Hawthorn, Casey, {et~al.}}]{buder2020galah+}
Buder, S., Sharma, S., Kos, J., {et~al.} 2020, Monthly Notices of the Royal Astronomical Society

\bibitem[{Buder {et~al.}(2022)Buder, Lind, Ness, Feuillet, Horta, Monty, Buck, Nordlander, Bland-Hawthorn, Casey, {et~al.}}]{buder2022galah}
Buder, S., Lind, K., Ness, M.~K., {et~al.} 2022, Monthly Notices of the Royal Astronomical Society, 510, 2407

\bibitem[{{Burbidge} {et~al.}(1957){Burbidge}, {Burbidge}, {Fowler}, \& {Hoyle}}]{Burbidge1957}
{Burbidge}, E.~M., {Burbidge}, G.~R., {Fowler}, W.~A., \& {Hoyle}, F. 1957, Reviews of Modern Physics, 29, 547, \dodoi{10.1103/RevModPhys.29.547}

\bibitem[{Campello {et~al.}(2013)Campello, Moulavi, \& Sander}]{campello2013density}
Campello, R.~J., Moulavi, D., \& Sander, J. 2013, in Advances in Knowledge Discovery and Data Mining: 17th Pacific-Asia Conference, PAKDD 2013, Gold Coast, Australia, April 14-17, 2013, Proceedings, Part II 17, Springer, 160--172

\bibitem[{Carrillo {et~al.}(2023)Carrillo, Deason, Fattahi, Callingham, \& Grand}]{carrillo2023can}
Carrillo, A., Deason, A.~J., Fattahi, A., Callingham, T.~M., \& Grand, R.~J. 2023, arXiv preprint arXiv:2306.00770

\bibitem[{Carrillo {et~al.}(2022)Carrillo, Hawkins, Jofr{\'e}, de~Brito~Silva, Das, \& Lucey}]{carrillo2022detailed}
Carrillo, A., Hawkins, K., Jofr{\'e}, P., {et~al.} 2022, Monthly Notices of the Royal Astronomical Society, 513, 1557

\bibitem[{{Chabrier}(2003)}]{Chabrier2003}
{Chabrier}, G. 2003, \apjl, 586, L133, \dodoi{10.1086/374879}

\bibitem[{Darwin(1859)}]{Darwin1859}
Darwin, C. 1859, On the Origin of Species by Means of Natural Selection (London: Murray)

\bibitem[{De~Brito~Silva {et~al.}(2022)De~Brito~Silva, Jofr{\'e}, Bourbert, Koposov, Prieto, \& Hawkins}]{deBritoSilva2022_j01020100}
De~Brito~Silva, D., Jofr{\'e}, P., Bourbert, D., {et~al.} 2022, Monthly Notices of the Royal Astronomical Society, 509, 4637

\bibitem[{Deal {et~al.}(2020)Deal, Goupil, Marques, Reese, \& Lebreton}]{deal2020chemical}
Deal, M., Goupil, M.-J., Marques, J., Reese, D., \& Lebreton, Y. 2020, Astronomy \& Astrophysics, 633, A23

\bibitem[{Di~Matteo {et~al.}(2007)Di~Matteo, Combes, Melchior, \& Semelin}]{di2007star}
Di~Matteo, P., Combes, F., Melchior, A.-L., \& Semelin, B. 2007, Astronomy \& Astrophysics, 468, 61

\bibitem[{Drummond \& Rambaut(2007)}]{Drummond2007}
Drummond, \& Rambaut. 2007, BMC Evolutionary Biology, 7

\bibitem[{Efstathiou(1990)}]{efstathiou1990galaxy}
Efstathiou, G. 1990, in Dynamics and Interactions of Galaxies: Proceedings of the International Conference, Heidelberg, 29 May--2 June 1989, Springer, 2--9

\bibitem[{Eyer {et~al.}(2022)Eyer, Audard, Holl, Rimoldini, Carnerero, Clementini, De~Ridder, Distefano, Evans, Gavras, {et~al.}}]{eyer2022gaia}
Eyer, L., Audard, M., Holl, B., {et~al.} 2022, arXiv preprint arXiv:2206.06416

\bibitem[{Felsenstein(2004)}]{FelsensteinBook2004}
Felsenstein, J. 2004, Inferring Phylogenies (Sinauer Associates)

\bibitem[{Freeman \& Bland-Hawthorn(2002)}]{Freeman2002}
Freeman, K., \& Bland-Hawthorn, J. 2002, Annual Review of Astronomy and Astrophysics, 40, 487

\bibitem[{{Gaia Collaboration} {et~al.}(2016{\natexlab{a}}){Gaia Collaboration}, {Brown}, {Vallenari}, {Prusti}, {de Bruijne}, {Mignard}, {Drimmel}, {Babusiaux}, {Bailer-Jones}, {Bastian}, \& et~al.}]{GaiaCollaboration+2016a}
{Gaia Collaboration}, {Brown}, A.~G.~A., {Vallenari}, A., {et~al.} 2016{\natexlab{a}}, \aap, 595, A2, \dodoi{10.1051/0004-6361/201629512}

\bibitem[{{Gaia Collaboration} {et~al.}(2016{\natexlab{b}}){Gaia Collaboration}, {Prusti}, {de Bruijne}, {Brown}, {Vallenari}, {Babusiaux}, {Bailer-Jones}, {Bastian}, {Biermann}, {Evans}, \& et~al.}]{GaiaCollaboration+2016b}
{Gaia Collaboration}, {Prusti}, T., {de Bruijne}, J.~H.~J., {et~al.} 2016{\natexlab{b}}, \aap, 595, A1, \dodoi{10.1051/0004-6361/201629272}

\bibitem[{{Gaia Collaboration} {et~al.}(2018){Gaia Collaboration}, Brown, Vallenari, Prusti, De~Bruijne, Babusiaux, Juh{\'a}sz, Marschalk{\'o}, Marton, Moln{\'a}r, {et~al.}}]{gaia2018gaia}
{Gaia Collaboration}, Brown, A., Vallenari, A., {et~al.} 2018, Astronomy \& Astrophysics, 616

\bibitem[{Gascuel \& Steel(2006)}]{Gascuel2006}
Gascuel, O., \& Steel, M. 2006, Molecular Biology and Evolution, 23, 1997

\bibitem[{Gray {et~al.}(2009)Gray, Drummond, \& Greenhill}]{gray2009language}
Gray, R.~D., Drummond, A.~J., \& Greenhill, S.~J. 2009, science, 323, 479

\bibitem[{Hall(2004)}]{Hall2004}
Hall, B.~G. 2004, Phylogenetic Trees Made Easy, 2nd edn. (Sinauer Associates)

\bibitem[{Hall(2013)}]{Hall2013}
---. 2013, Molecular Biology and Evolution, 30, 1229, \dodoi{10.1093/molbev/mst012}

\bibitem[{Hawkins {et~al.}(2015)Hawkins, Jofre, Masseron, \& Gilmore}]{hawkins2015using}
Hawkins, K., Jofre, P., Masseron, T., \& Gilmore, G. 2015, Monthly Notices of the Royal Astronomical Society, 453, 758

\bibitem[{Helmi {et~al.}(2018)Helmi, Babusiaux, Koppelman, Massari, Veljanoski, \& Brown}]{Helmi2018}
Helmi, A., Babusiaux, C., Koppelman, H.~H., {et~al.} 2018, Nature, 563, 85

\bibitem[{Hernquist(1990)}]{hernquist1990analytical}
Hernquist, L. 1990, Astrophysical Journal, Part 1 (ISSN 0004-637X), vol. 356, June 20, 1990, p. 359-364., 356, 359

\bibitem[{Ho(1995)}]{ho1995random}
Ho, T.~K. 1995, in Proceedings of 3rd international conference on document analysis and recognition, Vol.~1, IEEE, 278--282

\bibitem[{Holtzman {et~al.}(2018)Holtzman, Hasselquist, Shetrone, Cunha, Prieto, Anguiano, Bizyaev, Bovy, Casey, Edvardsson, {et~al.}}]{holtzman2018apogee}
Holtzman, J.~A., Hasselquist, S., Shetrone, M., {et~al.} 2018, The Astronomical Journal, 156, 125

\bibitem[{Horta {et~al.}(2022)Horta, Schiavon, Mackereth, Weinberg, Hasselquist, Feuillet, O'Connell, Anguiano, Allende-Prieto, Beaton, {et~al.}}]{horta2022chemical}
Horta, D., Schiavon, R.~P., Mackereth, J.~T., {et~al.} 2022, arXiv preprint arXiv:2204.04233

\bibitem[{{Iwamoto} {et~al.}(1999){Iwamoto}, {Brachwitz}, {Nomoto}, {Kishimoto}, {Umeda}, {Hix}, \& {Thielemann}}]{iwamoto99}
{Iwamoto}, K., {Brachwitz}, F., {Nomoto}, K., {et~al.} 1999, \apjs, 125, 439, \dodoi{10.1086/313278}

\bibitem[{Jackson {et~al.}(2021)Jackson, Jofr{\'e}, Yaxley, Das, de~Brito~Silva, \& Foley}]{jackson2021using}
Jackson, H., Jofr{\'e}, P., Yaxley, K., {et~al.} 2021, Monthly Notices of the Royal Astronomical Society, 502, 32

\bibitem[{{Jimenez} {et~al.}(2014){Jimenez}, {Tissera}, \& {Matteucci}}]{Jimenez2015}
{Jimenez}, N., {Tissera}, P.~B., \& {Matteucci}, F. 2014, ArXiv e-prints.
\newblock \doarXiv{1402.4137}

\bibitem[{Jofr{\'e} {et~al.}(2017)Jofr{\'e}, Das, Bertranpetit, \& Foley}]{jofre2017cosmic}
Jofr{\'e}, P., Das, P., Bertranpetit, J., \& Foley, R. 2017, Monthly Notices of the Royal Astronomical Society, 467, 1140

\bibitem[{Johnson {et~al.}(2022)Johnson, Conroy, Johnson, Peter, Cargile, Bonaca, Naidu, Woody, Ting, Han, {et~al.}}]{johnson2022dwarf}
Johnson, J.~W., Conroy, C., Johnson, B.~D., {et~al.} 2022, arXiv preprint arXiv:2210.01816

\bibitem[{Kuhner \& Felsenstein(1994)}]{Kuhner1994}
Kuhner, M.~K., \& Felsenstein, J. 1994, Molecular Biology and Evolution, 11, 459

\bibitem[{Langer(2012)}]{langer2012presupernova}
Langer, N. 2012, Annual Review of Astronomy and Astrophysics, 50, 107

\bibitem[{Lemey {et~al.}(2004)Lemey, Salemi, \& Vandamme}]{Lemey2004}
Lemey, P., Salemi, M., \& Vandamme, A.-M. 2004, The Phylogenetic Handbook, 2nd edn. (Cambridge University Press)

\bibitem[{{Leung} \& {Bovy}(2019)}]{LeungBovy}
{Leung}, H.~W., \& {Bovy}, J. 2019, \mnras, 483, 3255, \dodoi{10.1093/mnras/sty3217}

\bibitem[{MacKay \& Mac~Kay(2003)}]{mackay2003information}
MacKay, D.~J., \& Mac~Kay, D.~J. 2003, Information theory, inference and learning algorithms (Cambridge university press)

\bibitem[{Maddison \& Maddison(2009)}]{Maddison2009}
Maddison, \& Maddison. 2009, in Mesquite: a modular system for evolutionary analysis. Version 2.6

\bibitem[{Maiolino \& Mannucci(2019)}]{maiolino2019re}
Maiolino, R., \& Mannucci, F. 2019, The Astronomy and Astrophysics Review, 27, 3

\bibitem[{Majewski {et~al.}(2017)Majewski, Schiavon, Frinchaboy, Prieto, Barkhouser, Bizyaev, Blank, Brunner, Burton, Carrera, {et~al.}}]{majewski2017apache}
Majewski, S.~R., Schiavon, R.~P., Frinchaboy, P.~M., {et~al.} 2017, The Astronomical Journal, 154, 94

\bibitem[{Martins {et~al.}(2015)Martins, Herv{\'e}, Bouret, Marcolino, Wade, Neiner, Alecian, Grunhut, \& Petit}]{martins2015mimes}
Martins, F., Herv{\'e}, A., Bouret, J.-C., {et~al.} 2015, Astronomy \& Astrophysics, 575, A34

\bibitem[{Matsuno {et~al.}(2020)Matsuno, Aoki, Casagrande, Ishigaki, Shi, Takata, Xiang, Yong, Li, Suda, {et~al.}}]{matsuno2020star}
Matsuno, T., Aoki, W., Casagrande, L., {et~al.} 2020, arXiv preprint arXiv:2006.03619

\bibitem[{Matteucci(2012)}]{matteucci2012chemical}
Matteucci, F. 2012, Chemical evolution of galaxies (Springer Science \& Business Media)

\bibitem[{Meynet \& Maeder(2000)}]{meynet2000stellar}
Meynet, G., \& Maeder, A. 2000, arXiv preprint astro-ph/0006404

\bibitem[{Mihaescu {et~al.}(2009)Mihaescu, Levy, \& Pachter}]{Mihaescu2009}
Mihaescu, R., Levy, D., \& Pachter, L. 2009, Algorithmica, 54, 1

\bibitem[{Monachesi {et~al.}(2019)Monachesi, G{\'o}mez, Grand, Simpson, Kauffmann, Bustamante, Marinacci, Pakmor, Springel, Frenk, {et~al.}}]{monachesi2019auriga}
Monachesi, A., G{\'o}mez, F.~A., Grand, R.~J., {et~al.} 2019, Monthly Notices of the Royal Astronomical Society, 485, 2589

\bibitem[{{Mosconi} {et~al.}(2001){Mosconi}, {Tissera}, {Lambas}, \& {Cora}}]{mosconi2001}
{Mosconi}, M.~B., {Tissera}, P.~B., {Lambas}, D.~G., \& {Cora}, S.~A. 2001, MNRAS, 325, 34, \dodoi{10.1046/j.1365-8711.2001.04198.x}

\bibitem[{Navarro(1996)}]{navarro1996structure}
Navarro, J.~F. 1996, in Symposium-international astronomical union, Vol. 171, Cambridge University Press, 255--258

\bibitem[{{Ness} {et~al.}(2015){Ness}, {Hogg}, {Rix}, {Ho}, \& {Zasowski}}]{ness-15}
{Ness}, M., {Hogg}, D.~W., {Rix}, H.~W., {Ho}, A. Y.~Q., \& {Zasowski}, G. 2015, \apj, 808, 16, \dodoi{10.1088/0004-637X/808/1/16}

\bibitem[{{Nissen} \& {Gustafsson}(2018)}]{Nissen&Gustafsson2019}
{Nissen}, P.~E., \& {Gustafsson}, B. 2018, \aapr, 26, 6, \dodoi{10.1007/s00159-018-0111-3}

\bibitem[{Nissen \& Schuster(2010)}]{nissen2010two}
Nissen, P.~E., \& Schuster, W.~J. 2010, Astronomy \& Astrophysics, 511, L10

\bibitem[{{Raiteri} {et~al.}(1996){Raiteri}, {Villata}, \& {Navarro}}]{Raiteri1996}
{Raiteri}, C.~M., {Villata}, M., \& {Navarro}, J.~F. 1996, A$\&$A, 315, 105

\bibitem[{Recio-Blanco {et~al.}(2023)Recio-Blanco, Kordopatis, de~Laverny, Palicio, Spagna, Spina, Katz, Fiorentin, Poggio, McMillan, {et~al.}}]{recio2023gaia}
Recio-Blanco, A., Kordopatis, G., de~Laverny, P., {et~al.} 2023, Astronomy \& Astrophysics, 674, A38

\bibitem[{Ricker {et~al.}(2014)Ricker, Winn, Vanderspek, Latham, Bakos, Bean, Berta-Thompson, Brown, Buchhave, Butler, {et~al.}}]{ricker2014transiting}
Ricker, G.~R., Winn, J.~N., Vanderspek, R., {et~al.} 2014, Journal of Astronomical Telescopes, Instruments, and Systems, 1, 014003

\bibitem[{Robinson \& Foulds(1981)}]{robinson1981comparison}
Robinson, D.~F., \& Foulds, L.~R. 1981, Mathematical biosciences, 53, 131

\bibitem[{Saitou \& Nei(1987)}]{saitou1987neighbor}
Saitou, N., \& Nei, M. 1987, Molecular biology and evolution, 4, 406

\bibitem[{{Scannapieco} {et~al.}(2005){Scannapieco}, {Tissera}, {White}, \& {Springel}}]{scan05}
{Scannapieco}, C., {Tissera}, P.~B., {White}, S.~D.~M., \& {Springel}, V. 2005, MNRAS, 364, 552, \dodoi{10.1111/j.1365-2966.2005.09574.x}

\bibitem[{{Scannapieco} {et~al.}(2006){Scannapieco}, {Tissera}, {White}, \& {Springel}}]{scan06}
---. 2006, MNRAS, 371, 1125, \dodoi{10.1111/j.1365-2966.2006.10785.x}

\bibitem[{Smith(2020{\natexlab{a}})}]{Smith2020information}
Smith, M.~R. 2020{\natexlab{a}}, Bioinformatics, 36, 5007–5013, \dodoi{10.1093/bioinformatics/btaa614}

\bibitem[{Smith(2020{\natexlab{b}})}]{TreeDist}
---. 2020{\natexlab{b}}, TreeDist: Distances between Phylogenetic Trees. R package version 2.6.0, \dodoi{10.5281/zenodo.3528124}

\bibitem[{Smith(2022)}]{Smith2022robust}
---. 2022, Systematic Biology, 71, 1255, \dodoi{10.1093/sysbio/syab100}

\bibitem[{{Springel}(2005)}]{springel2005}
{Springel}, V. 2005, MNRAS, 364, 1105, \dodoi{10.1111/j.1365-2966.2005.09655.x}

\bibitem[{{Tinsley}(1979)}]{Tinsley1979}
{Tinsley}, B.~M. 1979, \apj, 229, 1046, \dodoi{10.1086/157039}

\bibitem[{Toomre(1977)}]{toomre1977mergers}
Toomre, A. 1977, in Evolution of Galaxies and Stellar Populations, 401

\bibitem[{Torrey {et~al.}(2012)Torrey, Cox, Kewley, \& Hernquist}]{torrey2012metallicity}
Torrey, P., Cox, T.~J., Kewley, L., \& Hernquist, L. 2012, The Astrophysical Journal, 746, 108

\bibitem[{Van~der Maaten \& Hinton(2008)}]{van2008visualizing}
Van~der Maaten, L., \& Hinton, G. 2008, Journal of machine learning research, 9

\bibitem[{{Wheeler} {et~al.}(2020){Wheeler}, {Ness}, {Buder}, {Bland-Hawthorn}, {Silva}, {Hayden}, {Kos}, {Lewis}, {Martell}, {Sharma}, {Simpson}, {Zucker}, \& {Zwitter}}]{Wheeler-20}
{Wheeler}, A., {Ness}, M., {Buder}, S., {et~al.} 2020, \apj, 898, 58, \dodoi{10.3847/1538-4357/ab9a46}

\bibitem[{{Woosley} \& {Weaver}(1995)}]{WW95}
{Woosley}, S.~E., \& {Weaver}, T.~A. 1995, ApJS, 101, 181, \dodoi{10.1086/192237}

\bibitem[{Yang(2014)}]{yang2014molecular}
Yang, Z. 2014, Molecular evolution: a statistical approach (Oxford University Press)

\bibitem[{Yaxley \& Foley(2019)}]{yaxley2019reconstructing}
Yaxley, K.~J., \& Foley, R.~A. 2019, Biological Journal of the Linnean Society, 128, 1021

\end{thebibliography}
\bibliographystyle{aasjournal}

\end{document}